\documentclass[12pt,aps,preprint,nofootinbib]{revtex4}

\usepackage{graphicx}
\usepackage{cancel}
\usepackage{amssymb}
\usepackage{textcomp}
\usepackage{amsmath}
\usepackage{bm}
\usepackage{times}
\usepackage{epsfig}
\usepackage{color}

\usepackage{comment}

\def\lsim{\mathrel{\raise.3ex\hbox{$<$\kern-.75em\lower1ex\hbox{$\sim$}}}}
\def\gsim{\mathrel{\raise.3ex\hbox{$>$\kern-.75em\lower1ex\hbox{$\sim$}}}}

\def\beq{\begin{equation}}
\def\eeq{\end{equation}}
\def\be{\begin{equation}}
\def\ee{\end{equation}}
\def\bea{\begin{eqnarray}}
\def\eea{\end{eqnarray}}

\def\etmiss{\cancel{E}_{T}}

\def\gev{\,{\rm GeV}}

\def\to{\rightarrow}

\newcommand{\minigraph}[5][0.25in]{\begin{minipage}{#2}\begin{center}\includegraphics[width=#2]{#5}\\\vspace{#3}\hspace{#1}{\footnotesize #4}\end{center}\end{minipage}}

\begin{document}
\preprint{~~PITT-PACC 1101}

\title{
Electroweakinos  in the Light of the Higgs Boson
}

\author{Tao Han$^{\bf a}$}
\email{than@pitt.edu}

\author{Sanjay Padhi$^{\bf b}$}
\email{Sanjay.Padhi@cern.ch}

\author{Shufang Su$^{\bf c}$}
\email{shufang@email.arizona.edu}

\affiliation{
$^{\bf a}$  Pittsburgh Particle physics, Astrophysics, and Cosmology Center, Department of Physics $\&$ Astronomy, University of Pittsburgh, 3941 O'Hara St., 
Pittsburgh, PA 15260, USA\\
$^{\bf b}$  Department of Physics, University of California--San Diego, 9500 Gilman Dr., La Jolla, CA 92093, USA\\
$^{\bf c}$  Department of Physics, University of Arizona, 1118 E. 4th St., Tucson, AZ 85721, USA
}

\begin{abstract}

Given the increasingly more stringent bounds on Supersymmetry (SUSY) from the LHC searches, 
we are motivated to explore the situation in which the only accessible SUSY states are the electroweakinos  (charginos and neutralinos). In the minimal SUSY framework, we systematically study the three general scenarios classified by the relative size of the gaugino mass parameters $M_{1},\ M_{2},$ and the Higgsino mass parameter $\mu$, with six distinctive cases, four of which would naturally result in a compressed spectrum of nearly degenerate LSPs. We present the relevant decay branching fractions and provide insightful understanding about the decay modes in connection with the Goldstone-boson Equivalence Theorem. We show the cross sections for electroweakino pair production at the LHC and ILC, and emphasize the unique signals involving the Standard Model-like Higgs boson as a new search reference. The electroweakino signal from pair production and subsequent decay to $Wh/Zh\ (h\to b\bar b)$ final state may yield a sensitivity of $95\%$ C.L.~exclusion (5$\sigma$ discovery) to the mass scale $M_{2},\ \mu \sim 350-400$ GeV ($220-270$ GeV) at the 14 TeV LHC with an luminosity of 300 fb$^{-1}$.
Combining with all the other decay channels, the $95\%$ C.L.~exclusion (5$\sigma$ discovery) may be extended to $M_{2},\ \mu \sim 480-700$ GeV ($320-500$ GeV). At the ILC, the electroweakinos could be readily discovered once the kinematical threshold is crossed, and their properties could be thoroughly studied. 

\end{abstract}

\maketitle

\section{Introduction}

The recent observations of a Standard Model (SM)-like Higgs boson ($h$) \cite{ATLASH,CMSH} have further strengthened the belief for a weakly coupled Higgs sector with Supersymmetry (SUSY)
% \cite{Nilles:1983ge} 
as the most compelling realization. 
If the weak-scale SUSY is realized in nature \cite{Nilles:1983ge}, 
the definitive confirmation will require the discovery of the supersymmetric partners of the electroweak (EW)
particles in the SM,  in particular the gauginos and Higgsinos,\footnote{We call the SUSY partners of the EW gauge bosons and the Higgs doublets the gauginos $(\tilde{B}, \tilde{W})$ and Higgsinos $(\tilde{H})$, respectively, the mass eigenstates the charginos $(\chi_{i}^{\pm})$ and neutralinos $(\chi_{i}^{0})$, and generically the electroweakinos (EWkinos) when no need for specification.} as recently stressed as the ``natural SUSY'' \cite{NSUSY0,NSUSY}.
The identification of the electroweak sector of the supersymmetric theory and the measurement of its parameters  
are especially important because it is commonly believed that the natural dark matter (DM) candidate, 
the ``Lightest Supersymmetric Particle''  (LSP),  resides in this sector, most likely the 
lightest neutralino \cite{Jungman:1995df}.

Given the current results on SUSY searches at the LHC \cite{ATLAS2013047,CMS2013012, ATLAS2013028, ATLAS2013035,ATLAS2013036, ATLAS2013049,ATLAS2013093,CMS2013006,CMS2013017} ,   the absence of the spectacular events of large 
hadronic activities plus substantial missing energy implies that 
new colored supersymmetric particles under QCD  strong interaction may not have been copiously produced. With some simple assumptions, the 
interpretation of the current LHC data leads to the mass bound for the gluino and light squarks as $m_{\tilde g}=m_{\tilde q}>1.8$ TeV, 
or $m_{\tilde g}> 1.3$ TeV with decoupled squark sector, $m_{\tilde q}>800$ GeV~\cite{ATLAS2013047, CMS2013012} with the other decoupled particles, based 
on the ATLAS/CMS analyses.    In anticipation of much heavier colored SUSY partners,  we are thus led to consider a more challenging search strategy, namely the SUSY signals only from the EW sector, the charginos and neutralinos.
On the other hand, the direct production of electroweak supersymmetric particles at the LHC suffers  from relatively small rates \cite{Baer:1994nr}. The current direct search bounds at the LHC are thus rather weak \cite{ATLAS2013028, ATLAS2013035,ATLAS2013036, ATLAS2013049,ATLAS2013093,CMS2013006, CMS2013017} and the future perspectives for the mass parameter coverage are limited \cite{ATLASTDR, CMSTDR}.
A further complication is that, some DM consideration favors a situation for nearly degenerate charginos and  neutralinos \cite{Arkani-Hamed:2006mb}, making their identification more challenging \cite{Giudice:2010wb}.

The deciding soft SUSY-breaking mass parameters for the Bino, Wino, and Higgsino are $M_1, M_{2}$ and $\mu$, respectively. Those parameters are related when adopting a specific SUSY-breaking mediation scenario, such as the minimal Super-gravity model  \cite{SUSYGUTs} and the minimal gauge-mediation \cite{GMSB}. 
Unfortunately, those minimal and predictive scenarios are disfavored by the current observation  of a 125 GeV SM-like Higgs boson \cite{MSUGRAbound, GMSBbound}.
In this work, we take a model-independent approach and study the SUSY signals with all possible relative values of these three SUSY-breaking mass parameters, which lead to six cases in the most general term. Among them, four cases would naturally result in a compressed spectrum of nearly degenerate LSPs.
We would like to address the question that to what extent in the parameter space, the SUSY signals only from the electroweakinos can be accessible. The answer to this question, in particular the accessible mass scale, is important not only for the current LHC experiments, but also for the planning of future collider programs. 

Given the intimate connection between the Higgs boson $(h)$ and the SUSY electroweak sector, it is evident that searching for SUSY may be greatly benefitted if one takes advantage of the existence of the Higgs boson.
The Higgs boson signal from SUSY cascade has been discussed via the heavy gluino and squark production  \cite{SUSYH} and via the electroweakino production \cite{Baer:2012ts}. More recently, ATLAS \cite{ATLAS2013093} and CMS \cite{CMS2013017} have also carried out some analysis for the $Wh$ final state, under the assumption that the decay of $\chi_2^0 \rightarrow \chi_1^0 h$ is $100\%$.  
Indeed, the Higgs boson often appears in one of the leading channels from neutralino and chargino decays  
$\chi_{2}^{0}\to \chi_{1}^{0} h, \ \chi_{2}^{\pm}\to \chi_{1}^{\pm} h$, and possibly $\chi_{3}^{0}\to \chi_{1,2}^{0} h$. By carefully exploring the by-now established channels from the decays of a 125 GeV Higgs boson $h\to b\bar b, WW^{*}, ZZ^{*}$, we find it promising to observe the robust electroweakino signals in the light of the Higgs boson. This is of critical importance: by constructing the Higgs boson in the complex events, one could confirm the existence of new physics beyond the SM associated with the Higgs sector. Overall, by exploiting the pair production of the electroweakinos via the Drell-Yan mechanism (DY) and their decays to the Higgs boson and to the leptons via $W^{\pm}/Z$, we may expect to reach up to an electroweakino mass about 700 GeV (500 GeV) for a $95\%$ C.L. exclusion (5$\sigma$ discovery), with 300 fb$^{-1}$ of integrated luminosity at the 14 TeV LHC. 

Our treatments are still conservative in two counts. First, we have not taken into account the possible contributions from the other electroweak states, namely the sleptons and the heavier Higgs bosons. Should the sleptons and the other Higgs bosons be light, comparable to or even lighter than the electroweakinos, they would be produced to enhance the signal both from their direct pair production and from the electroweakino decays.  
Also, we have not included the vector boson fusion (VBF) mechanism \cite{VBF} for the electroweakino production. The production rate for this mechanism is typically smaller than that of the DY processes by orders of magnitude depending on their masses. Its characteristics, however, for the forward-jet kinematics and the $t$-channel elecrtoweakino production may provide additional handles to complement the standard searches.

The rest of the paper is organized as follows. 
In Sec.~\ref{sec:model}, we present the electroweakino sector of the minimal Supersymmetric Standard Model (MSSM) and lay out the general scenarios for the relevant SUSY parameters in our study, resulting in six distinctive cases over all with respect to their mass relations. We outline their decay patterns and discuss in detail the decay branching fractions. Although as general as possible for the SUSY electroweak sector, we focus our attention to a situation where the colored SUSY states as well as the sleptons and other Higgs bosons are inaccessible at the LHC.
In Sec.~\ref{sec:pro_decay}, we first show the leading production channels of neutralinos and charginos at the LHC, and then explore the dominant final states from the decay of heavier electroweakino states. We show the cross sections at the 14 TeV LHC for all the six cases.  The results thus suggest the leading signals for the searches, particularly interesting of which is the SM-like Higgs boson in the final state. 
In Sec.~\ref{sec:analyses}, 
we first briefly summarize the relevant experimental bounds on the masses of the electroweakinos from the direct searches at the LEP2 and the LHC.
We then classify the signals according to their  
observable final states and emphasize the unique importance for taking advantage of the Higgs decay channels.  
We present the potential observability at the 14 TeV LHC in terms of our very general classification of the SUSY electroweak parameters. 
In Sec.~\ref{sec:ilc}, we discuss the dominant production modes of the electroweakinos at the International Linear Collider (ILC) and evaluate their production cross section at the 1 TeV C.M.~energy. We also comment on the physics potential for their property studies. 
Finally, we summarize our results in Sec.~\ref{sec:conclusion}.
Some approximate formulae for the NLSP decays are collected in an appendix. In particular, insightful understanding about the decay modes in connection with the Goldstone-boson Equivalence Theorem is provided.

%%%%%%%%%%%%%%%%%%%%%%%%%%%%%%%

\section{Model Parameters and Electroweakino Decays}
\label{sec:model}

\subsection{Model Specification}

We focus on the essential EW sector, namely, the electroweakinos. 
%gauginos and Higginos as the interaction eigenstates. 
Without assumptions for a SUSY-breaking mediation scenario, 
we consider the other SUSY particles, namely, gluinos, squarks and sleptons, to be inaccessible in the LHC searches. 
% ---------- old :
%Parametrically, we take 
%decouple from the spectrum, by taking 
%the soft SUSY breaking mass parameters to be \footnote{We do not need to keep track of the specific values, nor do we check the level of the fine-tune, as long as we assure the other heavy particles unobservable at the LHC, in accordance with our conservative treatment for the electroweakino sector. }
%\be
%M_3,\ M_{\tilde{f}}> 10~\text{TeV}, \quad A_{i} \simeq 0~\text{GeV}, \quad
%M_{A} > 1~\text{TeV},
%\ee
%where the heavy Higgs bosons governed by $M_{A}$ will also be decoupled from the theory. 
%
Parametrically, we set the gluino mass $M_{3}$, sfermon masses at multiple TeV and $A_{i} \simeq 0~\text{GeV}$,\footnote{We do not need to keep track of the specific values, nor do we check the level of the fine-tune, as long as we assure the other heavy particles unobservable at the LHC, in accordance with our conservative treatment for the electroweakino sector. } except for the third generation squarks mass parameters. 
Also,  we take $M_{A} \approx 1$ TeV, where the heavy Higgs bosons governed by $M_{A}$ will also be decoupled from the theory. 
We explicitly incorporate a SM-like Higgs boson of mass
\be
m_{h} = 125\ {\rm GeV},
\label{eq:mh}
\ee
which can be achieved by adjusting SUSY parameters in particular the stop 
mass parameters \cite{SMHiggs,MSUGRAbound, GMSBbound}. 
% \cite{SMHiggs} \cite{MSUGRAbound, GMSBbound}. 
For the gaugino and Higgsino sector, the mass matrix for the neutral components in the gauge-eigenstate basis of  
$\psi^0=(\tilde{B}, \tilde{W}^0, \tilde{H}_d^0, \tilde{H}_u^0)$ is
\begin{equation}
M_{\tilde{N}}=
\left(
\begin{array}{cccc}
M_1&0&-c_\beta s_W m_Z&s_\beta s_W m_Z \\
0&M_2&c_\beta c_W m_Z&-s_\beta c_W m_Z \\
-c_\beta s_W m_Z&c_\beta c_W m_Z&0&-\mu\\
s_\beta s_W m_Z&-s_\beta c_W m_Z&-\mu&0
\end{array}
\right),
\label{eq:mN}
\end{equation}
where we have used the abbreviations $s_W=\sin\theta_W, c_W=\cos\theta_W, s_\beta=\sin\beta$ and $c_\beta=\cos\beta$, for $\theta_W$ being the Weinberg angle and 
$\tan\beta=\langle H_u^0 \rangle /\langle H_d^0 \rangle$.  Similarly, the mass matrix of the charged components 
in the basis of $\psi^\pm=(\tilde{W}^+, \tilde{H}_u^+, \tilde{W}^-, \tilde{H}_d^-)$ is
\begin{equation}
M_{\tilde{C}}=
\left(
\begin{array}{cc}
0_{2 \times 2}&X^T_{2 \times 2} \\
X_{2 \times 2} &0_{2 \times 2}
\end{array}
\right),\ \ \ 
{\rm with} \ \ \ 
X_{2 \times 2}=
\left(
\begin{array}{cc}
M_2&\sqrt{2}s_\beta m_W \\
\sqrt{2} c_\beta m_W &\mu
\end{array}
\right).
\label{eq:mC}
\end{equation}
There are only four parameters involved in the mass matrices, two soft SUSY breaking mass parameters $M_1$ and $M_2$, 
the Higgs field mixing parameter $\mu$, and the electroweak symmetry breaking parameter $\tan\beta$. 
Diagonalization of the mass matrices gives the mass eigenstates (with increasing mass eigenvalues), namely, the Majorana fermions, neutralinos ${\chi}_i^0$ ($i=1 \ldots 4$),  and the Dirac fermions, charginos $ {\chi}_i^\pm$ ($i=1,2$).

The mixings among the gaugino states are induced by the electroweak symmetry breaking, as seen by the off-diagonal terms in Eqs.~(\ref{eq:mN}) and (\ref{eq:mC}). Relevant to our studies when $m_Z \ll |\mu \pm M_1|$ and $|\mu \pm M_2|$, the mixings between Bino (Wino) and Higgsinos are characteristically  suppressed by ${\cal O}(m_Z/|\mu \pm M_1|)$ (${\cal O}(m_Z/|\mu \pm M_2|)$). The mixings between Bino and Wino are further suppressed since they can only mix via Higgsino states.
Consequently, the four neutralinos are nearly a ``Bino-like",  a ``Wino-like", and a  ``Higgsino-like" pair 
$(\tilde{H}_d^0 \mp\tilde{H}_u^0)/\sqrt{2}$, with mass eigenvalues roughly $M_1$, $M_2$ and $\pm \mu$, respectively.  
In most of the parameter space under our consideration motivated by the current  lower bounds on the SUSY masses, 
this limit largely applies. The fundamental nature of the gauginos and Higgsinos prevails and the mixing effects are small. We can thus gain intuitive understanding about the behavior of production and decay patterns of the electroweakinos, as we will discuss in the following sections. 

For our phenomenological considerations, we work in the CP-conserving scenario and choose the usual sign convention $M_2>0$. Without assuming a unification scenario for the soft masses, $M_1$ and $\mu$ can still take $\pm$ sign. We adopt $M_{1}>0$\footnote{Flipping the sign of $M_1$ (or $M_2$) does not lead to a qualitatively different feature.} and consider both signs of $\mu$.  Note that $\mu>0$ is favored by muon $g-2$ consideration \cite{Baer:2001kn}.
In most of our discussion below, flipping the sign of $\mu$ does not lead to qualitatively different results.  We therefore use $\mu>0$ in most of the results presented below.  We will specify the cases in which the sign of $\mu$ matters, in particular, for Case AI and Case BI discussed below. 
We thus adopt the parameters in the broad range
\begin{equation}
 100\ {\rm GeV}  < M_{1},\ M_2,\ |\mu| < 1~{\rm TeV}, \quad 3 < \tan{\beta} < 50.
\label{mixi}
\end{equation}
While $M_2$ and $\mu$ are constrained to be above 100 GeV from the chargino searches at the LEP2 experiments \cite{LEPchargino1, LEPchargino2}.
$M_1$ could be much lower given the lack of model-independent limit on the Bino mass.   We note that our parameter choices are consistent with the current low energy bounds, 
 such as  the rare decay constraint from $b\to s \gamma$.  
 In a most general case, the mass parameters can be complex with CP-violating phases.  We do not consider such general CP-violating scenarios.   

%%%%%%%%%%%%%%%%%%%%%%%

\subsection{General Classification and the Electroweakino Decays}
\label{sec:cases}

To explore the phenomenological consequences in a most general approach, we present the three possible scenarios among the mass parameters of $M_{1},\ M_2,\ \mu$, and categorize them into six different cases. Each of those leads to characteristic phenomenology in their pair production and the decays of the electroweakinos. 
Since the sfermions are assumed to decouple, the heavier electroweakinos decay to the LSP ($\chi^0_1$) and a real or virtual electroweak gauge boson (generically denoted by $W,W^{*}$ or $Z,Z^{*}$, for either on-shell or off-shell) and a Higgs boson ($h$).   The decay via an off-shell Higgs boson is highly suppressed due to the small Yukawa couplings,  for modest values of $\tan\beta$. 
We will stress the situation when the Higgs boson plays a crucial role if kinematically accessible.  We have set $m_h=125$ GeV as stated in Eq.~(\ref{eq:mh}) 
throughout our numerical evaluations. 

%%%%%%%%%%%%%%%%%%%%%%%

 \begin{itemize}
{\bf \item{Scenario A:}  $M_{1} < M_{2},\ |\mu|$ }
\end{itemize}

This is the usual canonical scenario, which is strongly motivated by the Bino-like (LSP) dark matter \cite{Jungman:1995df} and by the grand unified theories with gaugino mass unification \cite{SUSYGUTs}. 
 There are two qualitatively different physics cases we would like to explore, namely
\bea
&& {\rm Case\ AI:}\quad M_{2} < |\mu|,\quad   {\chi}_1^\pm,  {\chi}_2^0 {\rm \  \ are\ Wino-like};\   {\chi}_2^\pm,  {\chi}_{3,4}^0 {\rm \ \ are\ Higgino-like}; \\
&& {\rm Case\ AII:}\quad |\mu| < M_{2},\quad   {\chi}_1^\pm,  {\chi}_{2,3}^0 {\rm \ \ are\ Higgino-like};\  {\chi}_2^\pm,  {\chi}_4^0 {\rm \  \ are\ Wino-like}.
\eea
For Case AI, the Winos are  lighter than Higgsinos, and thus are the next to the LSP (denoted by NLSPs), while for Case AII, it is the reverse and thus the Higgsino NLSPs.
Without losing much generality, for illustrative purposes in Sections II and III, we vary $M_{2}$ while fixing $|\mu|=1$ TeV for Case AI, and vary $\mu$ while fixing $M_2=$ 1 TeV for Case AII, along with $\tan\beta=10$. 
We will explore the characteristic differences for the observable signals in these two cases. Whenever appropriate, we will also
illustrate the features with different values of $\tan\beta$.  

\begin{figure}
\minigraph{8.1cm}{-0.2in}{(a)}{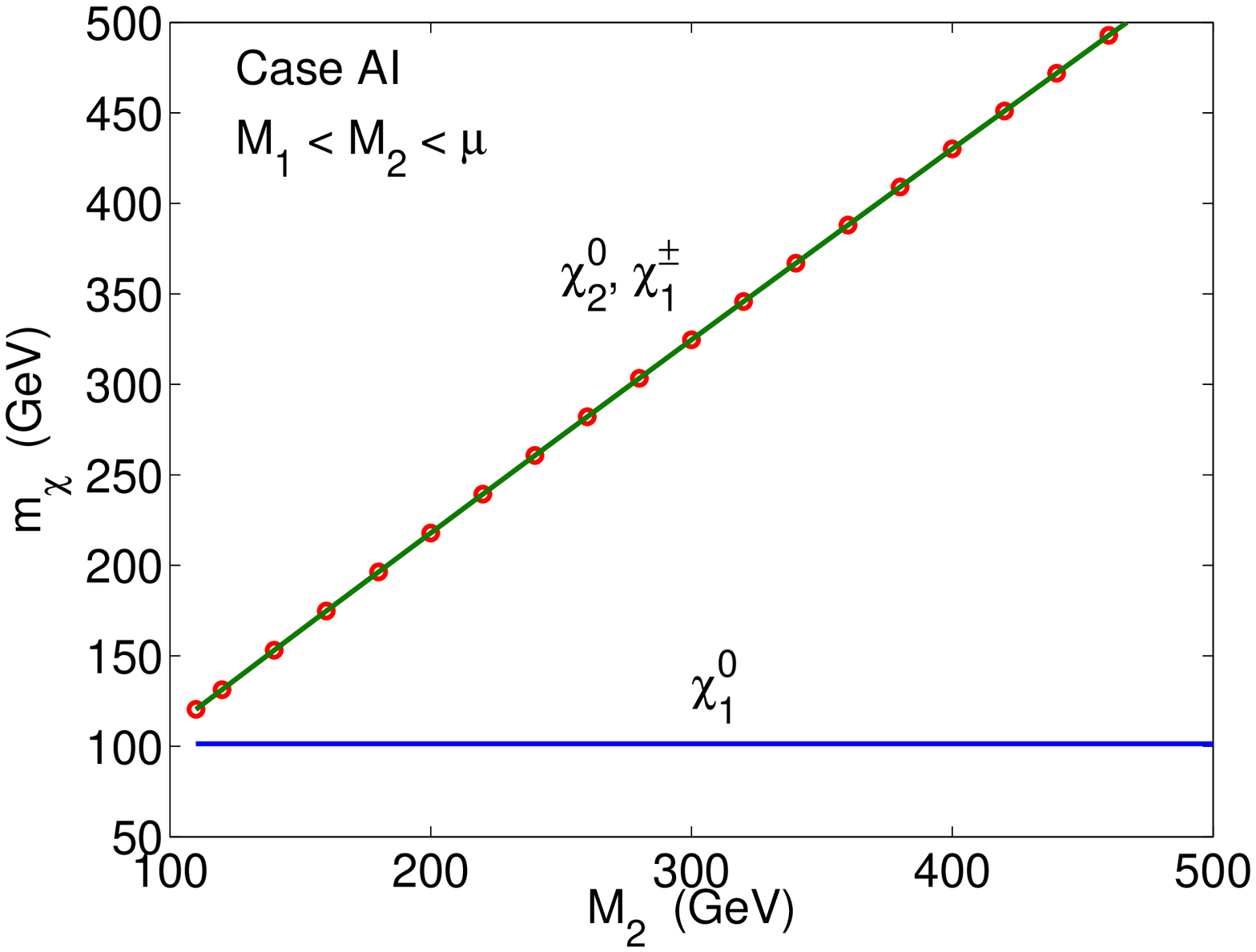}
\minigraph{8.1cm}{-0.2in}{(b)}{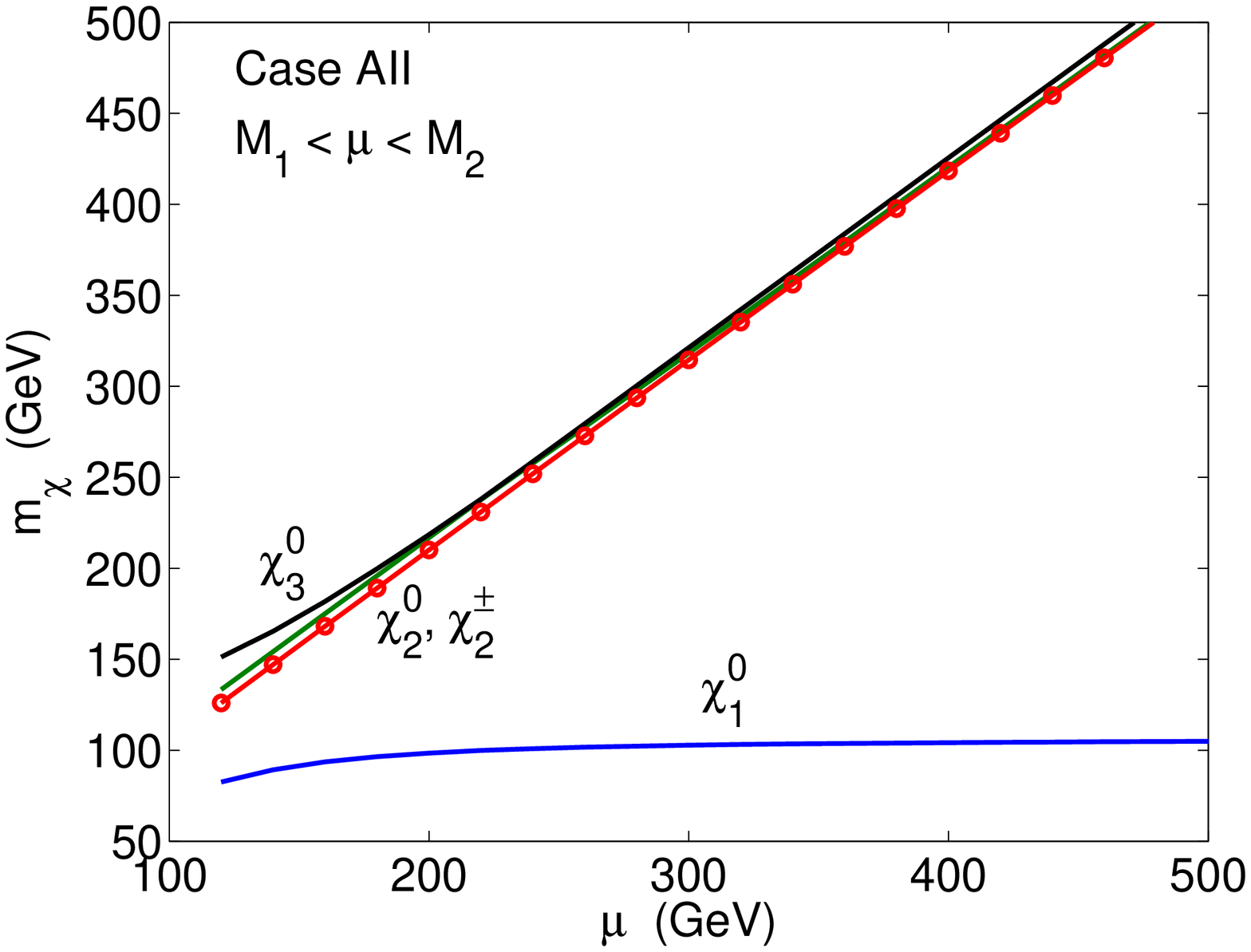}
\minigraph{8.1cm}{-0.2in}{(c)}{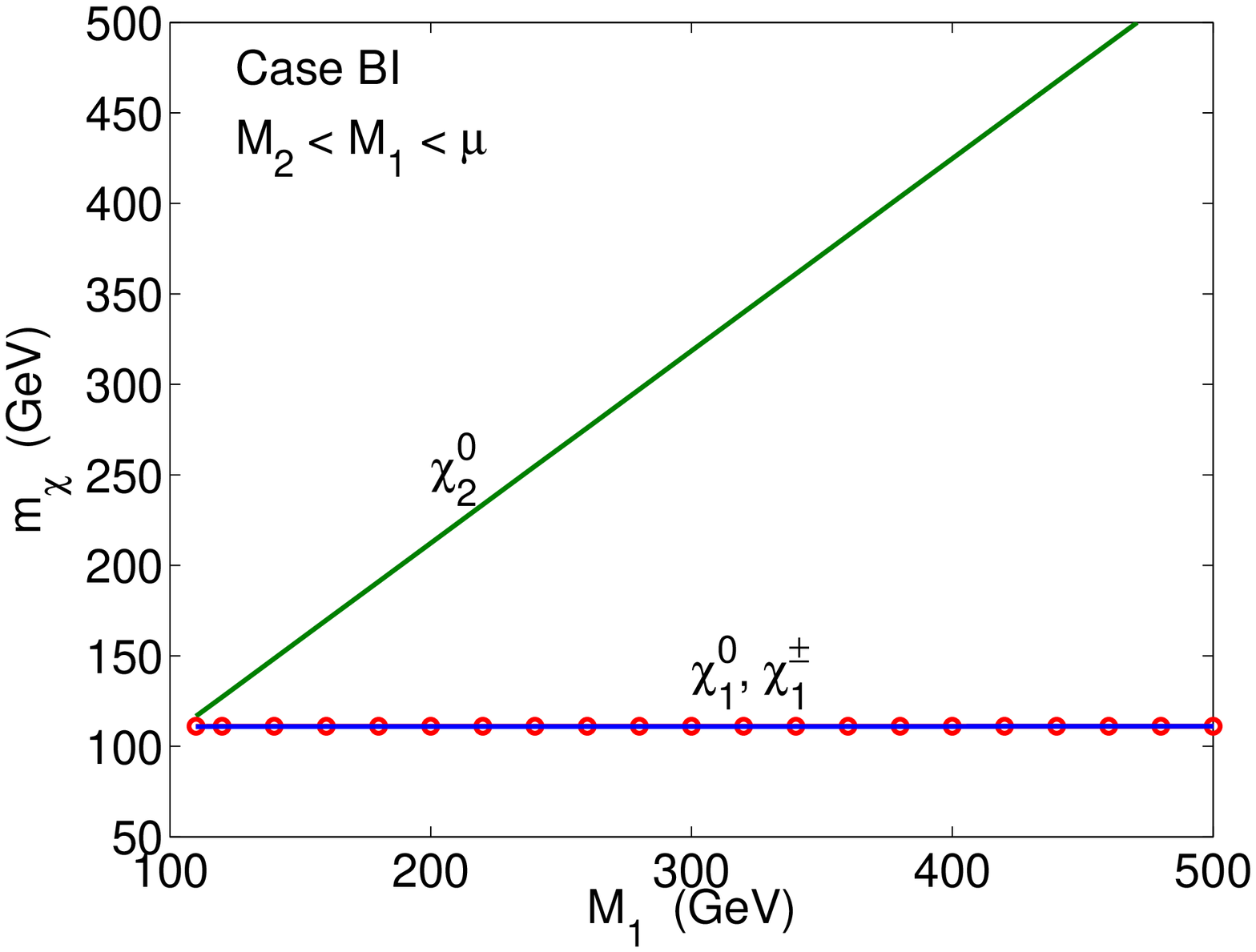}
\minigraph{8.1cm}{-0.2in}{(d)}{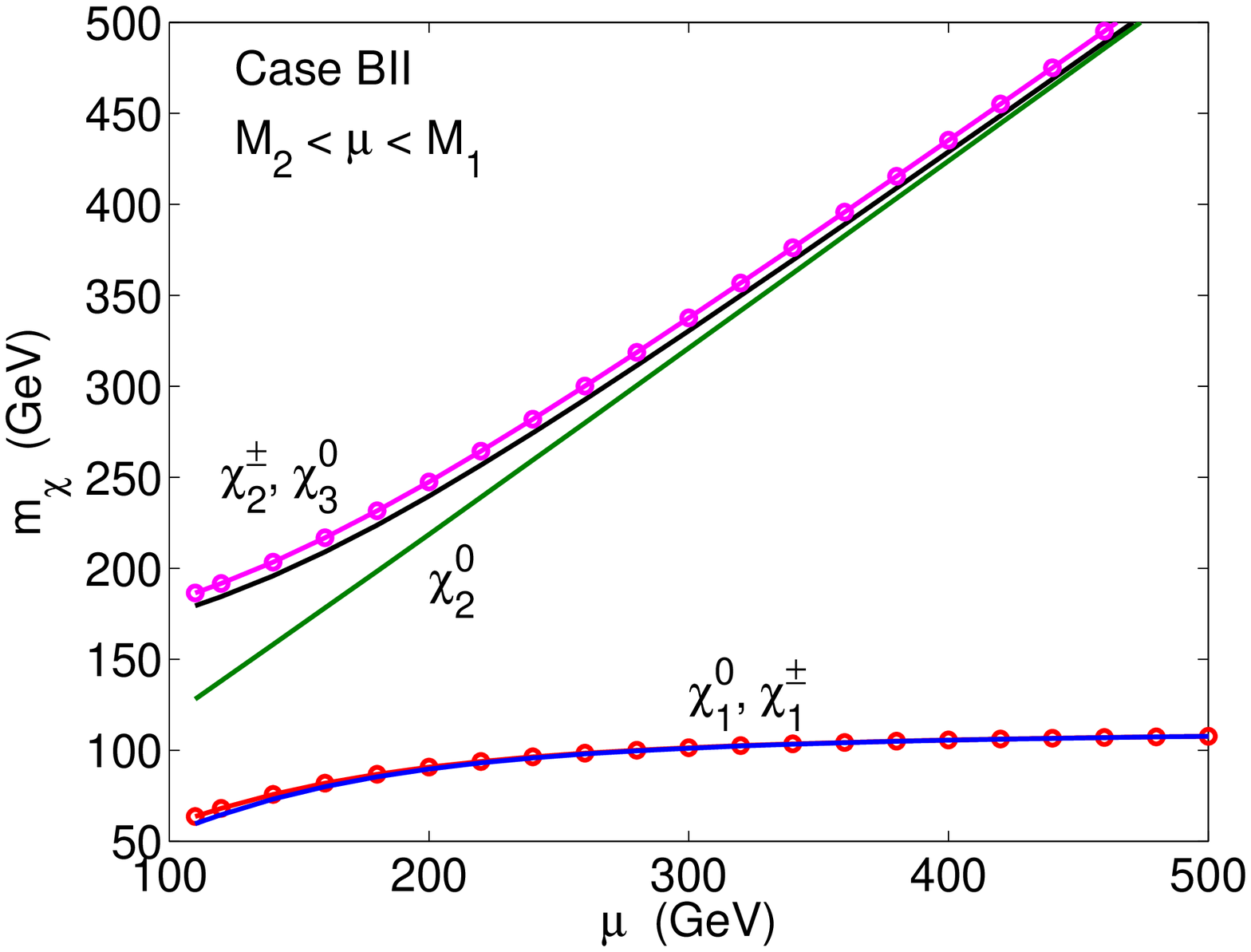}
\minigraph{8.1cm}{-0.2in}{(e)}{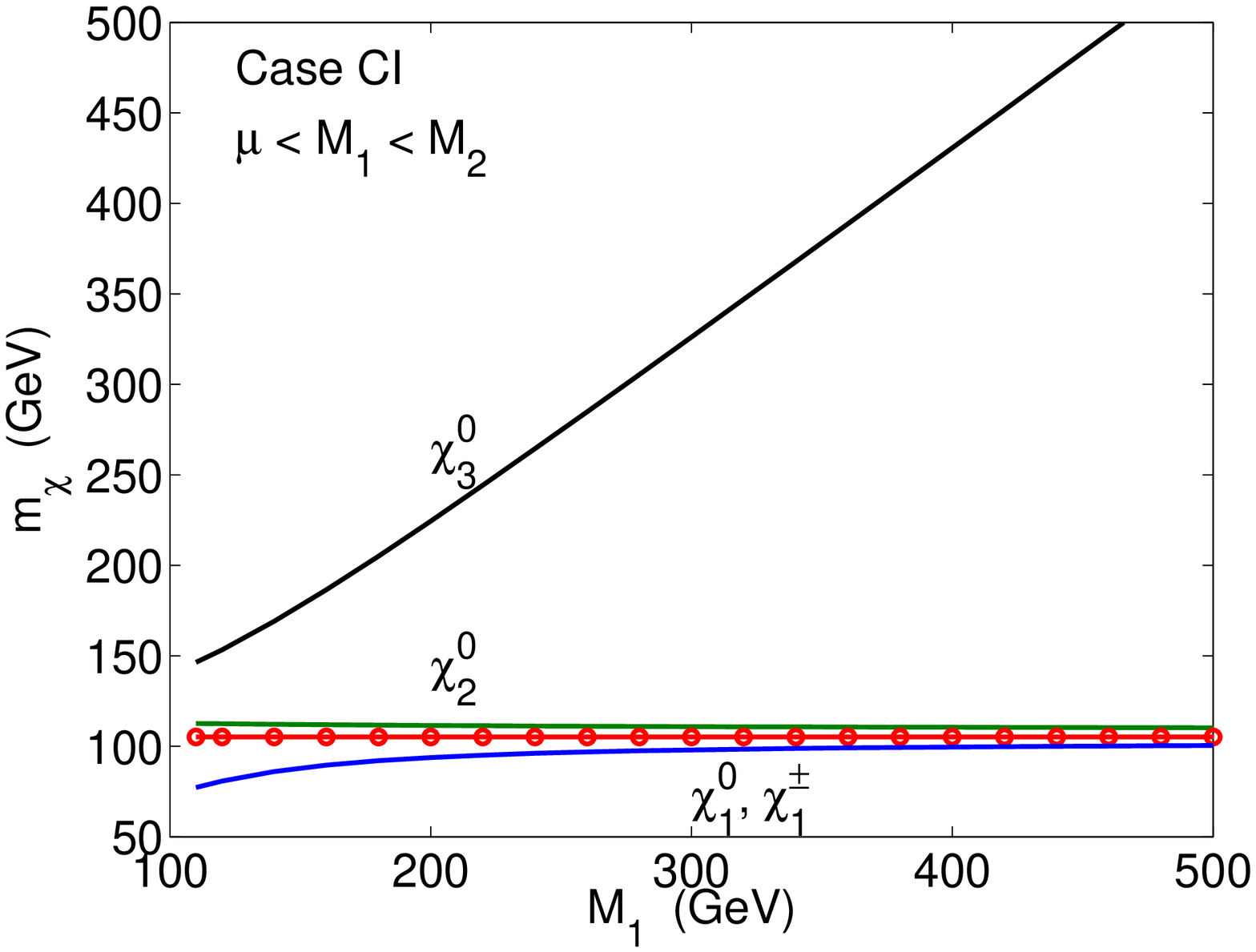}
\minigraph{8.1cm}{-0.2in}{(f)}{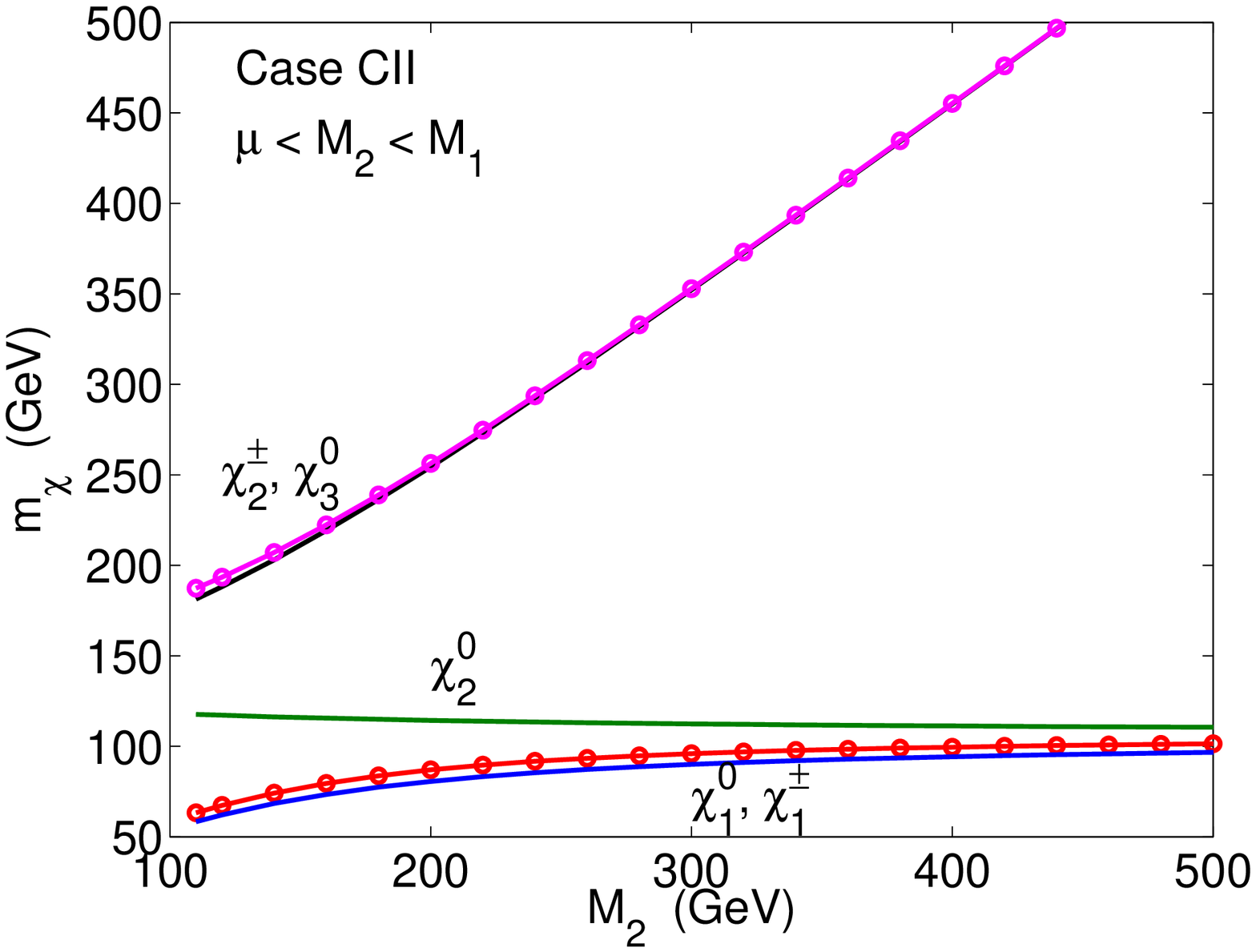}
\caption{ Lower-lying neutralino and chargino masses for the six cases: AI$-$CII. Solid curves are for neutralino states and those with circles are for chargino states.  The mass parameter for the LSP is set to be 100 GeV,  that for the heaviest gaugino or Higgsino is set to be 1 TeV, and $\tan\beta =10$.  
} 
\label{fig:masses}
\end{figure}

\begin{figure}
\includegraphics[scale=1.5,width=5 in]{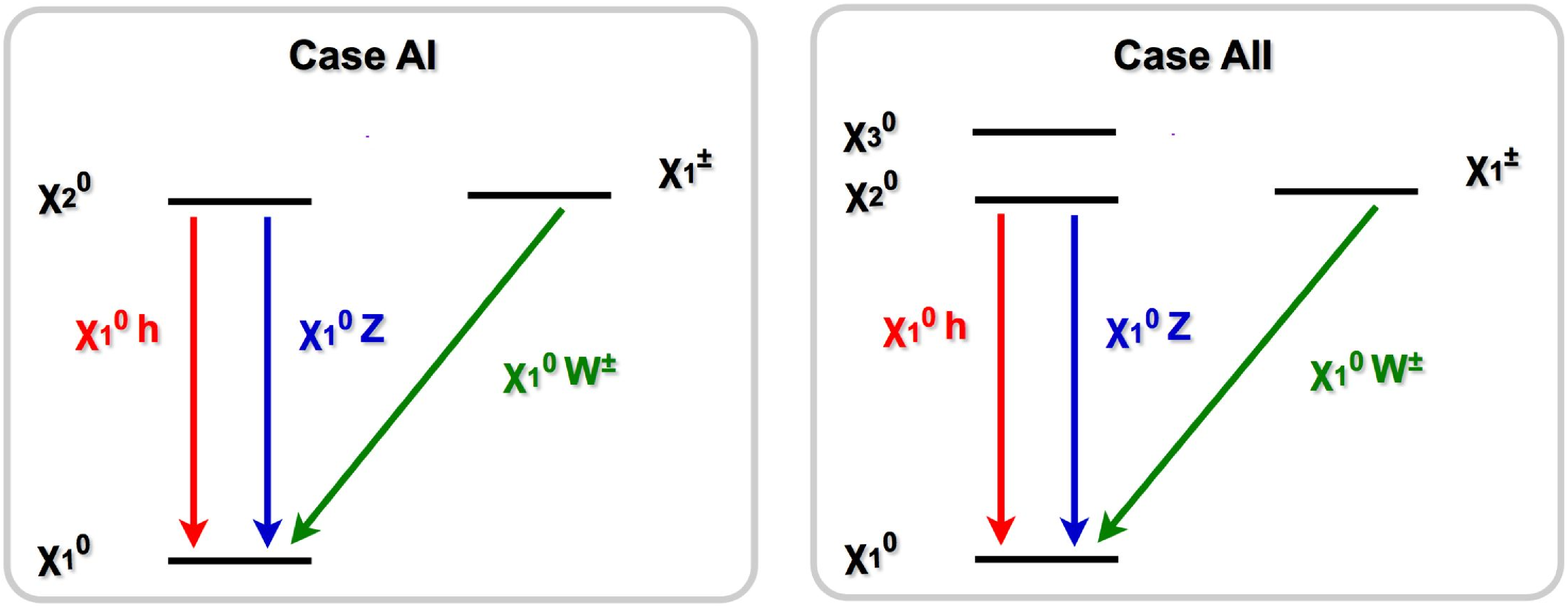}
\includegraphics[scale=1.5,width=5 in]{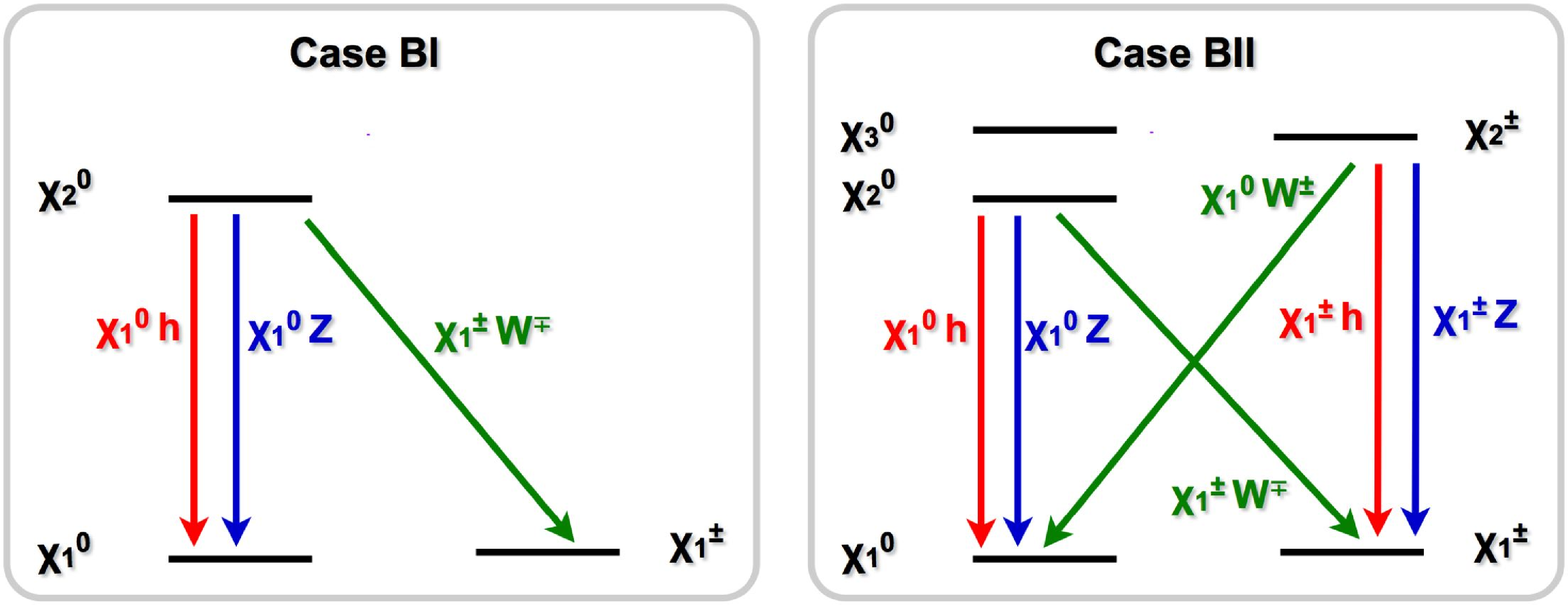}
\includegraphics[scale=1.5,width=5 in]{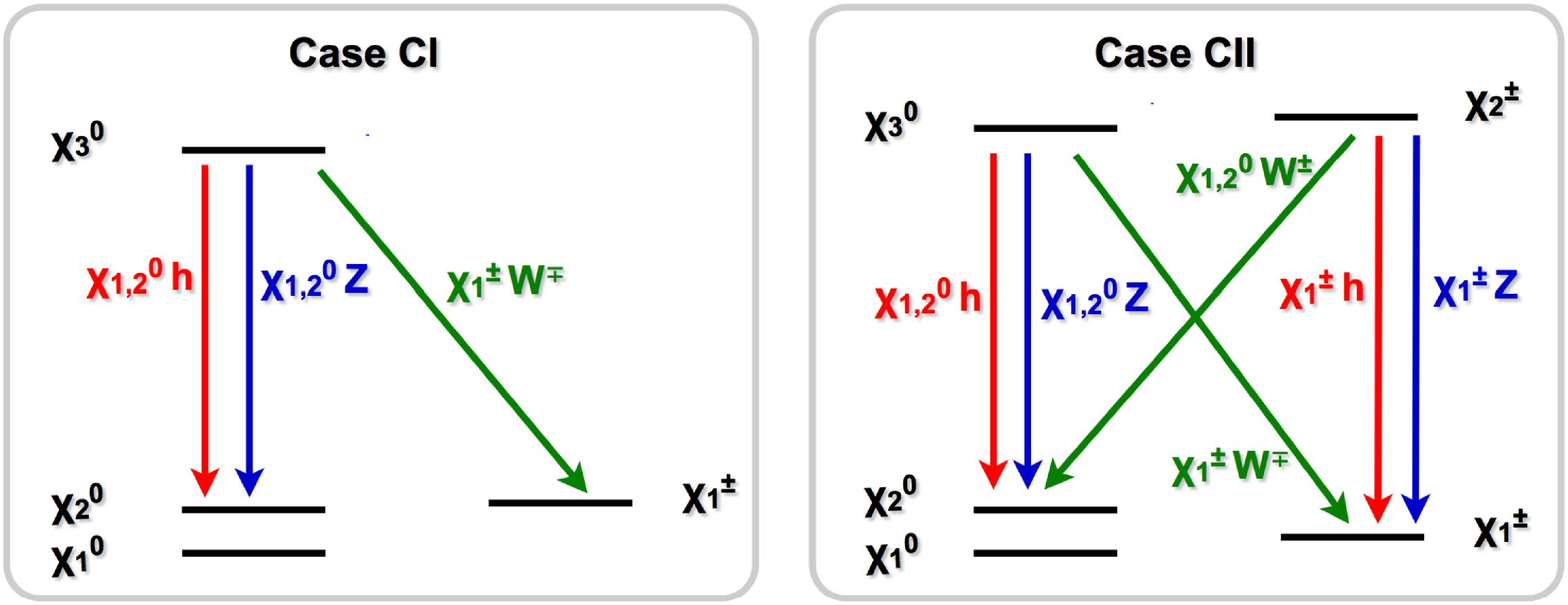}
\caption{Decay patterns of NLSP's for all the six cases AI$-$CII. } 
\label{fig:CaseABC_decay}
\end{figure}

In Fig.~\ref{fig:masses}, we present the physical masses of the lower lying neutralinos and charginos.
The mass spectrum, as well as decay branching fractions for neutralinos and charginos are calculated using 
SUSY-HIT 1.3 \cite{Djouadi:2006bz}.  
Figures \ref{fig:masses}(a) and (b) are for Case AI versus the mass parameters $M_{2}$  and for Case AII versus $\mu$   with $M_{1}=100$ GeV. The LSP, $ {\chi}_1^0$, is mostly Bino for both cases with mass close to $M_1$. 
The sub-leading mixing component in the LSP is at the order of ${\cal O}(m_{Z}/\mu)$ for the Higgsino component, and ${\cal O}(m_{Z}^{2}/\mu^{2})$ for the Wino component. 
The Higgsino component in Case AII, on the other hand, is less suppressed in particular at the smaller values of $\mu$, as shown in  Fig.~\ref{fig:masses}(b).  
For Case AI,  $ {\chi}_1^\pm$ and $ {\chi}_2^0$ are mostly Winos, with mass around $M_2$. The mass splitting between $\chi_2^0$ and $\chi_1^\pm$ is very small.
In fact, the nearly degeneracy of these states calls for a new convention to call them NLSPs altogether. The convenience will be seen more clearly later when 
discussing the decays.
 For Case AII, both the light chargino $ {\chi}_1^\pm$ and the second and the third neutralinos $ {\chi}_{2,3}^0$ are mostly Higgsinos, with mass around $|\mu|$.  
The mass splittings between those Higgsino-like states are small for $\mu$ larger than about 200 GeV. For small values of $\mu$ however, mass splittings as large 
as 20$-$30 GeV could occur, as seen in  Fig.~\ref{fig:masses}(b). These differences in masses gets smaller as $\mu$ increases, thus referred to as naturally compressed spectra \cite{Martin:2007gf}.
 In particular, this would lead to unsuppressed decays of $ {\chi}_3^0$ to ${\chi}_2^0/\chi_1^\pm$ in the 
small $\mu$ case. Heavier states, ${\chi}_2^\pm$ and ${\chi}_4^0$,  become out of reach.  

%%%%%%%%%%%%%%%%%%%%%

\begin{figure}[tb]
\minigraph{8.1cm}{-0.2in}{(a)}{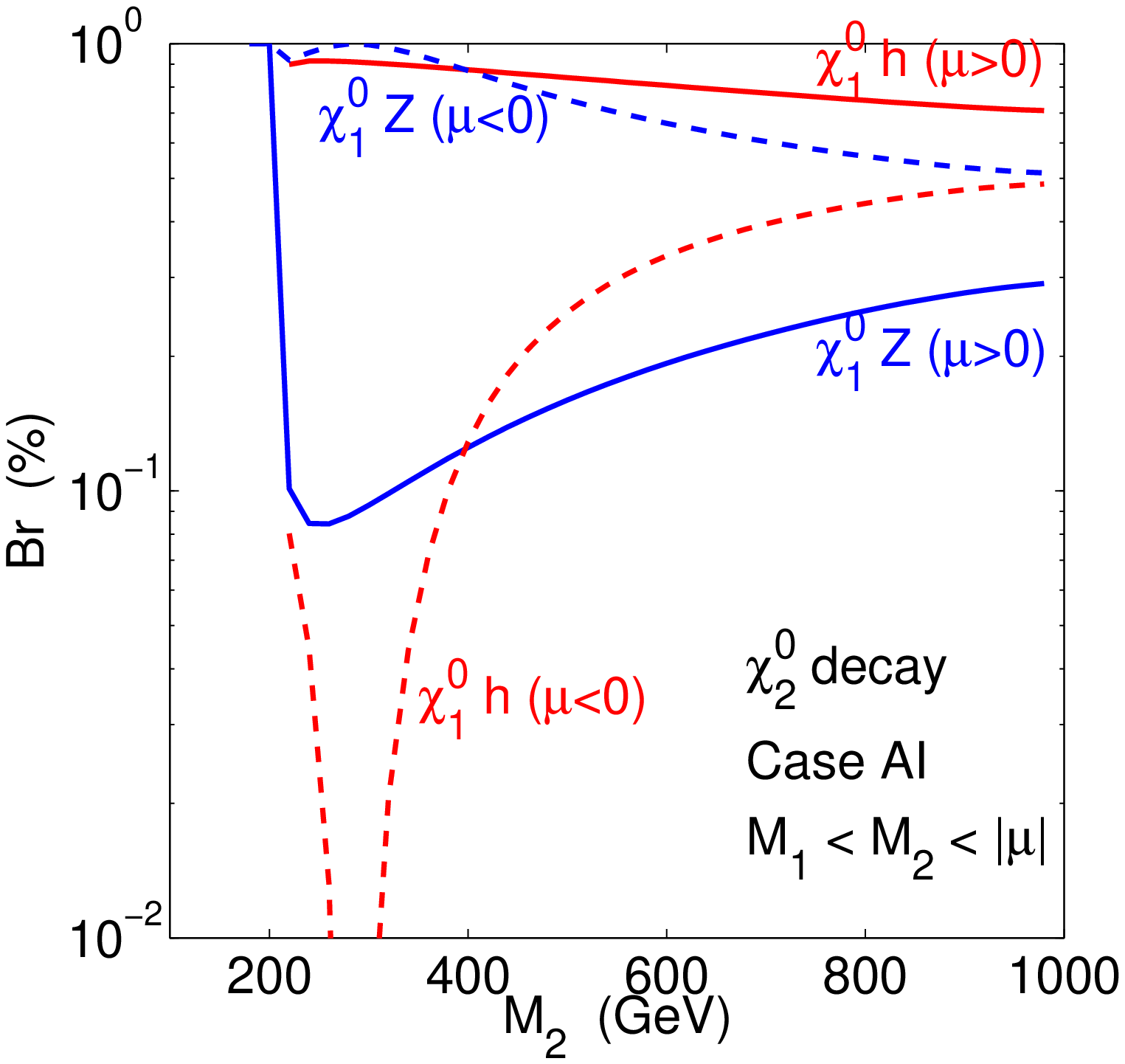}
\minigraph{8.1cm}{-0.2in}{(b)}{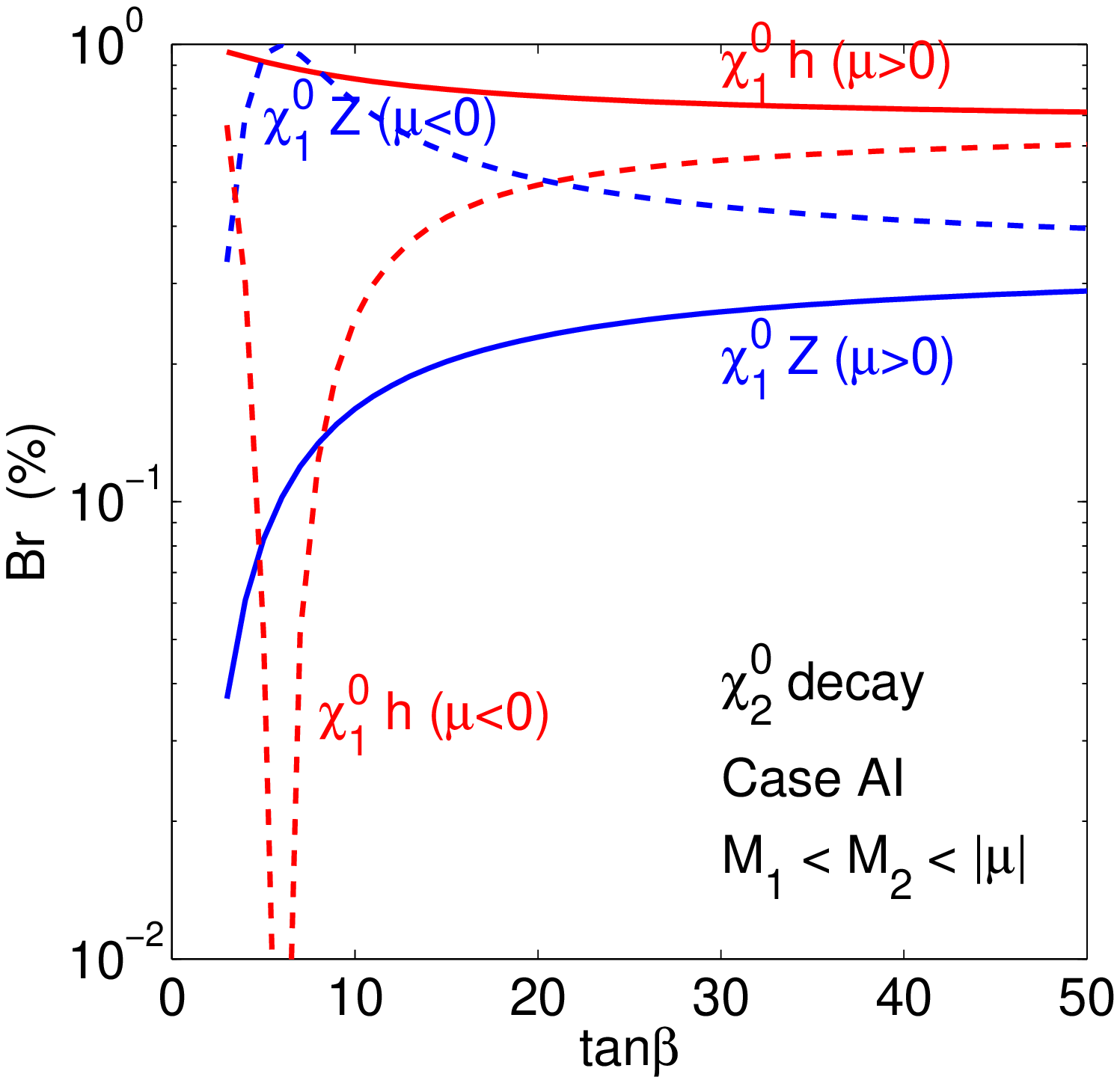}
 \caption{Case AI  with Wino-like NLSPs and Bino-like LSP: Decay branching fractions of   ${\chi}_2^0$ (a) versus $M_2$   for $\tan\beta=10$, and (b) versus $\tan\beta$ for $M_2=500$ GeV.  Two-body on-shell inclusive decays are labelled by 
 ${\chi}_1^0 Z$  and ${\chi}_1^0 h$.   Solid lines are for $\mu>0$ and dashed lines are for $\mu<0$.
Other parameters are set as $M_1=100$ GeV, $|\mu|=1$ TeV.    }
\label{fig:decays_winoNLSP}
\end{figure}

%%%%%%%%%%%%%%%%%%%%%%%%%%%%%%%%%%%%%%%%%%%%%%%

To a large extent, the electroweakino phenomenology is governed by the NLSP decays. We depict the NLSP decay patterns for all the six cases in Fig.~\ref{fig:CaseABC_decay}, and their corresponding  decay branching fractions in Figs.~\ref{fig:decays_winoNLSP}$-$\ref{fig:decays_M2NLSP_muLSP}. The partial width formulae are collected in the Appendix. 
The transitional decays among the degenerate Winos or Higgsinos NLSPs ({\it e.g.} ${\chi}^{0}_{2}\leftrightarrow {\chi}^{\pm}_{1}$) are almost always suppressed due to the small mass splitting among the multiplets. Dominant decay modes for NLSPs are always those directly down to the Bino-like LSP.  

For Cases AI and AII with Wino and Higgsino NLSPs, respectively, 
 the two-body decay of ${\chi}_1^\pm \to {\chi}_1^0 W$ dominates 
 leading to $f\bar{f}^\prime {\chi}_1^0$ of about a $100\%$ branching fraction.
 Leptonic and hadronic final states are essentially governed by the $W$ decay branching fractions to the SM fermions, namely about 67\% 
for ${\chi}^{0}_{1} qq^\prime$, and $11\%$ for ${\chi}^{0}_{1} \ell\nu_\ell$ for each lepton flavor. 

For ${\chi}_2^0$ decay in Case AI, there are two competing channels as  in shown in  Fig.~\ref{fig:CaseABC_decay}:
\beq
{\chi}_{2}^0 \rightarrow {\chi}_1^0 Z, ~{\chi}_1^0 h, 
\label{eq:AI_chi20}
\eeq
once both modes are kinematically open. Both partial decay widths are suppressed by a factor of ${\cal O}(m_Z^2/\mu^2)$ comparing to other cases discussed below (except Case BI), since such decays occur via the mixture of Higgsino states in either ${\chi}_1^0$ or ${\chi}_2^0$. 
The decay branching fractions are shown in Figs.~\ref{fig:decays_winoNLSP}  (a) versus $M_2$, 
and  (b) versus $\tan\beta$, respectively.    Solid lines are for $\mu>0$ and dashed lines are for $\mu<0$.
It is important to see that once ${\chi}_2^0 \rightarrow {\chi}_1^0 h$ channel is open, it quickly dominates when $\mu>0$: ${\rm Br}({\chi}_2^0 \rightarrow {\chi}_1^0 h)$ 
is about 82\% for $M_2=500$ GeV, while ${\rm Br}({\chi}_2^0 \rightarrow {\chi}_1^0 Z)$ is about 18\%. 
The branching fractions of $Z$ and $h$ modes are reversed for $\mu<0$, about 50$-$100\% for $\chi_1^0 Z$ and $\lesssim 50$\% for $\chi_1^0 h$ with $\tan\beta=10$.  The dependence on the sign of $\mu$ comes from  $(2 s_{2 \beta}+M_2/\mu)$ term in Eq.~(\ref{eq:AI_chi20h}).   In particular, the cancellation between these two terms for $\mu<0$ case leads to the dip in ${\rm Br}(\chi_2^0 \rightarrow \chi_1^0 h)$, as shown in Fig.~\ref{fig:decays_winoNLSP}.
For relatively large $\tan\beta$, the branching fractions for $\chi_1^0 h$ and $\chi_1^0 Z$ channels approach a constant.   While for small $\tan\beta$, the sign of $\mu$ have a large impact on the   branching fractions for the    $\chi_1^0 h$ and $\chi_1^0 Z$ channels, as shown in Fig.~\ref{fig:decays_winoNLSP}(b).

Below the threshold of the Higgs channel $M_2<M_1 + m_{h}$, the branching fractions for various  final states  follow the $Z$ decays to the SM fermions, about 55\% into light quarks, 15\% into $bb$, 20\% into neutrinos, and 3.3\% into each lepton flavor.  For  $M_2$ slightly above $M_{1}$, loop induced radiative decay ${\chi}_2^0\rightarrow {\chi}_1^0 \gamma$ reaches about $10\%$, while the final state photon will be very soft, making its identification difficult.   
The phase space suppression near the threshold for  ${\chi}_1^0 bb$ and ${\chi}_1^0 \tau\tau$ channels are also appreciable.   

\begin{figure}[tb]
\minigraph{8.1cm}{-0.2in}{(a)}{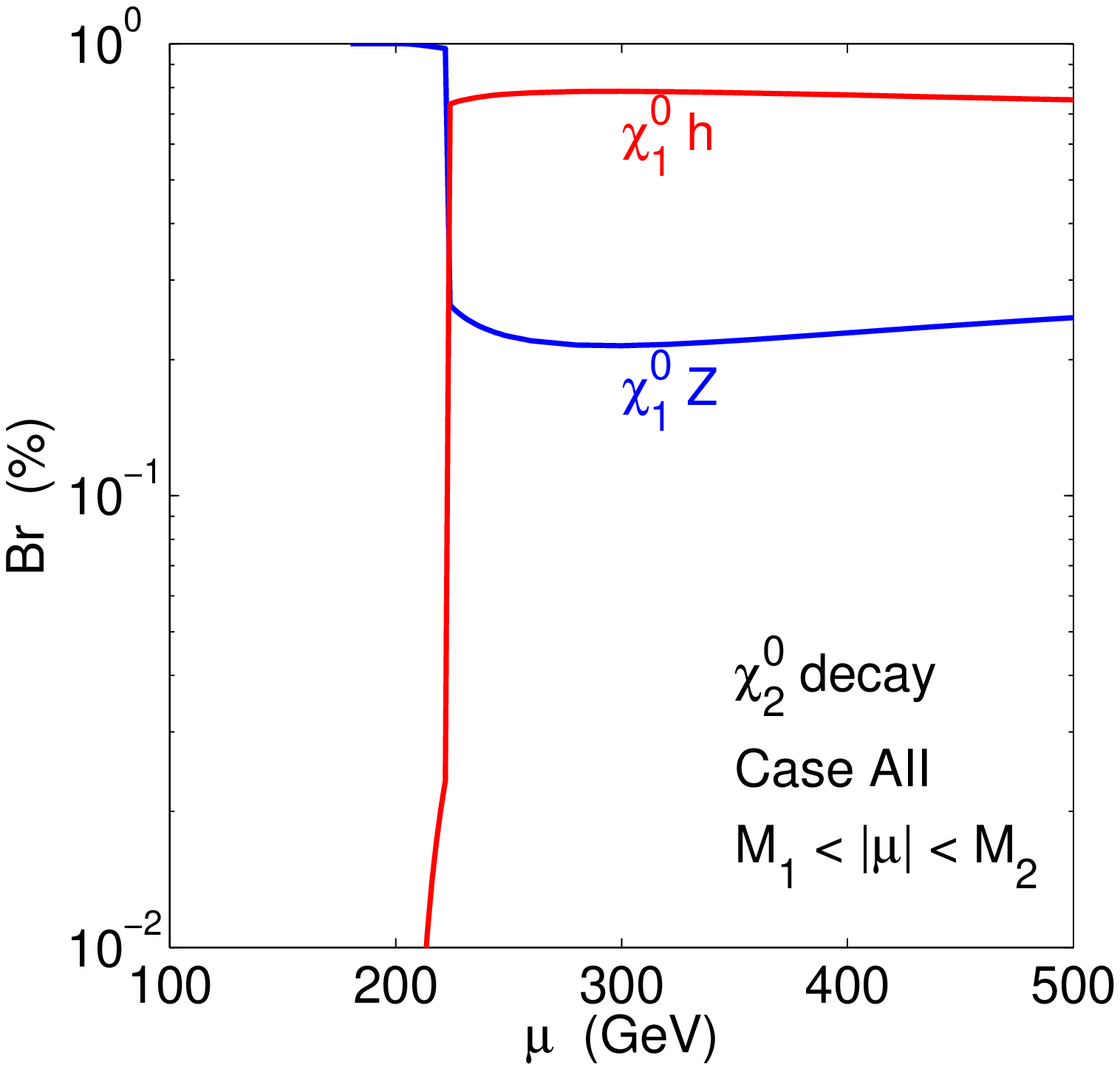}
\minigraph{8.1cm}{-0.2in}{(b)}{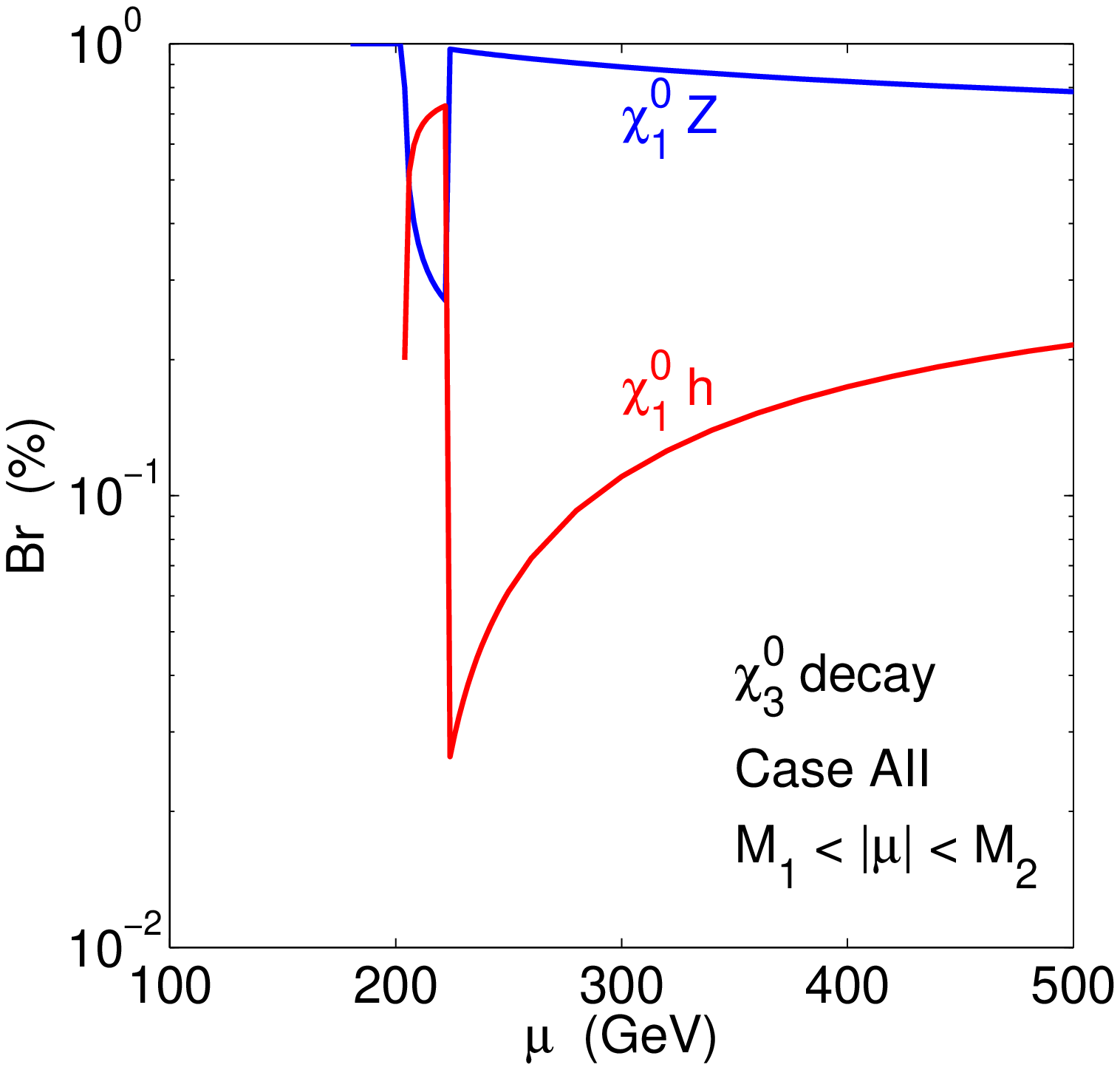}
  \caption{
Case AII with Higgsino-like NLSPs and Bino-like LSP: Decay branching fractions of (a) ${\chi}_2^0$, and 
(b) ${\chi}_3^0$ versus $\mu$, for $M_1=100$ GeV, $M_{2}=1$ TeV and $\tan\beta=10$.  }
\label{fig:decays_higgsinoNLSP}
\end{figure}

Figures \ref{fig:decays_higgsinoNLSP} show the decay branching fractions of  (a)  $\chi^{0}_{2}$   and (b) $\chi_3^0$, respectively, versus $\mu$ for  the Higgsino NLSP Case AII,  with $M_2$ fixed to be 1 TeV.   For $\mu \gtrsim 250$ GeV, the decay pattern for $\chi_2^0$ is   qualitatively similar to that of the light Wino Case AI with $\mu>0$.  Branching fraction of $\chi_2^0 \rightarrow \chi_1^0 h$ and $\chi_2^0 \rightarrow \chi_1^0 Z$ is about 75\% and 25\% for $\mu=500$ GeV, respectively.   The decays of $\chi_3^0$, however, are more preferable to $\chi_1^0 Z$.  The difference in the decay pattern of $\chi_2^0$ and $\chi_3^0$ is due to the different composition of $\chi_{2,3}^0$ as  $\frac{1}{\sqrt{2}} ({\tilde{H}}_d^0 \mp {\tilde{H}}_u^0)$. 
Note that in Fig.~\ref{fig:decays_higgsinoNLSP} the branching fraction of  $\chi_3^0 \rightarrow \chi_1^0 h$ shows a sudden drop around 230 GeV, coming from  the level crossing of the two Higgsino-like mass eigenstates.   
For $m_{\chi_{2,3}^0} - m_{\chi_1^0} < m_Z$, off-shell decay via $Z^*$ again dominates, with the branching fraction of fermion final states similar to that of $\chi_2^0$ in Case AI.

 In the limit of large $\tan\beta$ and  $|\mu\pm M_1| \gg m_Z$ such that all final states particles are effectively massless comparing to the parent particle,  ${\rm Br}(\chi_{2,3}^0 \rightarrow \chi_1^0 h) \approx {\rm Br}(\chi_{2,3}^0 \rightarrow \chi_1^0 Z) \approx $ 50\%.  While for $\tan\beta \rightarrow 1$, one of the $h$ or $Z$ channel is highly suppressed while the other channel is greatly enhanced since $ {\rm Br}(\chi_{2,3}^0 \rightarrow \chi_1^0 h):{\rm Br}(\chi_{2,3}^0 \rightarrow \chi_1^0 Z)\approx (s_\beta \pm c_\beta)^2:(s_\beta \mp c_\beta)^2$.  
 
Flipping the sign of $\mu$ also lead to the reversal of branching fractions into $h$ and $Z$ modes for  large $\tan\beta$.  However, since $\chi_2^0$ and $\chi_3^0$ are either pair produced  at colliders as $\chi_2^0\chi_3^0$ or they are produced in associated with $\chi_1^\pm$ with similar cross sections at the LHC, changing the sign of $\mu$ has little impact on the overall cross sections of the  observed final states.

For small $|\mu\pm M_1 | \sim m_Z$, the mass splittings between the Higgsino multiplets $\chi_3^0$ and $\chi_2^0/\chi_1^\pm$ could reach 20 $-$ 30 GeV. Although not shown in the figures, there are leading decay modes between Higgsino states:
\bea
\chi^{0}_{3}\to \chi^{\pm}_{1} W^{*},\ \chi^{0}_{2} Z^{*}.
\label{eq:chi3}
\eea
 Even with the phase space suppression comparing to the decay of $\chi_3^0$ directly down to $\chi_1^0$,  the branching fractions for $\chi_3^0 \rightarrow \chi_1^\pm W^*$ could dominate over $\chi_3^0 \rightarrow \chi_1^0 Z^*$ since the coupling $\chi_3^0 \chi_1^\pm W$ is unsuppressed, while $\chi_3^0 \chi_1^0 Z$ suffers from Bino-Higgsino mixing.  
 It should be noted, however, that the decay products will be very soft due to the small mass difference, so that it renders the experimental observation difficult at hadron colliders. At an ILC, however, the clean experimental environment may allow the observation of those decay modes.   
 
%%%%%%%%%%%%%%%%%%%%%%%%

\begin{itemize}
\item{\bf Scenario B:} $M_{2} < M_{1},\ |\mu|$
\end{itemize}

This is the situation of Wino LSP, as often realized in anomaly mediated SUSY breaking scenarios \cite{amsb}.
The lightest states $\chi_1^0$ and $\chi_1^\pm$ are nearly degenerate in mass close to $M_{2}$. It thus makes more sense to follow the newly introduced convention to call them all  ``LSPs''.\footnote{Note that in the usual convention, the neutral Wino $\chi_1^0$ is called the LSP and the charged Wino $\chi_1^\pm$ is called the NLSP.}
In this scenario, there are two possible mass relations we will explore
\bea
&& {\rm Case\ BI:}\quad M_{1} < |\mu|,\quad 
\ \chi_2^0 {\rm \  \ Bino-like}; \ \chi_2^\pm,\ \chi_{3,4}^0 {\rm \  \ Higgsino-like}; \\
&& {\rm Case\ BII:}\quad |\mu| < M_{1}, \quad  
 \ \chi_2^\pm,\ \chi_{2,3}^0 {\rm \  \ Higgsino-like}; \
 \chi_{4}^0 {\rm \ \  Bino-like}.~~
\eea
 
In Figs.~\ref{fig:masses}(c) and (d), we present the physical masses of the lower-lying neutralinos and charginos with $M_{2}=100$ GeV, for Case BI versus the mass parameters $M_{1}$ while fixing $\mu=1$ TeV;  and for Case BII versus $\mu$ while fixing $M_{1}=1$ TeV. Similar to Scenario A, there is almost no mixing in Wino- and Bino-like states for large $\mu$ as in Case AI. The Bino-like $\chi_2^0$ is NLSP, and the Higgsinos are heavy and decoupled. In Case BII on the other hand, a large mixing could occur between Wino- and Higgsino-like states when $\mu$ is relatively small, less than 200 GeV. Above that, the Higgsinos group together as the NLSPs. 

%%%%%%%%%%%%%%%
\begin{figure}[tb]
\minigraph{8.1cm}{-0.2in}{(a)}{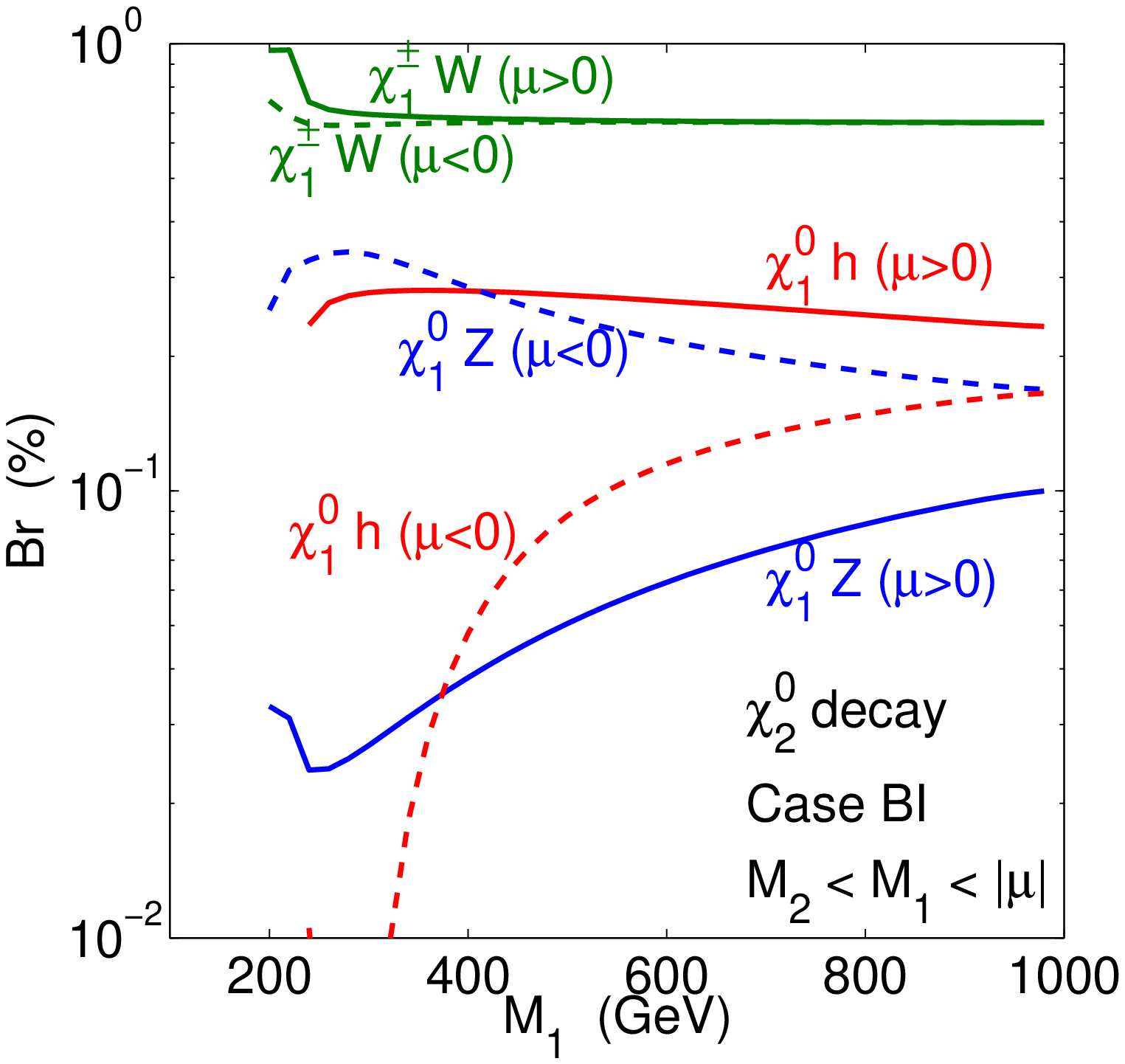}
\minigraph{8.1cm}{-0.2in}{(b)}{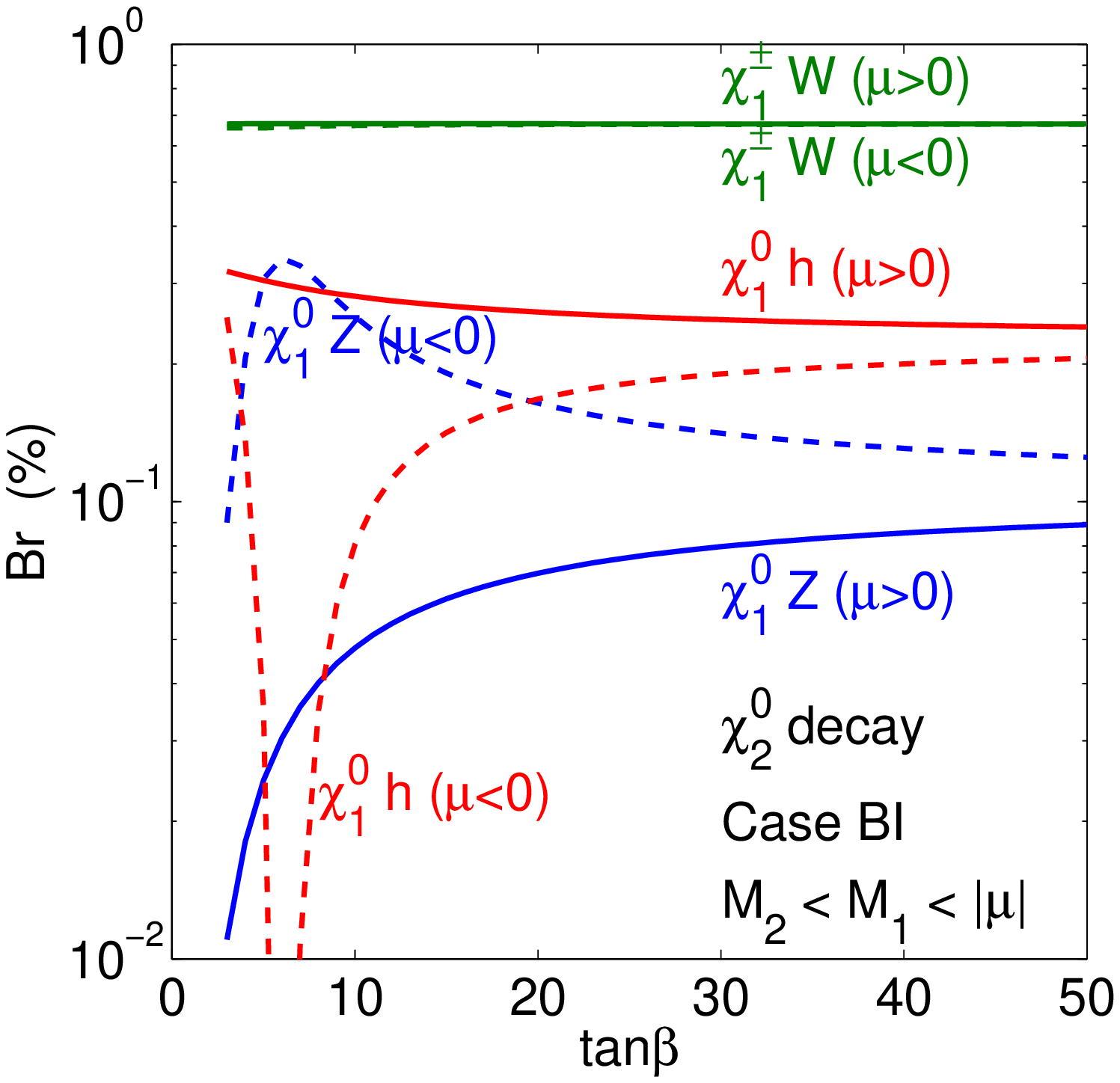}
\caption{Case BI  with Bino-like NLSP and Wino-like LSPs: Decay branching fractions of   ${\chi}_2^0$ (a) versus $M_1$  for $\tan\beta=10$, and (b) versus $\tan\beta$  for $M_1=500$ GeV.    Solid lines are for $\mu>0$ and dashed lines are for $\mu<0$.  Other parameters are set as $M_2=100$ GeV, $|\mu|=1$ TeV.
 }\label{fig:decays_M1NLSP_M2LSP}
\end{figure}

Fig.~\ref{fig:CaseABC_decay}(c)  presents the decay patterns of the NLSP $\chi_2^0$ in Case BI, and their corresponding  decay branching fractions are shown in Fig.~\ref{fig:decays_M1NLSP_M2LSP}.  The leading decay modes are 
\beq
\chi_{2}^0 \rightarrow \chi_1^\pm W^\mp, \chi_1^0 Z, ~\chi_1^0 h.
\label{eq:BI_chi20}
\eeq
The partial decay widths for those channels are suppressed by ${\cal O}(m_Z^2/\mu^2)$, similar to Case AI, as the decay occurs via the   Bino/Wino-Higgsino mixing.  

The decay branching fractions for  $\chi_2^0$ in Case BI are shown in Fig.~\ref{fig:decays_M1NLSP_M2LSP}(a) versus $M_1$ and (b) versus $\tan\beta$.     Under the limit of $M_1 - M_2 \gg m_Z$, $|\mu \pm M_{1,2}| \gg m_Z$, and large $\tan\beta$,  the partial decay widths to various final states in Case BI satisfy the approximate relations
\beq
\Gamma_{\chi_1^+W^-}=\Gamma_{\chi_1^-W^+}\approx\Gamma_{\chi_1^0Z}+\Gamma_{\chi_1^0h}.
\eeq
For $\mu=500$ GeV, the branching fraction of $\chi_2^0$ is 68\%, 27\%, and 5\% for $W$, $h$ and $Z$ channels, respectively.  
It is interesting to note that $\chi_2^0$ is more likely to decay into $h$ than to $Z$ for $\mu>0$ and more likely to decay to $Z$ than to $h$ for $\mu<0$ at small $\tan\beta$.   The effect of the sign of $\mu$ can be explained using the approximate formulae   Eq.~(\ref{eq:BI_chi20h}) in the Appendix.  The decay branching fraction to $W^\pm$, on the other hand, depends little on the sign of $\mu$. 
 
%%%%%%%%%%%%%%%%%%%%%%%

\begin{figure}[tb]
\minigraph{8.1cm}{-0.2in}{(a)}{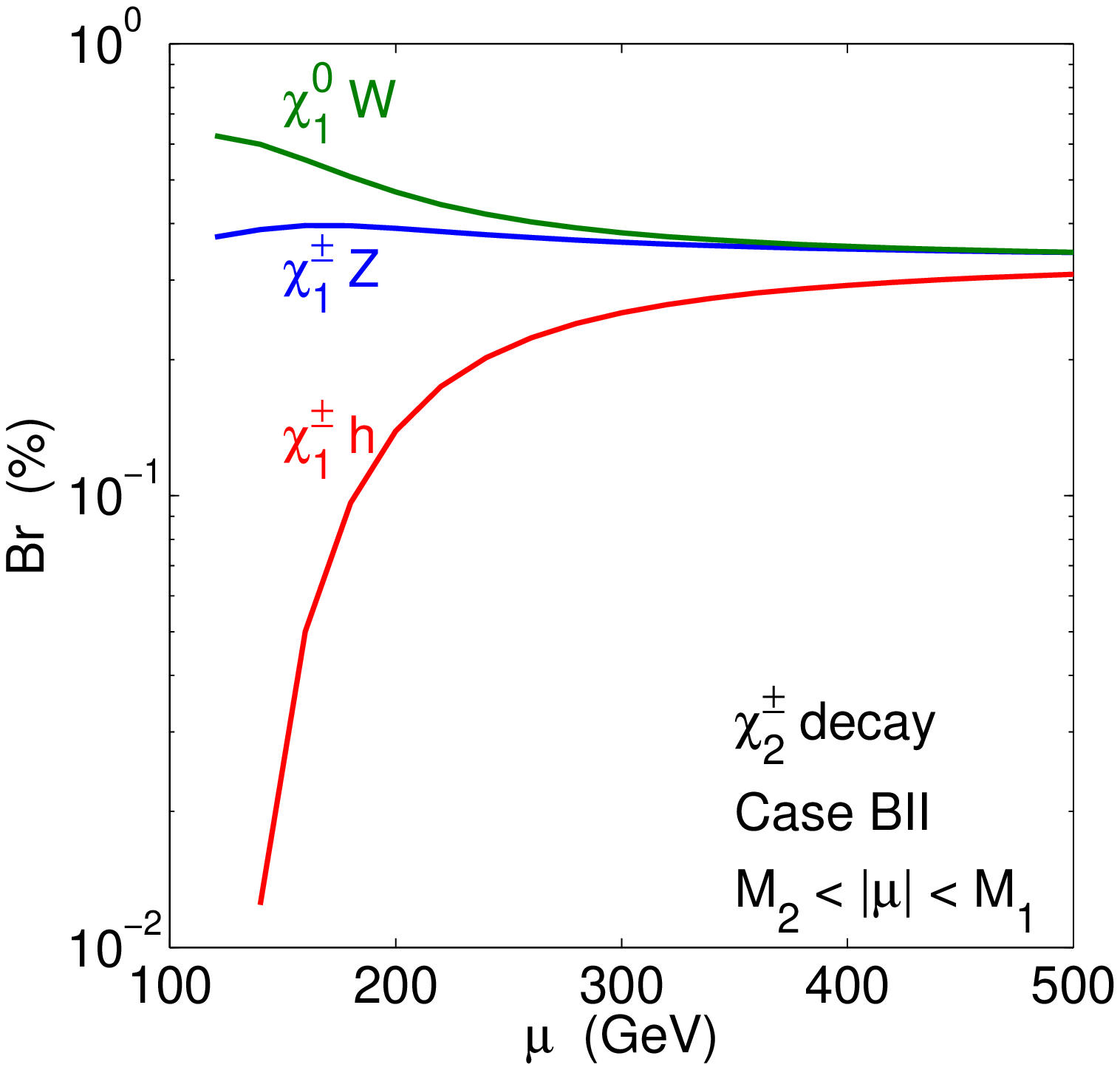}
\minigraph{8.1cm}{-0.2in}{(b)}{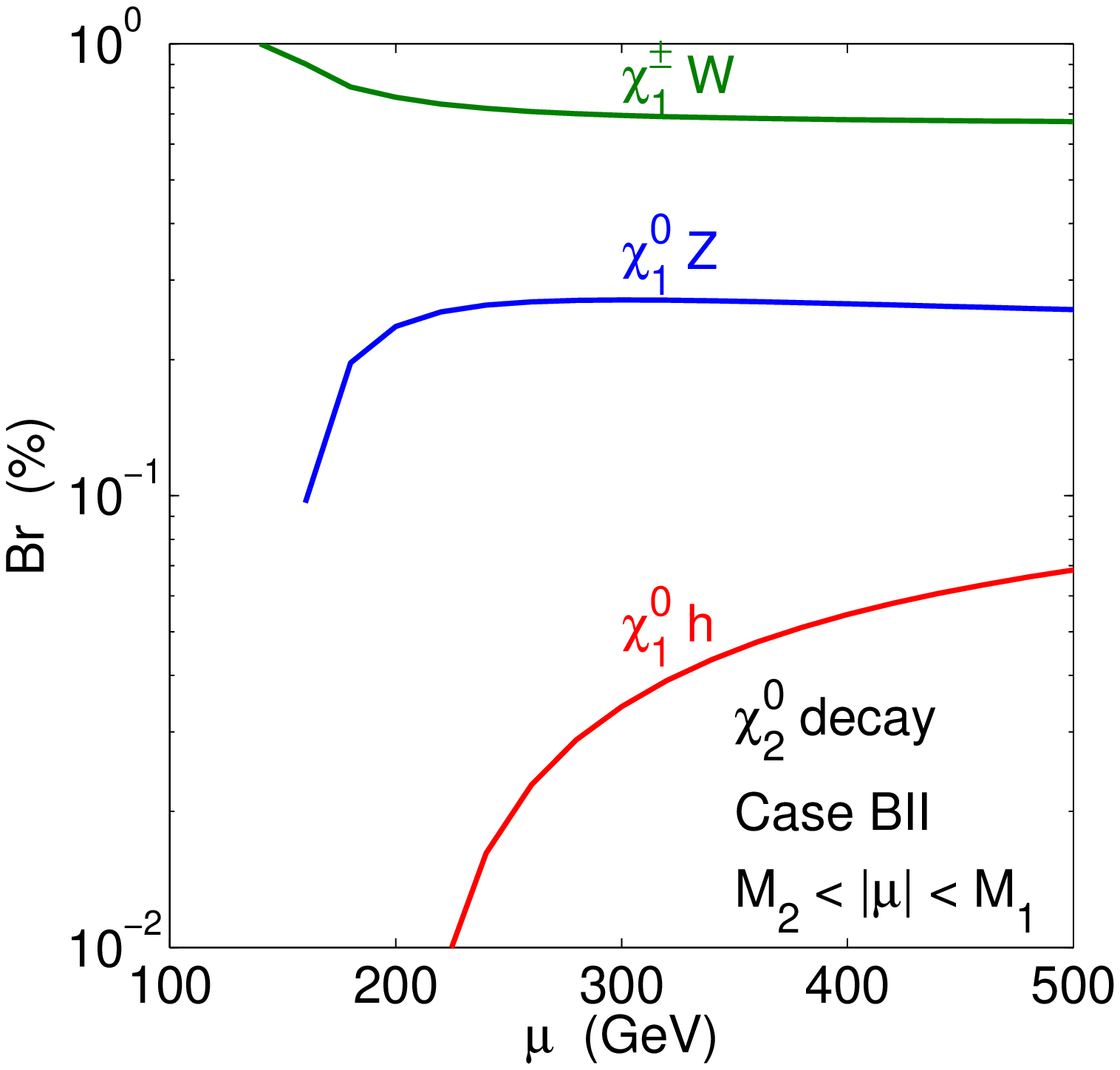}
\minigraph{8.1cm}{-0.2in}{(c)}{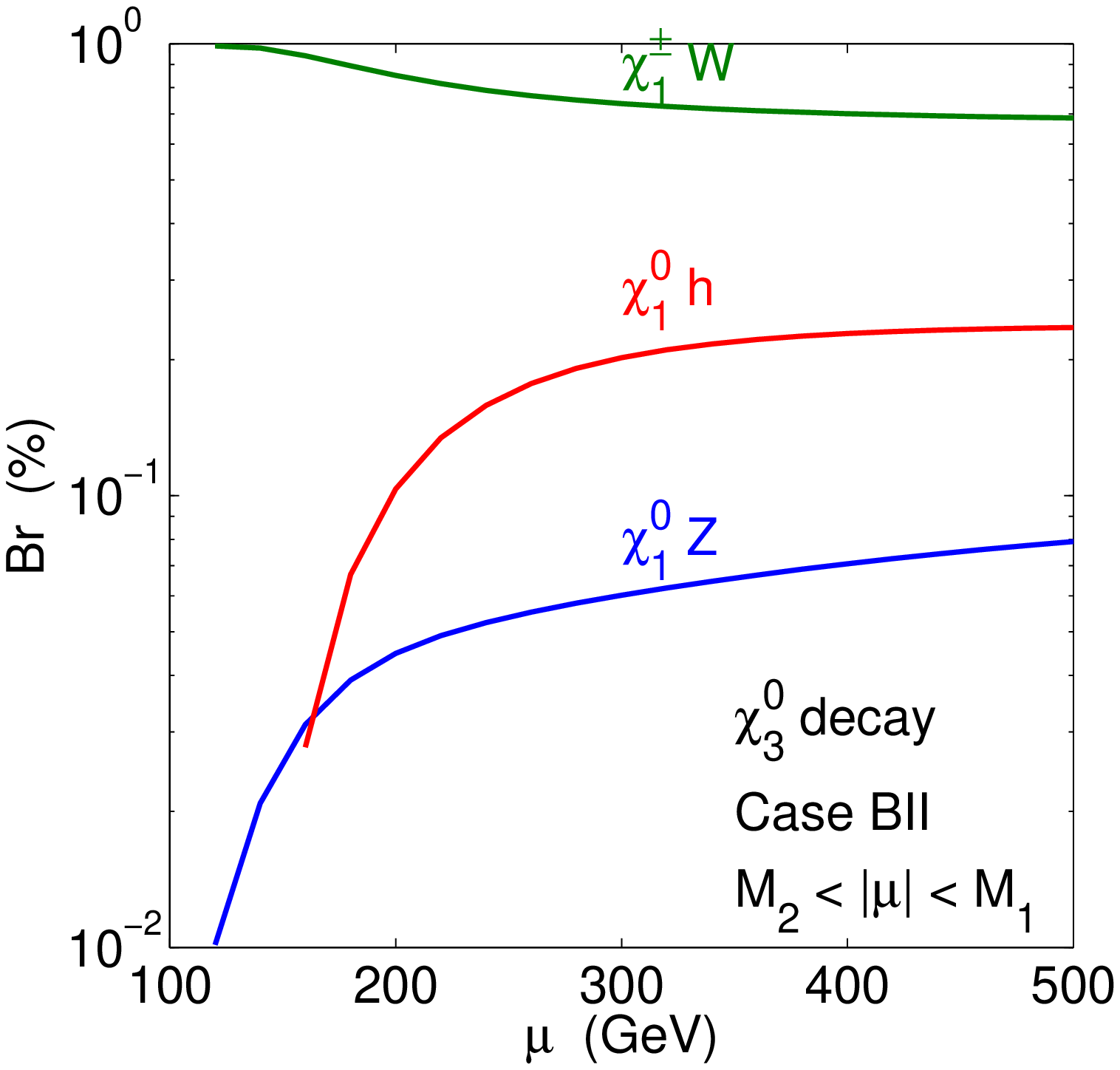}
\minigraph{8.1cm}{-0.2in}{(d)}{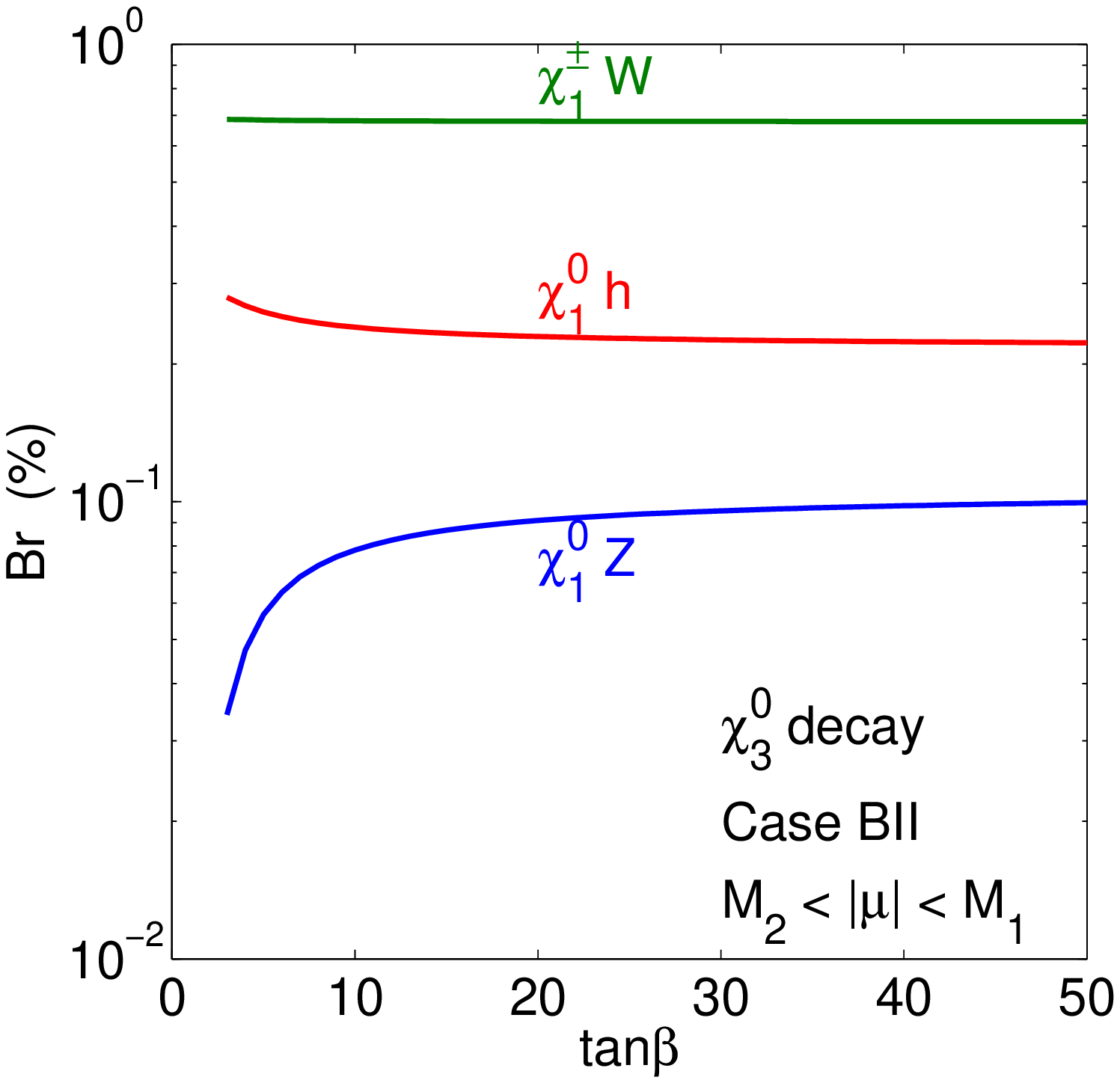}
 \caption{Case BII Higgsino NLSPs and Wino LSPs:  Decay branching fractions of 
 (a) $\chi_{2}^{\pm}$, (b) $\chi_{2}^{0}$ and (c) $\chi_{3}^{0}$ versus $\mu$ with $\tan\beta=10$, and (d) $\chi_{3}^{0}$ versus $\tan\beta$, for $\mu=500$ GeV.  Other parameters are chosen as  $M_2=100$ GeV, $M_{1}=1$ TeV.
 }
\label{fig:decays_muNLSP_M2LSP}
\end{figure}

The decay branching fractions for the NLSPs $\chi_2^\pm$, $\chi_2^0$ and $\chi_3^0$ in Case BII are shown in Fig.~\ref{fig:decays_muNLSP_M2LSP}.  Given the LSPs being nearly degenerate neutral and charge Winos $\chi_1^0$, $\chi_1^\pm$, more decay channels open for the Higgsino NLSPs. 

For $\chi_2^\pm$, the dominant decay modes are 
\beq
\chi_2^\pm \rightarrow \chi_1^0 W, \chi_1^\pm Z, ~\chi_1^\pm h.
\label{eq:BII_chi2pm}
\eeq
Under the limit of $  |\mu \pm M_2| \gg m_Z$, 
the ratios of the partial decay widths is roughly $\Gamma_{\chi_1^0 W}:\Gamma_{\chi_1^\pm Z}:\Gamma_{\chi_1^\pm h} \approx 1:1:1$, with small deviation caused by phase space effects.  The $\tan\beta$ dependence is   very weak, especially for large $\mu$.  For $\mu=500$ GeV, the branching fractions of $\chi_2^\pm$ to $W$, $Z$ and $h$ channels are roughly 35\%, 35\%, and 30\%, respectively.

The decay channels for the second and the third neutralinos\footnote{Note that the composition of $\chi_{2,3}^0$ in Case BII is opposite to that of $\chi_{2,3}^0$ in Case AII.}  $\chi_{2,3}^0 \approx \frac{1}{\sqrt{2}} ({\tilde{H}}_d^0 \pm {\tilde{H}}_u^0)$, with $+$ sign for $\chi_{2}^0$ and $-$ sign for $\chi_{3}^0$,  are 
\beq
\chi_{2,3}^0 \rightarrow \chi_1^\pm W^\mp, \chi_1^0 Z, ~\chi_1^0 h.
\label{eq:BII_chi2030}
\eeq
Under the limit of $ |\mu \pm M_2| \gg m_Z$, the following simplified relation holds for the partial decay widths (and decay branching fractions as well) of $\chi_{2,3}^0$:
\beq
\Gamma_{\chi_1^+W^-}=\Gamma_{\chi_1^-W^+}\approx\Gamma_{\chi_1^0Z}+\Gamma_{\chi_1^0h}.
\eeq
For both $\chi_2^0$ and $\chi_3^0$, decay to $W$ dominates since both $\chi_1^+W^-$ and  $\chi_1^-W^+$ contribute.
 $\chi_2^0$ is more likely to decay to $Z$ while $\chi_3^0$ is more likely to decay to $h$ for $\mu>0$. 

The $\tan\beta$ dependence of the branching fractions into $Z$ and $h$ channels is similar to that of Case BII.
${\rm Br} (\chi_2^0 \rightarrow \chi_1^0Z (h))$ varies between 30\% $-$ 24\% (3\% $-$ 9\%) for $\tan\beta$ between 3 $-$ 50, and similarly for $\chi_3^0$ decay with the branching fraction for the $Z$ and $h$ modes switched. 
${\rm Br}(\chi_{2,3}^0\rightarrow \chi^\pm  W^\mp)$, however, is almost independent of $\tan\beta$.   For $\mu=500$ GeV, the branching fraction of $\chi_2^0 (\chi_3^0)$ is 67\% (68\%), 26\% (8\%), and 7\% (24\%) for $W,\ Z$ and $h$ channels, respectively.    In the limit of large $\tan\beta$ and very heavy Higgsino mass, $ {\rm Br}(\chi_{2,3}^0 \rightarrow \chi_1^\pm W^\mp)   \approx 4 {\rm Br}(\chi_{2,3}^0 \rightarrow \chi_1^0 h) \approx 4 {\rm Br}(\chi_{2,3}^0 \rightarrow \chi_1^0 Z) \approx 68 \%$.   Flipping the sign of $\mu$ has similar effects on the $\chi_{2,3}^0$ decay branching fractions as in Case AII for the $Z$ and $h$ modes, while affects little of the $W$ mode. 

%%%%%%%%%%%%%%%%%%%%%%%%%

\begin{itemize}
\item{\bf Scenario C:} $|\mu| < M_{1},\ M_{2}$
\end{itemize}

This is the situation of Higgsino LSP \cite{NSUSY}, with the lightest states $\chi_{1,2}^0$ and $\chi_1^\pm$ being Higgsino-like.  The two possible mass relations here are 
\bea
&& {\rm Case\ CI:}\quad M_{1} < M_{2},\quad 
\  
\chi_{3}^0 {\rm \ \  Bino-like};\ \chi_2^\pm, \ \chi_4^0{\rm \  \ Wino-like};  
\\
&& {\rm Case\ CII:}\quad M_{2} < M_{1}, \quad  
 \chi_2^\pm,\ \chi_3^0{\rm \  \ Wino-like}; \  \chi_{4}^0 {\rm \ \  Bino-like}.~~
\eea
 
In Figs.~\ref{fig:masses}(e) and (f), we present the physical masses of the lower lying neutralinos and charginos with $\mu=100$ GeV, for Case CI versus the mass parameters $M_{1}$ while fixing $M_{2}=1$ TeV;  and for Case CII versus $M_{2}$ while fixing $M_{1}=1$ TeV. 
In both cases, relatively large mixing occurs for smaller values $M_1<200$ GeV in (e) and $M_2<300$ GeV in (f).
For larger values, the Higgsinos again group together as the LSPs. 

%%%%%%%%%%%%%%%%%%%%%%%%%%%%%%%%%%%

\begin{figure}[tb]
\minigraph{8.1cm}{-0.2in}{(a)}{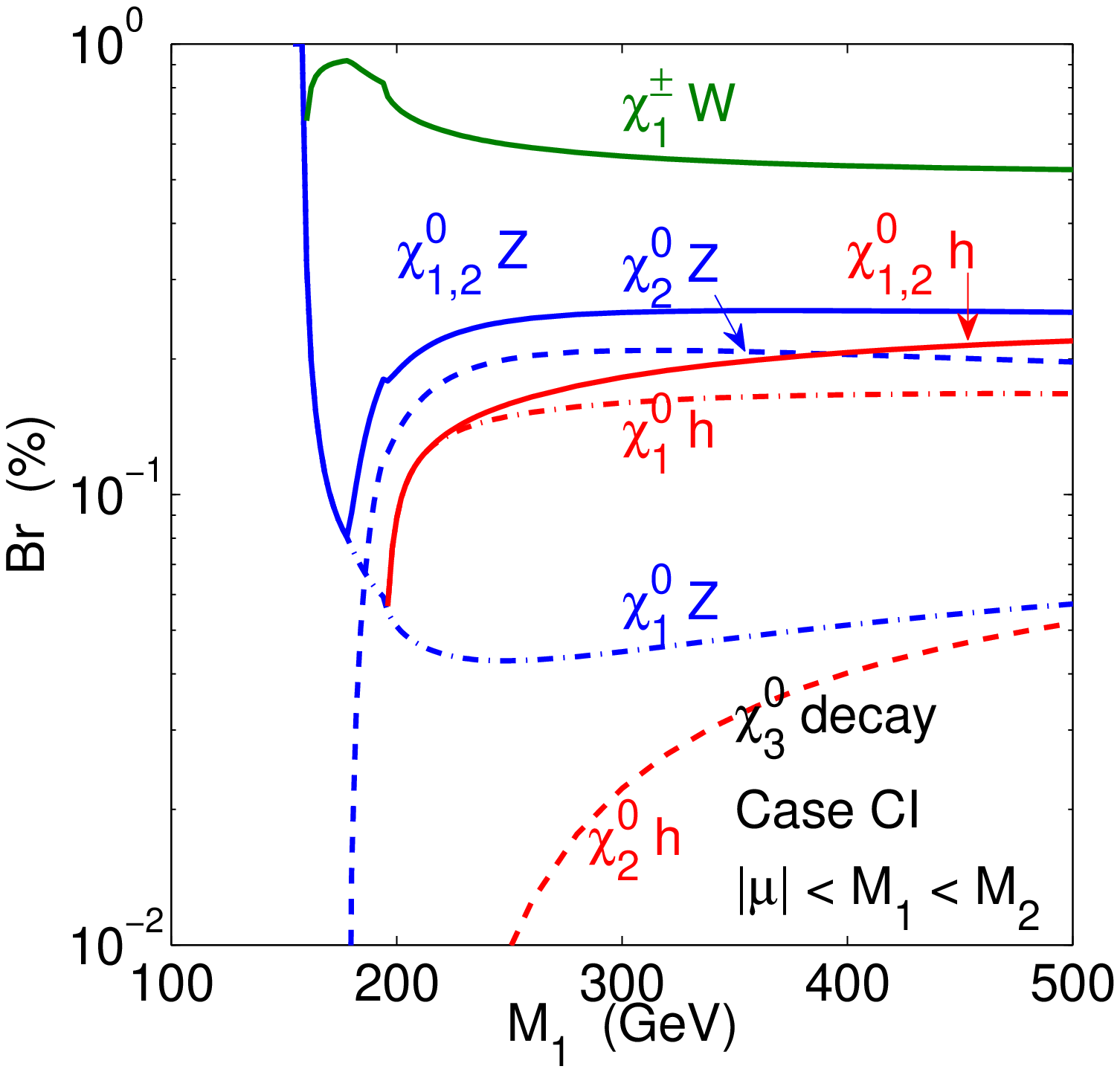}
\minigraph{8.1cm}{-0.2in}{(b)}{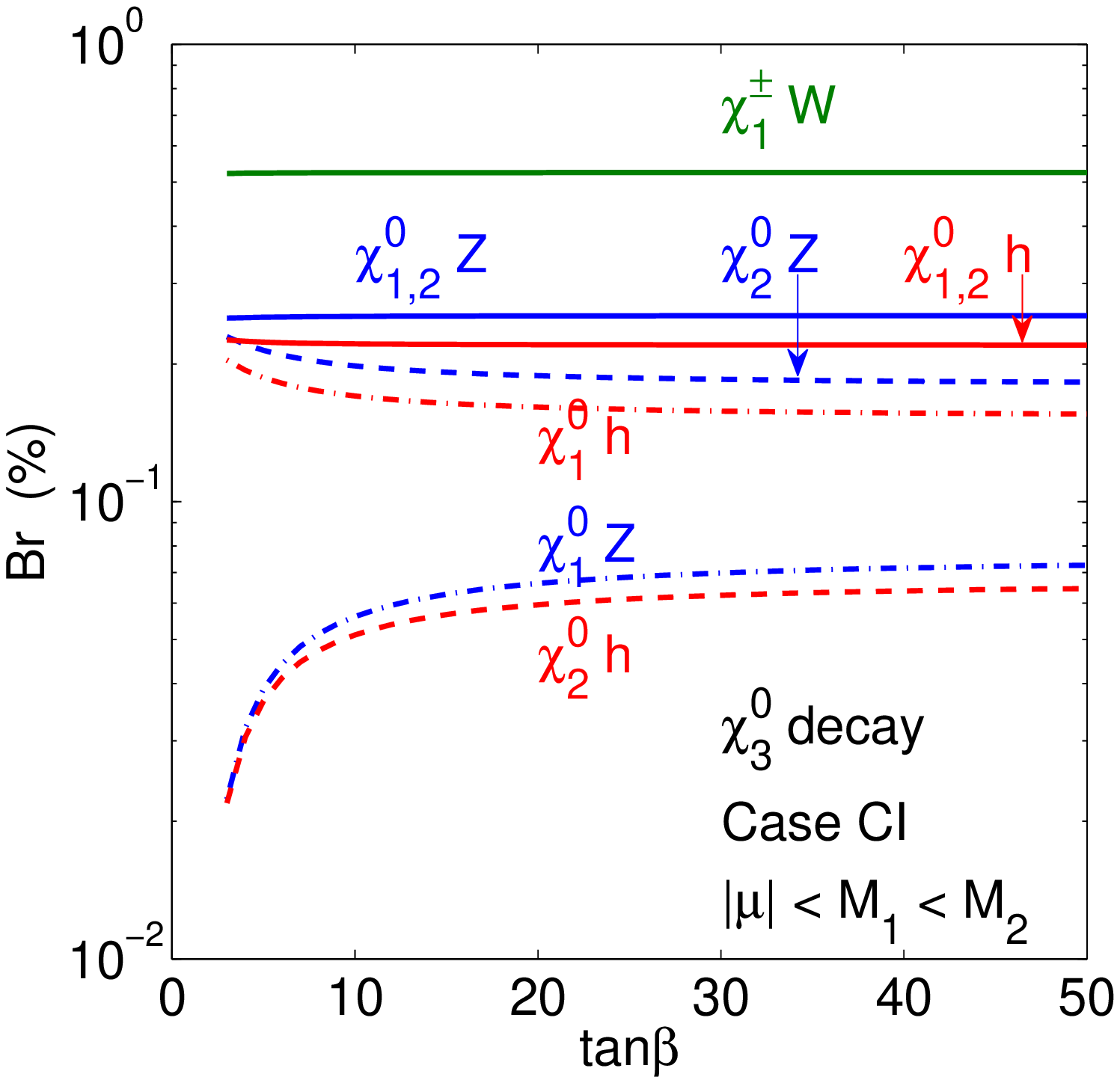}
  \caption{Case CI Bino-like NLSP and Higgsino-like LSPs:  Decay branching fractions of $\chi_{3}^{0}$ (a) versus $M_1$ for $\tan\beta=10$, (b) versus $\tan\beta$ for $M_1=500$ GeV, where $\chi_{1,2}^{0}$ indicates the sum over the $\chi_{1}^{0}$ and $\chi_{2}^{0}$ channels. Other parameters are chosen as $\mu=100$ GeV and $M_2=1$ TeV.  }
\label{fig:decays_M1NLSP_muLSP}
\end{figure}

Given the LSPs being the nearly degenerate neutral and charged Higgsinos $\chi_{1,2}^0$, $\chi_1^\pm$, more decay channels open for the Bino-like NLSP. 
The decay channels for $\chi_{3}^0$ in Case CI are depicted in Figure~\ref{fig:CaseABC_decay}(e) as 
\beq
\chi_{3}^0 \rightarrow \chi_1^\pm W^\mp, ~\chi_1^0 Z, ~\chi_2^0 Z, ~\chi_1^0 h, ~\chi_2^0 h.
\label{eq:CI_chi30}
\eeq
The decay branching fractions for the NLSP $\chi_3^0$ are shown in Fig.~\ref{fig:decays_M1NLSP_muLSP}, with the approximate formulae  for  the partial decay widths to various final states given in Eqs.~(\ref{eq:CaseCI_chi30h})$-$(\ref{eq:CaseCI_chi30W}). The following relation between the partial decay widths (and decay branching fractions as well)  holds 
\beq
\Gamma_{\chi_1^+W^-}=\Gamma_{\chi_1^-W^+}\approx\Gamma_{\chi_1^0Z}+\Gamma_{\chi_1^0h}
\approx \Gamma_{\chi_2^0Z}+\Gamma_{\chi_2^0h} \approx \Gamma_{\chi_1^0h}+\Gamma_{\chi_2^0h} \approx \Gamma_{\chi_1^0Z}+\Gamma_{\chi_2^0Z}. 
\eeq 

Since $\chi_1^0$ and $\chi_2^0$ are hard to distinguish experimentally due to its small mass splitting, 
$\chi_{1}^0h$ and $\chi_{2}^0h$ shall be combined as far as experimentally observation goes, and similarly for $\chi_{1}^0Z$ and $\chi_{2}^0Z$.
While the decay branching fraction of individual channel $\chi_{1}^0h$, $\chi_{2}^0h$, $\chi_{1}^0 Z$, and  $\chi_{2}^0 Z$ varies with $\tan\beta$,   ${\rm Br}_{\chi_{1,2}^0 h}={\rm Br}_{\chi_{1}^0 h}+{\rm Br}_{\chi_{2}^0 h}$,  ${\rm Br}_{\chi_{1,2}^0 Z}$, as well as  ${\rm Br}_{\chi_{1}^\pm W^\mp}$ are almost independent of $\tan\beta$, as shown in Fig.~\ref{fig:decays_M1NLSP_muLSP}(b). For $\mu=500$ GeV, the branching fractions of $\chi_3^0$ are 52\%, 26\%, and 22\% for  $W$, $Z$ and $h$ channels, respectively. 
 %%%%%%%%%%%%%%%%%%%%%%%%%%%%%%%%

\begin{figure}[tb]
\minigraph{8.1cm}{-0.2in}{(a)}{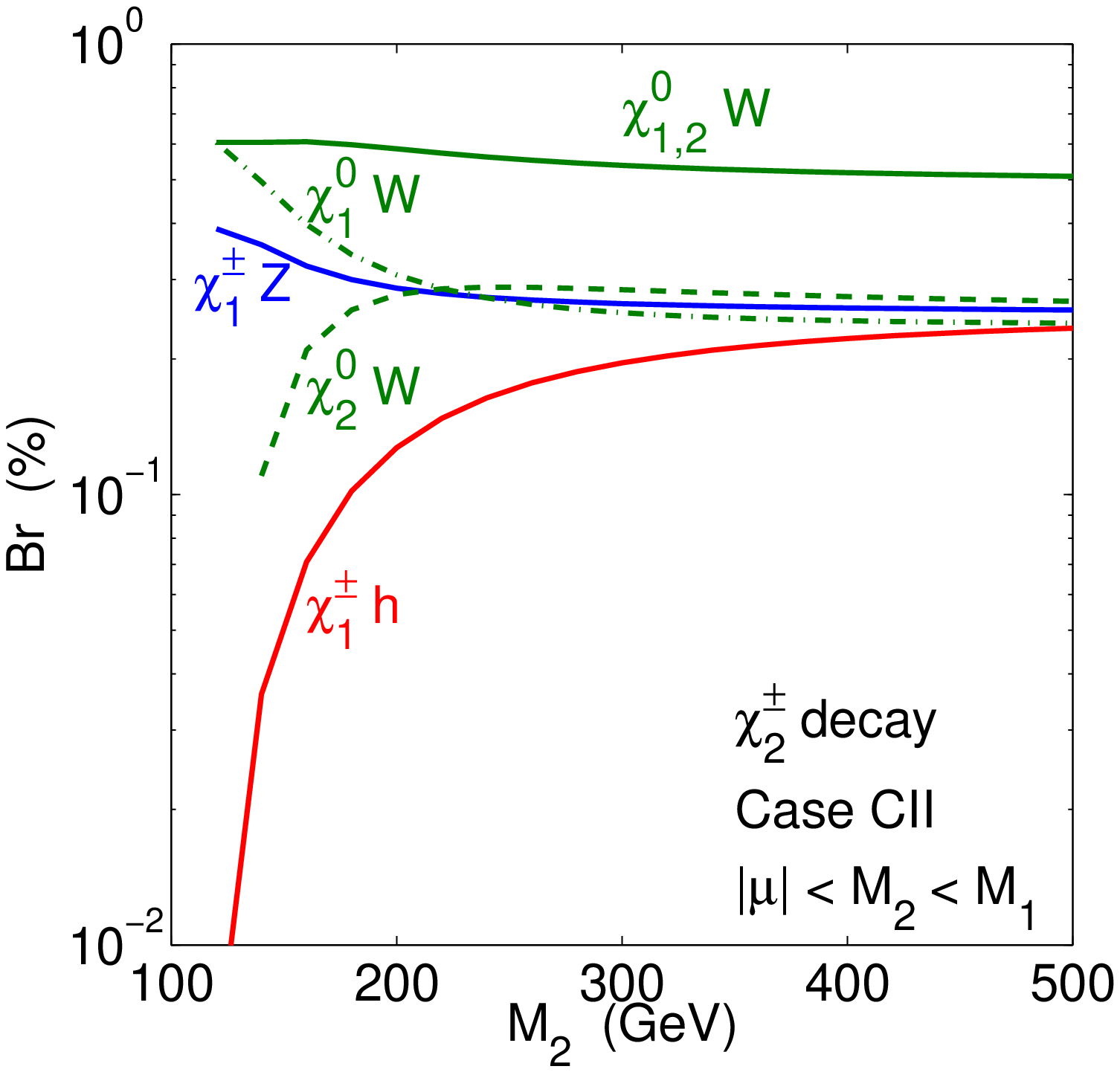}
\minigraph{8.1cm}{-0.2in}{(b)}{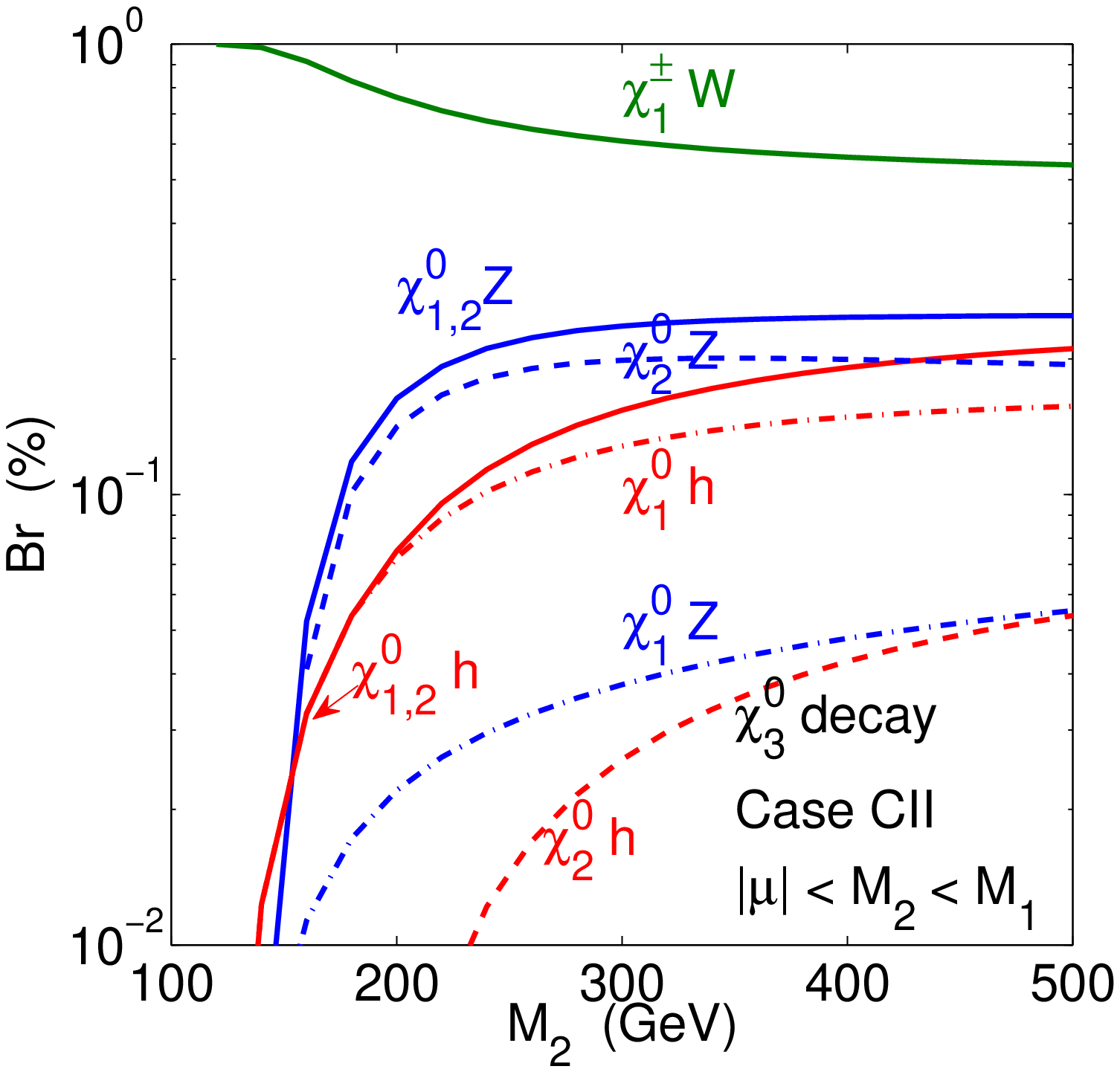}
\caption{Case CII Wino NLSPs and Higsino LSPs: Decay branching fractions of (a) ${\chi}_2^\pm$, and 
(b) ${\chi}_3^0$  versus $M_{2}$, for $\mu=100$ GeV, $M_{1}=1$ TeV and $\tan\beta=10$. Note that $\chi_{1,2}^{0}$ indicates the sum over the $\chi_{1}^{0}$ and $\chi_{2}^{0}$ channels. } 
\label{fig:decays_M2NLSP_muLSP}
\end{figure}
 
The decay branching fractions for the NLSPs $\chi_2^\pm$  and $\chi_3^0$ in Case CII are shown in Fig.~\ref{fig:decays_M2NLSP_muLSP}.   
For $\chi_2^\pm$, the dominant decay modes are 
\beq
\chi_2^\pm \rightarrow \chi_1^0 W,  ~\chi_2^0 W, ~\chi_1^\pm Z, ~\chi_1^\pm h.
\label{eq:CII_chi2pm}
\eeq
Under the limit of $|M_2 \pm \mu|\gg m_Z$, the ratios of the partial decay widths is roughly $\Gamma_{\chi_1^0W}:\Gamma_{\chi_2^0W}: \Gamma_{\chi_1^\pm Z}:\Gamma_{\chi_1^\pm h} \approx 1:1:1:1$.  The $\tan\beta$ dependence is   very weak, especially for large $M_2$.   Due to the near degeneracy of $\chi_1^0$ and $\chi_2^0$, $\chi_1^0W$ and $\chi_2^0W$ final states can not be distinguished experimentally.  Combining these two channels,   the branching fractions of $\chi_2^\pm$ to $W$, $Z$ and $h$ channels are roughly 51\%, 26\%, and 23\%, respectively.  In the limit of large $M_2$, the branching fractions approach the asymptotic limit
 ${\rm Br}(\chi_{2}^\pm \rightarrow \chi_{1,2}^0 W) \approx 2 {\rm Br}(\chi_{2}^\pm \rightarrow \chi_1^\pm h) \approx 2 {\rm Br}(\chi_{2}^\pm \rightarrow \chi_1^\pm Z) \approx   $ 50 \%.

The decay pattern for  $\chi_{3}^0$ in Case CII are  very similar to $\chi_3^0$ decay in Case CI:
\beq
\chi_{3}^0 \rightarrow \chi_1^\pm W^\mp, ~\chi_1^0 Z, ~\chi_2^0 Z, ~\chi_1^0 h, ~\chi_2^0 h.
\label{eq:CII_chi30}
\eeq
Under the limit of $|M_2 \pm \mu|\gg m_Z$, the partial decay widths to various final states follow similar formulae as Eqs.~(\ref{eq:CaseCI_chi30h})$-$(\ref{eq:CaseCI_chi30W}), with the replacement of $M_1$ by $M_2$.
Combining $\chi_1^0$ and $\chi_2^0$ final states, the branching fraction of $Z$ channel is almost the same as the $h$ channel, which is about half of the branching fraction of the $W$ final states. 
For $\mu=500$ GeV, the branching fractions of $\chi_3^0$ are 54\%, 24\%, and 22\% for  $W$, $Z$,  and $h$ channels, respectively.    
  
%%%%%%%%%%%%%%%

\section{Electroweakino production at the LHC}
\label{sec:pro_decay}

 \begin{figure}[t]
 \includegraphics[scale=2,width=6in]{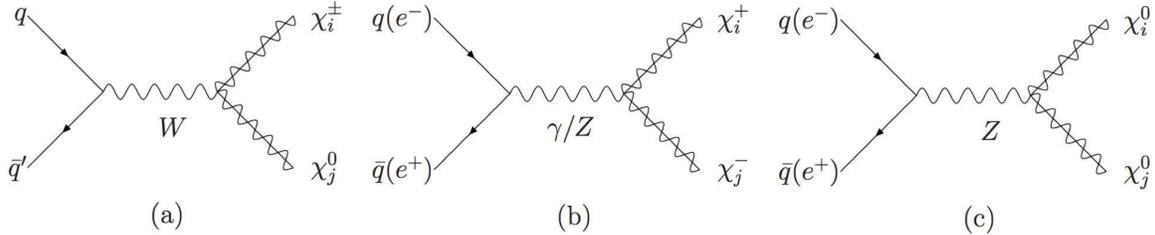}
  \caption{Feynman diagram for neutralino/chargino pair production.}
\label{fig:Feyn_diam} 
\end{figure}

Without the contributions of production and the cascade decays from the gluinos, squarks, nor sleptons and heavy Higgs bosons, the electroweakinos are pair-produced by the standard electroweak processes.  The leading contributions under our consideration are the Drell-Yan (DY) processes via the $s$-channel exchange of $W/Z/\gamma$, as shown in Fig.~\ref{fig:Feyn_diam}.
\bea
 pp &\to& \chi_i^\pm \chi_j^0 \ X ,\ \ \chi_i^+ \chi_j^- \ X, \ \ \chi_i^0 \chi_j^0 \ X,
\label{eq:DY}
\eea
where $i,j=1\ldots 4$ for neutralinos and $i,j=1\ldots 2$ for charginos, and $X$ generically denotes the hadronic remnants. Dominant processes are typically those that involves two Wino-like or two Higgsino-like states, since their relevant couplings to 
$W$, $Z$ and $\gamma$ are unsuppressed.  Furthermore, the electroweakino pair production via $W$-exchange in Fig.~\ref{fig:Feyn_diam}(a) has the largest cross section due to the large ${\rm SU(2)}_{L}$ coupling.
There could also be $t$-channel contributions with the exchange of $u$- and $d$-squarks.
 In our current treatment, we will neglect those effects under the assumption of  heavy squarks.
 
The electroweakinos could also be produced via weak vector boson fusion processes (VBF) \cite{VBF}
\beq
qq' \to qq'  \chi_i^+ \chi_j^0,  \ \  qq' \chi_i^+ \chi_j^-,  \ \  qq' \chi_i^0 \chi_j^0\ .
\eeq
The production rate for this mechanism is typically smaller than that of the DY processes by one to two orders of magnitude depending on their masses.  Thus these channels do not contribute much to the inclusive signal of our consideration \cite{Giudice:2010wb}. On the other hand, if a signal is observed via the DY processes, 
the unique kinematics of the forward-backward jets \cite{Barger:1990py} make the signal quite characteristic to study \cite{VBF}.
 
We now present the signal production rates via the DY processes as a function of a relevant mass parameter, in all the cases discussed in the last section. We show these in Fig.~\ref{fig:m2mu} at the 14 TeV LHC, including the next-to-leading oder (NLO) QCD corrections, which is about $20\% -30\%$ increase to the overall cross sections
 comparing to the LO results \cite{Beenakker:1999xh}. The cross sections at the 8 TeV LHC is about a factor of two smaller in the low gaugino mass region $\sim 200-300$ GeV, while they become smaller by about one order of magnitude at a high mass near 1 TeV. For the sake of illustration, we have taken 
\be
\tan\beta=10,\quad {\rm min}(M_{1},\ M_{2},\ |\mu|) = 100~\gev, \quad {\rm max}(M_{1},\ M_{2},\ |\mu|) = 1000~\gev,
\label{eq:paras}
\ee
unless stated otherwise.  The results for the leading NLSP pair production channels presented here are rather insensitive to the choice of this   values.   The numerical results presented below are always for $\mu>0$. 
 Here and henceforth, we adopt the parton distribution functions CTEQ6 \cite{cteq}.
We now present the production cross sections for all the cases and also discuss the leading decays of the electroweakinos to the SM final states. 

%%%%%%%%%%%%%%%%%%%%%%%%%

\subsection{Scenario A: $M_{1} < M_{2},\ |\mu|$}

%%%%%%%%%%%%%%%%%%%%%%%%%

\begin{itemize}
\item{Case AI:}  $M_{1} < M_{2}< |\mu|$
\end{itemize}

\begin{figure}
\minigraph{8.1cm}{-0.2in}{(a)}{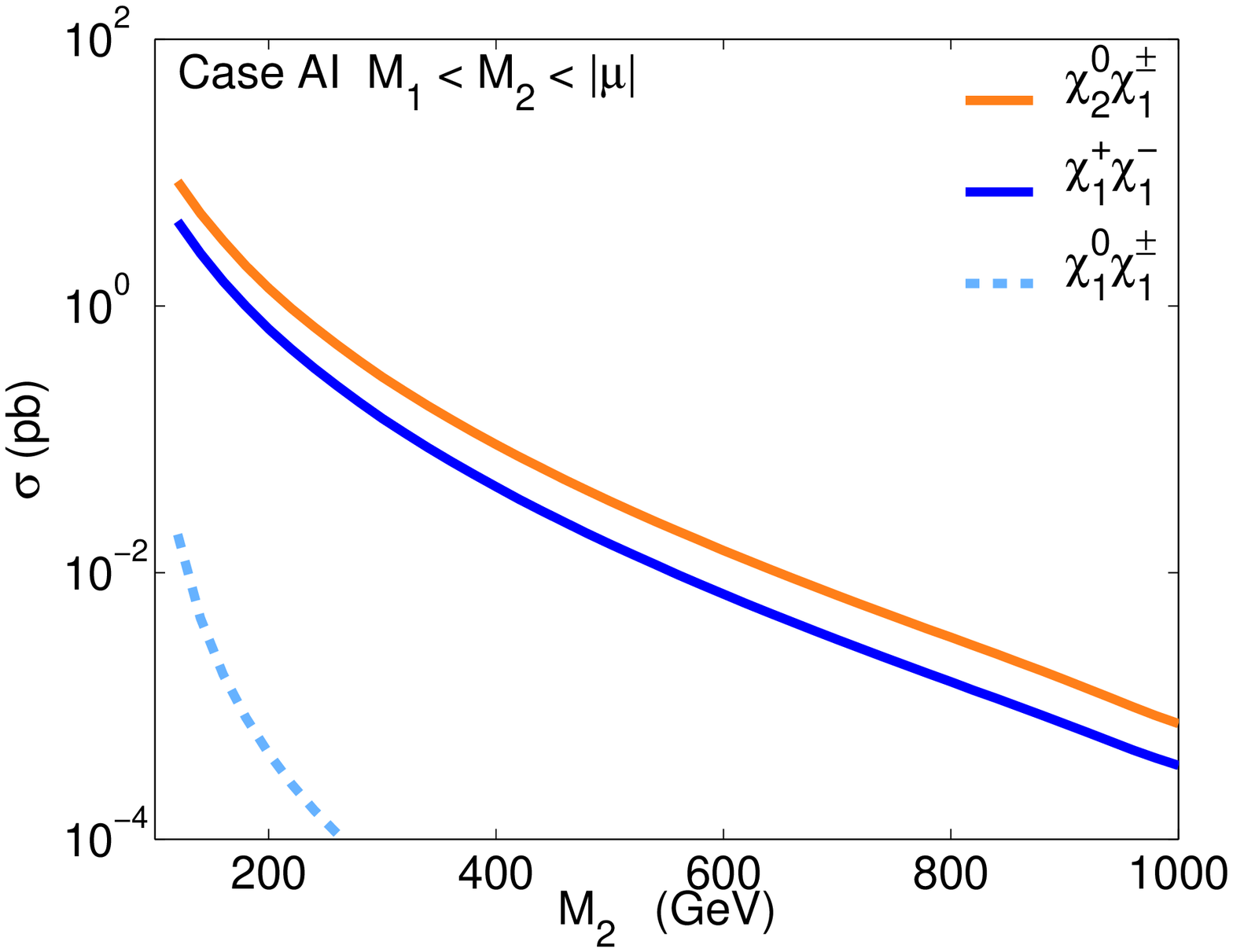}
\minigraph{8.1cm}{-0.2in}{(b)}{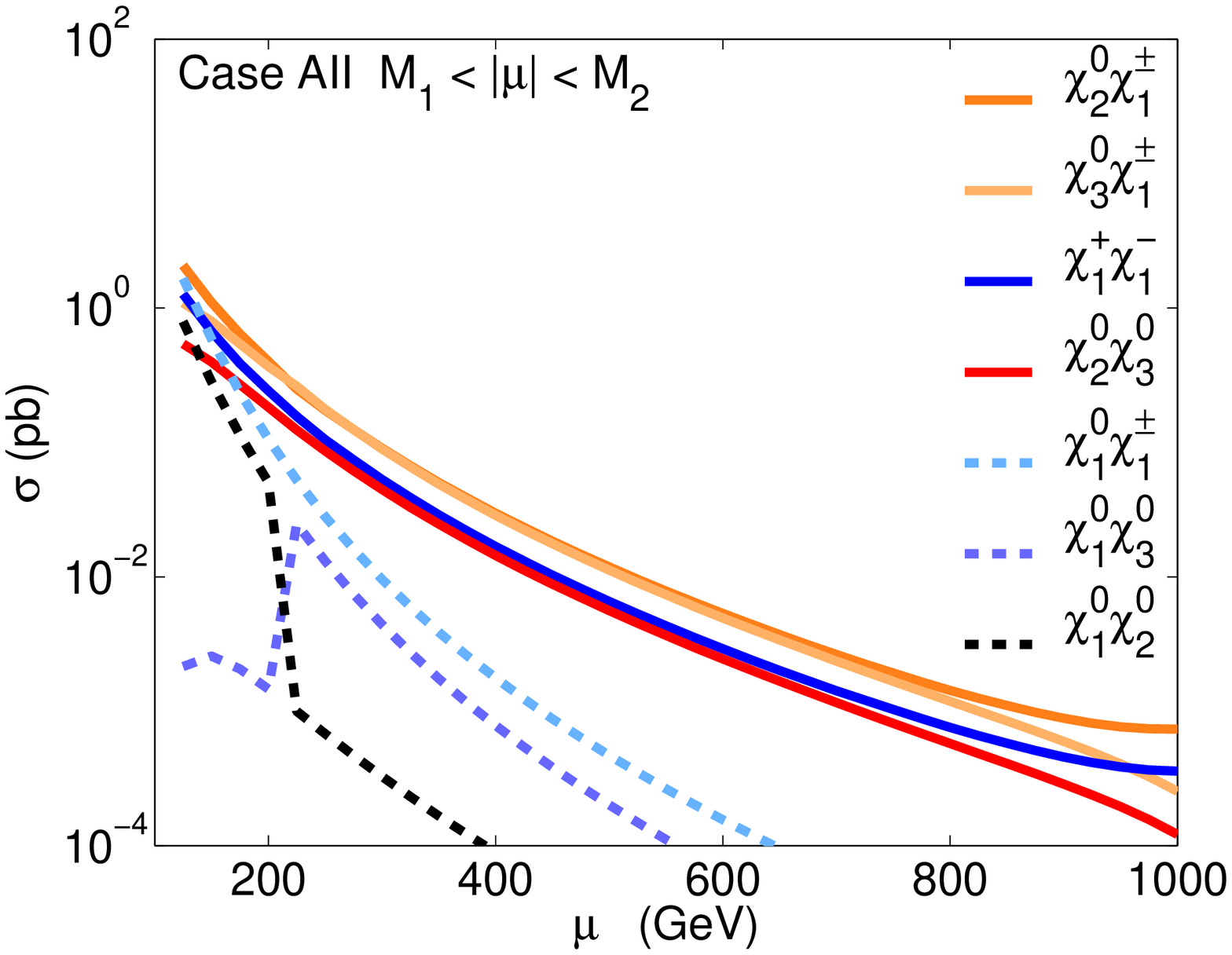}
\minigraph{8.1cm}{-0.2in}{(c)}{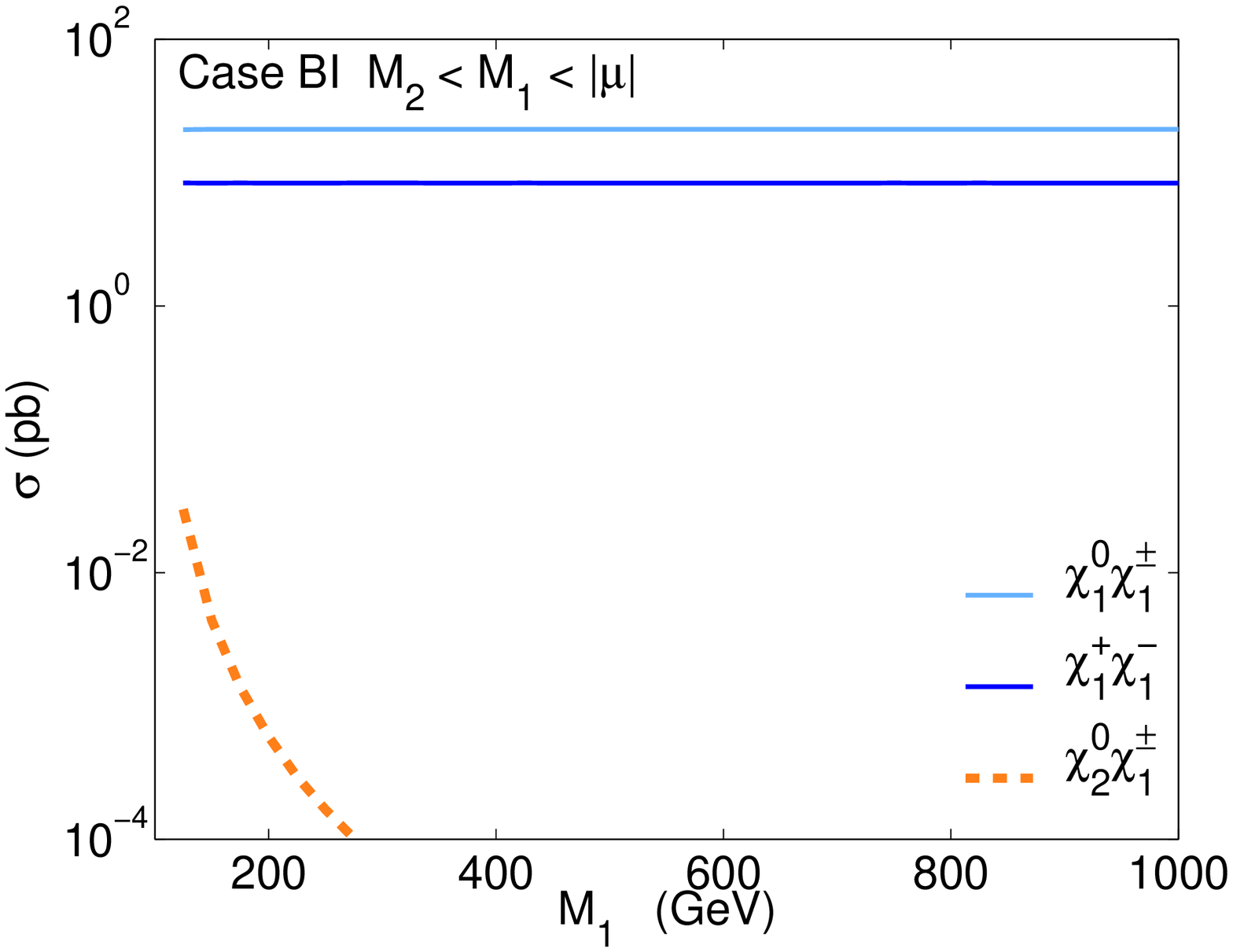}
\minigraph{8.1cm}{-0.2in}{(d)}{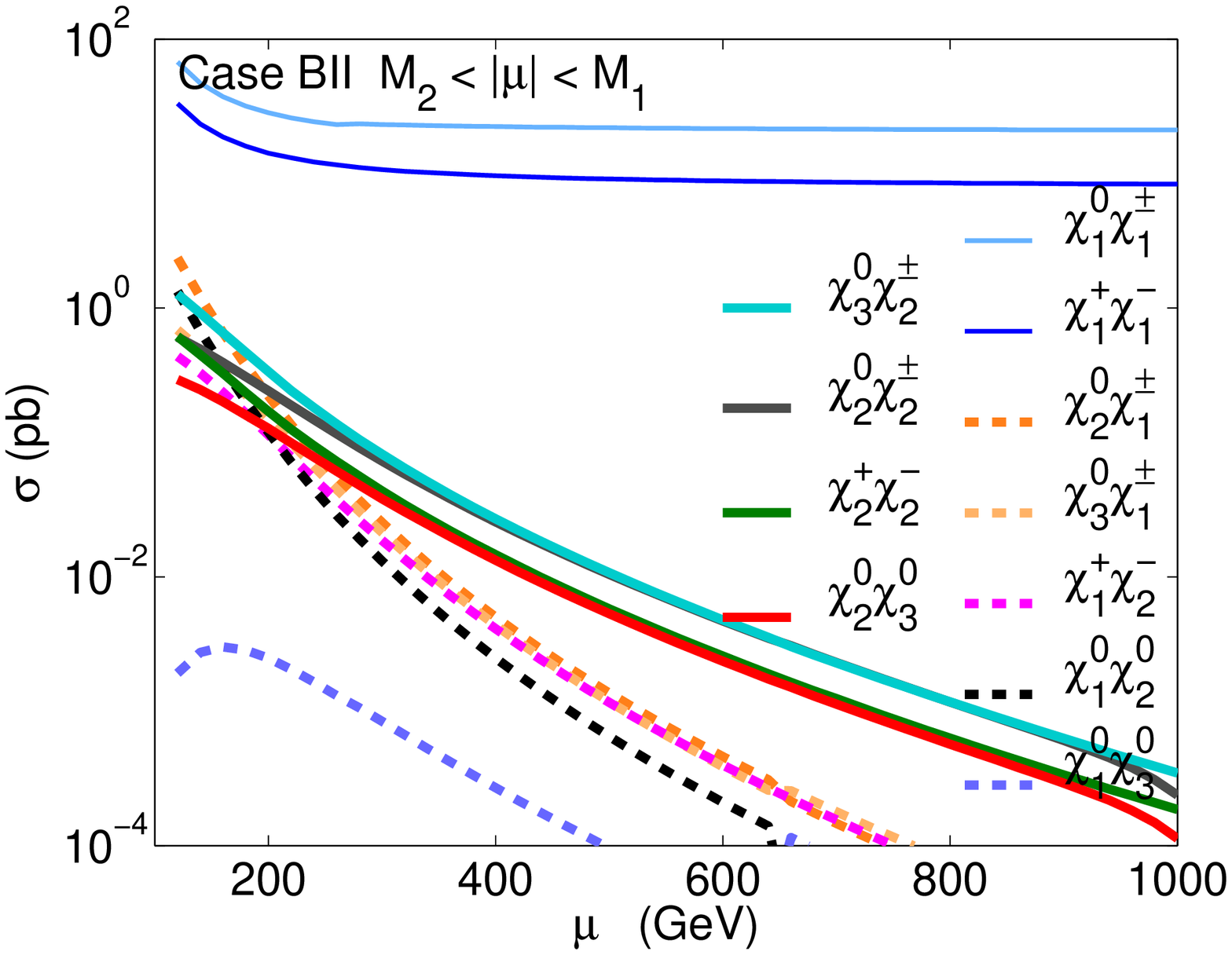}
\minigraph{8.1cm}{-0.2in}{(e)}{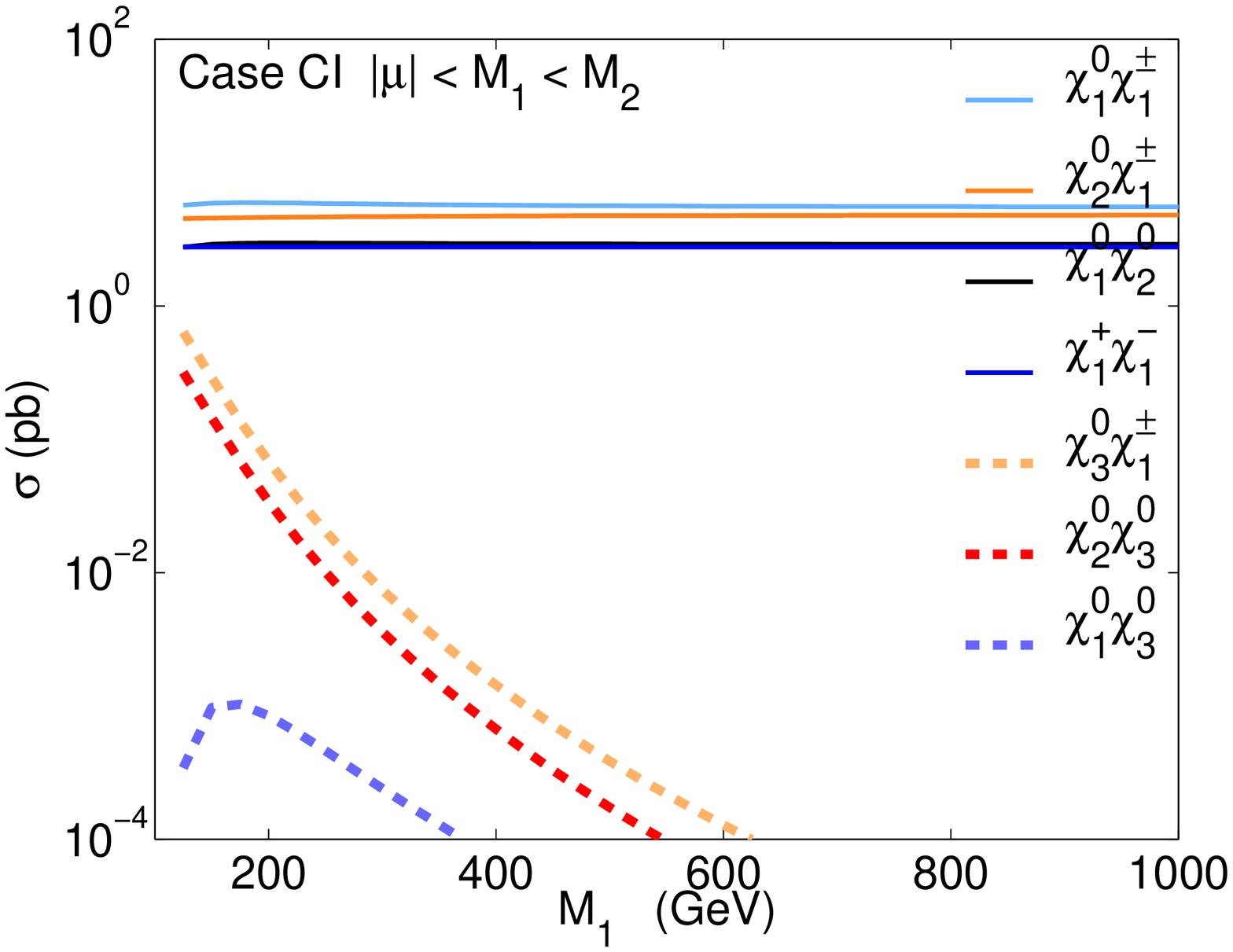}
\minigraph{8.1cm}{-0.2in}{(f)}{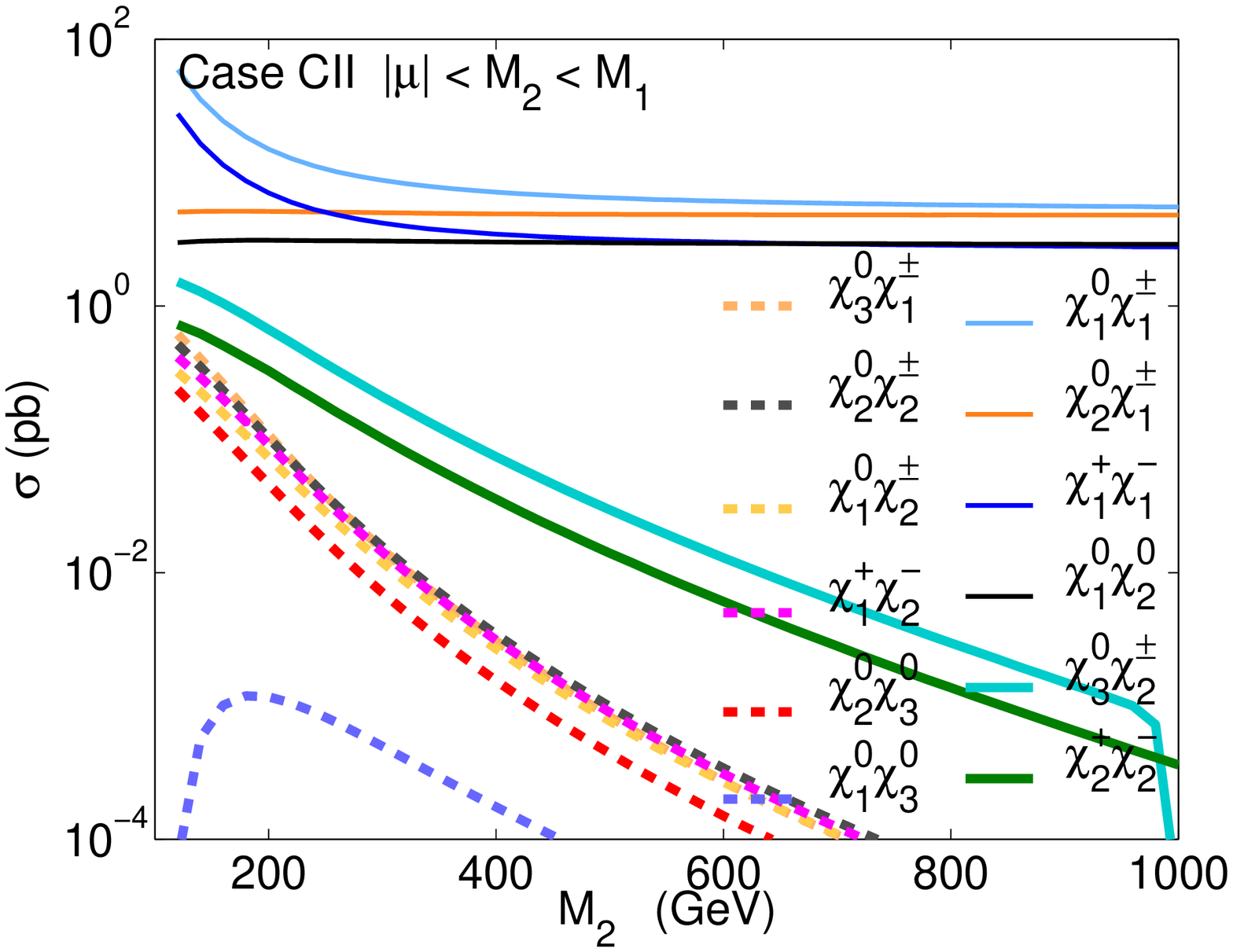}
\caption{
Total cross sections for the chargino and neutralino pair production at the NLO in QCD at the 14 TeV LHC for all the six cases: 
(a) Case AI: versus $M_2$ for $M_{1}=100$ and $\mu= 1$ TeV, (b) Case AII: versus $\mu$ for $M_{1}=100$ and $M_{2}=1$ TeV,
(c) Case BI: versus $M_1$ for $M_{2}=100$ and $\mu= 1$ TeV, (e) Case BII: versus $\mu$ for $M_{2}=100$ and $M_{1}=1$ TeV,
(e) Case CI: versus $M_1$ for $\mu =100$ and $M_{2}= 1$ TeV, (b) Case CII: versus $M_{2}$ for $\mu=100$ and $M_{1}=1$ TeV.
 }
\label{fig:m2mu} 
\end{figure}

This case is characterized by a Bino-LSP and three Wino-NLPs.
The   cross sections at the NLO in QCD for the 14 TeV LHC are shown in Fig.~\ref{fig:m2mu}(a)  versus  $M_2$.  The leading production channels are
\be 
{\rm Case\ AI:}\quad  
pp \to \chi_1^\pm \chi_2^0 \ X ,\ \ \chi_1^+ \chi_1^- \ X .
\label{eq:CaseA1}
\ee
These are the typical case for ``Wino-like'' production, with the unsuppressed ${\rm SU}(2)_L$ couplings.
The cross section summing over the leading channels is typically at the order of 1 pb for $M_2$ at about 200 GeV, and it drops to about 1 fb at 1 TeV.   The dominant cross sections have very weak dependence on $M_{1}$, only through the state mixing. 
The next potentially relevant channel,  $ \chi_1^\pm \chi_1^0 X$ production, is suppressed by almost three orders of magnitude, since it involves a small Bino-Wino mixing in $\chi_1^0$ or two orders of Bino/Wino-Higgsino mixings. All the other channels, especially those involving a Higgsino, are negligibly small. 

%%%%%%%%%%%%%%%%%%%%%%%%%%
 
\begin{itemize}
\item{Case AII:}  $M_{1} < |\mu| < M_2$
\end{itemize}

For Case AII with a Bino-like LSP and four Higgsino-like NLSPs,   cross sections at the NLO in QCD for the 14 TeV LHC are shown in Fig.~\ref{fig:m2mu}(b)  versus  $\mu$.  
The leading channels are more involved, as lower-lying NLSPs  are the four ``Higgsino-like'' states: $\chi_1^\pm$, $\chi_2^0$ and $\chi_3^0$. We thus have, in turn, 
\be
{\rm Case\ AII:}\quad  
 pp \to \chi_1^\pm \chi_2^0  X,\ \  \chi_1^\pm \chi_3^0  X,\  \ \chi_1^+ \chi_1^- X,\  \ {\rm and}\ \ \chi_2^0\chi_3^0  X, 
\label{eq:CaseAII}
\ee
again with unsuppressed ${\rm SU}(2)_L$ couplings. The next group of channels involves in the Bino-like LSP, such as $\chi_1^0  \chi_1^\pm, \  \chi_1^0 \chi_3^0,\  \chi_1^0 \chi_2^0$. They fall off faster at higher $\mu$ due to the ${\cal O}(m_Z/\mu)$ Bino-Higgsino mixing suppression. 
Contributions from $ \chi_2^0\chi_2^0  X$ and $ \chi_3^0\chi_3^0  X$ are small since 
$Z \chi_2^0 \chi_2^0$ and $Z \chi_3^0 \chi_3^0$ coupling vanish in the pure Higgsino mass eigenstate limit. 
 The total production rates for the Higgsino cross section in Case AII are slightly smaller than that in Case AI: about 400 fb for $\mu$ at about 200 GeV, and drops to 1 fb for $\mu$ around 1 TeV.

%%%%%%%%%%%%%%%%%%%%%%%%%

\subsection{Scenario B: $M_{2} < M_{1},\ |\mu|$}

\begin{itemize}
\item{Case BI:  $M_{2} < M_1< |\mu|$}
\end{itemize}

This case is characterized by three Wino-like LSPs and a Bino-like NLSP. 
The total cross sections at the NLO in QCD for the 14 TeV LHC are shown in Fig.~\ref{fig:m2mu}(c) versus  $M_1$.  The leading channels $\chi_1^\pm \chi_1^0 \ X$ and $\chi_1^+ \chi_1^- \ X$ are the pair production of the LSPs, which is almost unobservable via conventional searches given the small mass splitting of $m_{\chi_1^\pm} - m_{\chi_1^0}$. The subdominant channel $\chi_1^\pm \chi_2^0 \ X$ is suppressed by either the small Bino-Wino mixing or two powers of Bino/Wino-Higgsino mixing.  The cross section is only about 4 fb for $M_1$ around 150 GeV, and quickly drops down to 0.1 fb for $M_1 \sim 250$ GeV.  
 The search for the nearly degenerate Wino-like LSPs in Case BI at the LHC could be very challenging \cite{WinoLSP_collider,VBF} and we will not discuss it in this work. Instead, we will comment on its straightforward observability at an ILC. 

%%%%%%%%%%%%%%%%%%%%%%%%%%%%%%%%%%%%%%

\begin{itemize}
\item{Case BII:  $M_{2} < |\mu| < M_1$}
\end{itemize}

For Case BII with three Wino LSPs and four Higgsino NLSPs, total cross sections at the NLO in QCD for the 14 TeV LHC are shown in Fig.~\ref{fig:m2mu}(d) versus  $\mu$.  Similar to Case BI, the top two production channels 
$\chi_1^\pm \chi_1^0 \ X$ and $\chi_1^+ \chi_1^- \ X$ are those of the LSPs, therefore essentially unobservable at hadron colliders. The next set of production channels is similar to those of Case AII for Higgsino pair production as NLSPs: 
\be
{\rm Case\ BII:}\quad  
 pp \to \chi_2^\pm \chi_2^0  X,\ \  \chi_2^\pm \chi_3^0  X,\  \ \chi_2^+ \chi_2^- X,\  \ {\rm and}\ \ \chi_2^0\chi_3^0  X, 
\label{eq:CaseBII}
\ee
with unsuppressed ${\rm SU}(2)_L$ couplings. 
Contributions from $\chi_1^0 \chi_2^\pm X$, $\chi_1^\pm \chi_{2,3}^0X$, $\chi_1^0\chi_{2,3}^0X$, and $\chi_1^\pm \chi_2^\mp X$ are only comparable for $\mu \lsim 200$ GeV, and become small due to the suppressed ${\cal O}(m_Z/\mu)$ Higgsino components in $\chi_1^{0,\pm}$ for $\mu \gtrsim 200 -300$ GeV.

%%%%%%%%%%%%%%%%%%%

\subsection{Scenario C: $\mu < M_{1},\ M_{2}$}

\begin{itemize}
\item{Case CI:  $|\mu| < M_1< M_2$}
\end{itemize}

This is a case with four Higgsino-like LSPs and a Bino-like NLSP.
The total cross sections at the NLO in QCD for the 14 TeV LHC are shown in Fig.~\ref{fig:m2mu}(e) versus  $M_1$.  The leading channels $\chi_1^\pm \chi_{1,2}^0 \ X$, $\chi_1^+ \chi_1^- \ X$, and $\chi_1^0 \chi_2^0 \ X$ are those of pair production of the nearly degenerate Higgsino LSPs, which are hard to observe at hadron colliders as in Case BI previously discussed.  
The subdominant channels of Higgsino-Bino pair production $\chi_1^\pm \chi_3^0 \ X$, $\chi_{1,2}^0 \chi_3^0 \ X$ are suppressed by the small Bino-Higgsino mixing ${\cal O}(m_Z/M_1)$. The suppression factor is milder than that of Case BI. The cross section is about 300 fb for $M_1$ around 150 GeV, and quickly drops down to 0.1 fb for $M_1 \sim 600$ GeV.  
Similar to Case BI as discussed above, the search for the nearly degenerate Higgsino-like LSPs at the LHC could be very challenging \cite{HiggsinoLSP_collider,VBF} and we will not discuss it in this work. We will again comment on its straightforward observability at an ILC. 

%%%%%%%%%
\begin{itemize} 
\item{Case CII:  $|\mu| < M_2 < M_1$}
\end{itemize}

For the four Higgsino LSPs and three Wino NLSPs, total cross sections at the NLO in QCD for the 14 TeV LHC are shown in Fig.~\ref{fig:m2mu}(f) versus  $M_2$.  Similar to Case CI, the leading channels of pair production of nearly degenerate Higgsino LSPs are hard to observe at the LHC.  
The next set of processes is similar to that of Case AI for Wino pair production
\be
{\rm Case\ CII:}\quad  
 pp \to \chi_2^\pm \chi_3^0  X, \  \ \chi_2^+ \chi_2^- X, 
\label{eq:CaseCII}
\ee
with unsuppressed ${\rm SU}(2)_L$ couplings.
Note that for small $M_2$, the cross sections for the those subprocesses are smaller than Wino pair productions in Bino-like LSP - Wino-like NLSPs  Case AI.  This is because at low $M_2$, relatively large Wino-Higgsino mixing pushes up the mass spectrum of the Winos $\chi_2^\pm$ and $\chi_3^0$ much more than the small Bino-Wino mixing does in Case AI, as shown in the mass spectrum Fig.~\ref{fig:masses}.   Contributions from subleading processes $\chi_{1,2}^0 \chi_2^\pm X$,   $\chi_1^\pm\chi_3^0 X$, $\chi_{1,2}^0 \chi_3^0 X$ and $\chi_1^\pm \chi_2^\mp X$  are typically small due to the ${\cal O}(m_Z/M_2)$ suppression of Wino-Higgsino mixing except  for small $M_2$.   The total cross section is about 700 fb for $M_2$ around 200 GeV, and it drops to about 1 fb for $M_2$ around 1 TeV. 

%%%%%%%%%%%%%%%%%%%%%%%%%%%%%%%%%%

\subsection{Summary for the Signals at the LHC}

\begin{table}
%{\small
\begin{tabular}{|c| ll| c| c|c|c|c|c|c |c|}    \hline
    
   & \multicolumn{2}{|c|}{NLSP decay Br's }  & Production &\multicolumn{7}{|c|}{Total Branching Fractions (\%)}\\ \cline{5-11}
  
  &&&&$W^+W^-$&$W^\pm W^\pm$&$WZ$&$Wh$&$Zh$&$ZZ$&$hh$ \\
    \hline \hline
 Case AI & $\chi_1^\pm \rightarrow  \chi_1^0 W^\pm$ &100\%& $ \chi_1^\pm \chi_2^0 $  & 
 &&18&82&&&\\
$M_{1} < M_{2}< \mu$ & $\chi_2^0 \rightarrow \chi_1^0 h$ & 82\%(96$-$70\%) & $ \chi_1^+ \chi_1^-  $ & 
100 &&&&&&\\
\hline 
 Case AII  &$\chi_1^\pm \rightarrow  \chi_1^0 W^\pm$&100\%& $\chi_1^\pm \chi_{2}^0$&
 &&26&74&&&\\
$M_{1} <\mu< M_{2}$  & $\chi_2^0 \rightarrow \chi_1^0 h$  &74\%(90$-$70\%)& $\chi_1^\pm \chi_{3}^0$ &
 &&78&23&&&\\
  
 & $\chi_3^0 \rightarrow \chi_1^0 Z$&78\%(90$-$70\%)&$\chi_1^+ \chi_1^-$&
 100 &&&&&&\\
 &  &  &
$ \chi_2^0\chi_3^0 $&
 &&&&63&20&17\\
 
  \hline \hline

Case BI &\multicolumn{10}{|l|}{ }\\ 
 $M_{2} < M_{1}< \mu$ &\multicolumn{10}{|l|}{$\chi_2^0\rightarrow \chi_1^\pm W^\mp, \chi_1^0 h, \chi_1^0 Z$, \ \ \ 68\%, 27\%($31-24\%$), 5\%($1-9\%$), \ \ \ production suppressed.}\\ \hline
 Case BII &$\chi_2^\pm \rightarrow  \chi_1^0 W^\pm$ &35\%   &$\chi_2^\pm \chi_{2}^0$&
12&12&32&23&10&9&2
\\   
 $M_{2} < \mu < M_{1}$ &$\chi_2^\pm \rightarrow \chi_1^\pm Z$&35\%   &$\chi_2^\pm \chi_{3}^0$ &
 12&12&26&29&11&3&7
  \\   
  &$\chi_2^\pm \rightarrow \chi_1^\pm h$ & 30\%   &$\chi_2^+ \chi_{2}^-$&
 12&&25&21&21&12&9
  \\  
    
  &$\chi_2^0 \rightarrow \chi_1^\pm W^\mp$ &67\%   &$\chi_2^0 \chi_{3}^0$&
23&23&23&21&7&2&2
 \\   
  &$\chi_2^0 \rightarrow \chi_1^0 Z$ & 26\%(30$-$24\%)  &&
&&&&&&
  \\
 &$\chi_3^0 \rightarrow \chi_1^\pm W^\mp$ &68\%   &&  
 &&&&&&\\
&$\chi_3^0 \rightarrow \chi_1^0 h$ &24\%(30$-$23\%) &&
 &&&&&&\\
 \hline \hline
Case CI &\multicolumn{10}{|l|}{ }\\ 
$\mu < M_{1} < M_{2}$  &\multicolumn{10}{|l|}{$\chi_3^0\rightarrow \chi_1^\pm W^\mp, \chi_{1,2}^0 Z, \chi_{1,2}^0 h$, \ \ \ 52\%, 26\%, 22\%, \ \ \ production suppressed.}\\ \hline
 Case CII &$\chi_{2}^\pm \rightarrow  \chi_{1,2}^0 W^\pm$ &51 \%   &$\chi_2^\pm \chi_{3}^0$&
14&14&27&23&11&6&5
\\   
$\mu < M_{2} < M_{1}$  &$\chi_{2}^\pm \rightarrow  \chi_{1}^\pm Z$ &26 \%   &$\chi_2^+ \chi_{2}^-$&
26&&26&24&12&7&5
\\   
  &$\chi_{2}^\pm \rightarrow  \chi_{1}^\pm h$ &23 \%   & &
&&&&&&
\\   
  &$\chi_{3}^0 \rightarrow  \chi_{1}^\pm W^\mp$ &54 \%   & &
&&&&&&
\\   
  &$\chi_{3}^0 \rightarrow  \chi_{1,2}^0 Z$ &24 \%   & &
&&&&&&
\\   
  &$\chi_{3}^0 \rightarrow  \chi_{1,2}^0 h$ &22 \%   & &
&&&&&&
\\   
 \hline \hline
 
    \end{tabular}
   \caption{Dominant production and decay channels for the NLSPs. The mass parameter for NLSPs is taken to be 500 GeV  and $\tan\beta=10$, $\mu>0$  is used a benchmark point.  Numbers in parentheses show the variation of the decay branching fractions for $\tan\beta$ varying between 3 to 50.  For signals listed in the last 7 columns, there are always MET + possible soft jets/leptons.   
  }
\label{table:signal} 
\end{table}
%%%%%%%%%%%%%%%%%%%%%%%%%%%%%%%%%%%%%%%%%%%%

We have laid out the most general electroweakino scenarios based on the relations among the gaugino soft mass parameters $M_{1},\ M_{2}$ and the Higgsino mass parameter $\mu$. In the absence of substantial mixing when all the mass parameters are of the similar size, the three sets of multiplets (namely a Bino, 3 Winos and 4 Higgsinos) are each nearly degenerate in mass, respectively. 

The three scenarios with six distinctive cases are summarized in Table \ref{table:signal}. 
 For each case, we show the dominant pair production channels for the NLSP electroweakinos and their decay modes with branching fractions, which are given for the parameters of benchmark values as in Eq.~(\ref{eq:paras}),
 and the mass parameter corresponding to the NLSP mass taken to be 500 GeV.  For the decay branching fractions,  
most of them are insensitive to the particular value of $\tan\beta$. For those that do have $\tan\beta$ dependence, we show the variation in the parenthesis with $\tan\beta$ in the range of $3-50$. 
 
Generally speaking, the Wino-like electroweakinos are of the highest values of the production cross section. The next are the Higgsino-like ones. The Bino-like states are of the smallest production rate. 
Thus, Case A presents the idealistic cases with leading production of Wino-like NLSPs (Case AI) and Higgsiino-like NLSPs (Case AII), both dominantly decay via the Bino-like LSP.
For the rest of cases, they all naturally result in a compressed spectrum of nearly degenerate LSPs.
The leading production channels are the Wino-like LSPs in Case B and the Higgsino-like LSPs in Case C. As discussed earlier, the LSP multiplet production will be difficult to observe at hadron colliders because of the mass degeneracy and the soft decay products \cite{WinoLSP_collider,HiggsinoLSP_collider}. This possesses significant difficulty for their searches at the LHC and we will thus leave Cases BI and CI for the future exploration. Instead, we will comment on them for the ILC studies in a  later section. 
On the other hand, the situation of the observability may be improved if the sub-leading production cross sections via the NLSPs are not small. These are indeed what happens as in Case BII for Higgsino-like NLSPs production and in Case CII for Wino-like NLSPs production.

\begin{figure}
\minigraph{8.1cm}{-0.2in}{(a)}{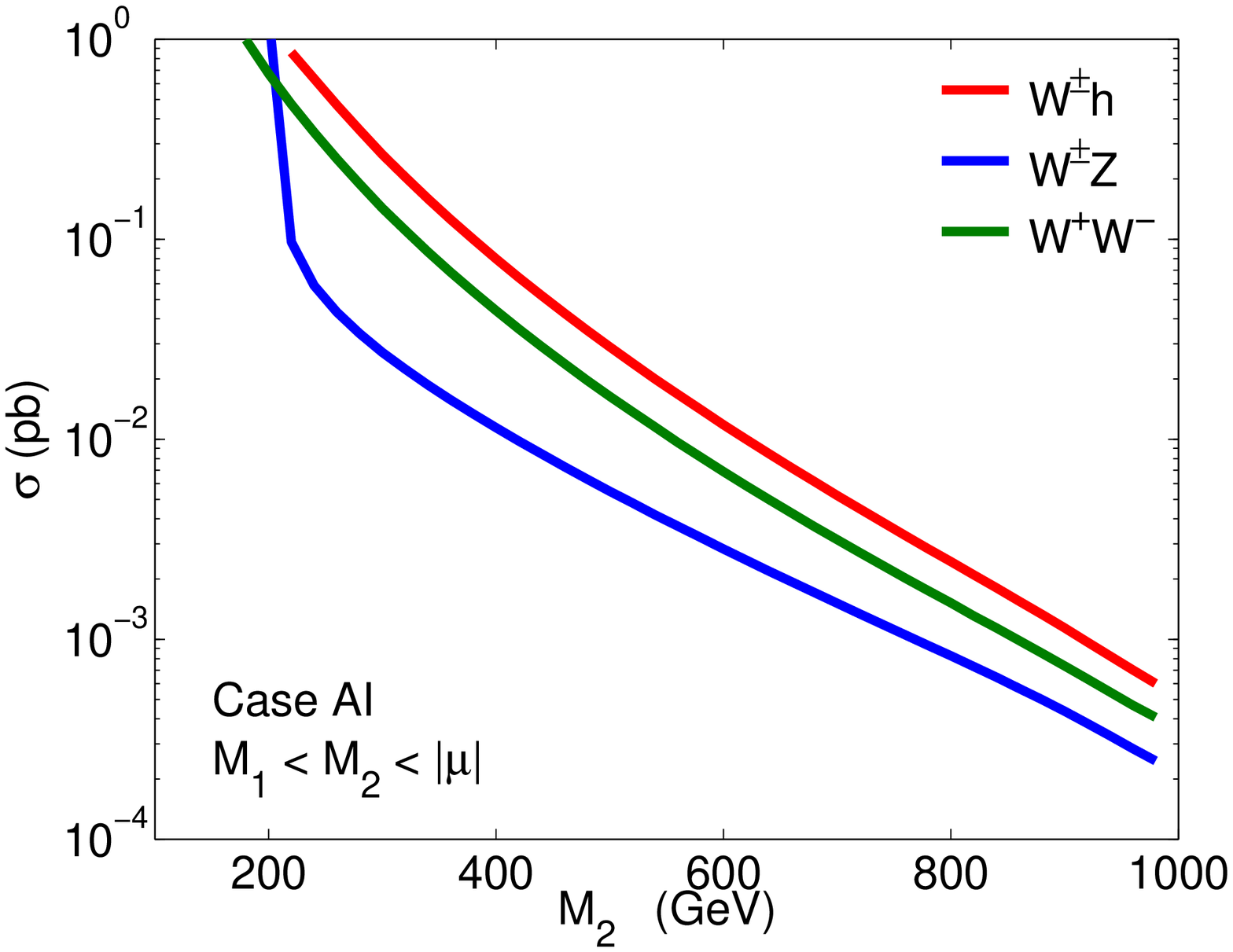}
\minigraph{8.1cm}{-0.2in}{(b)}{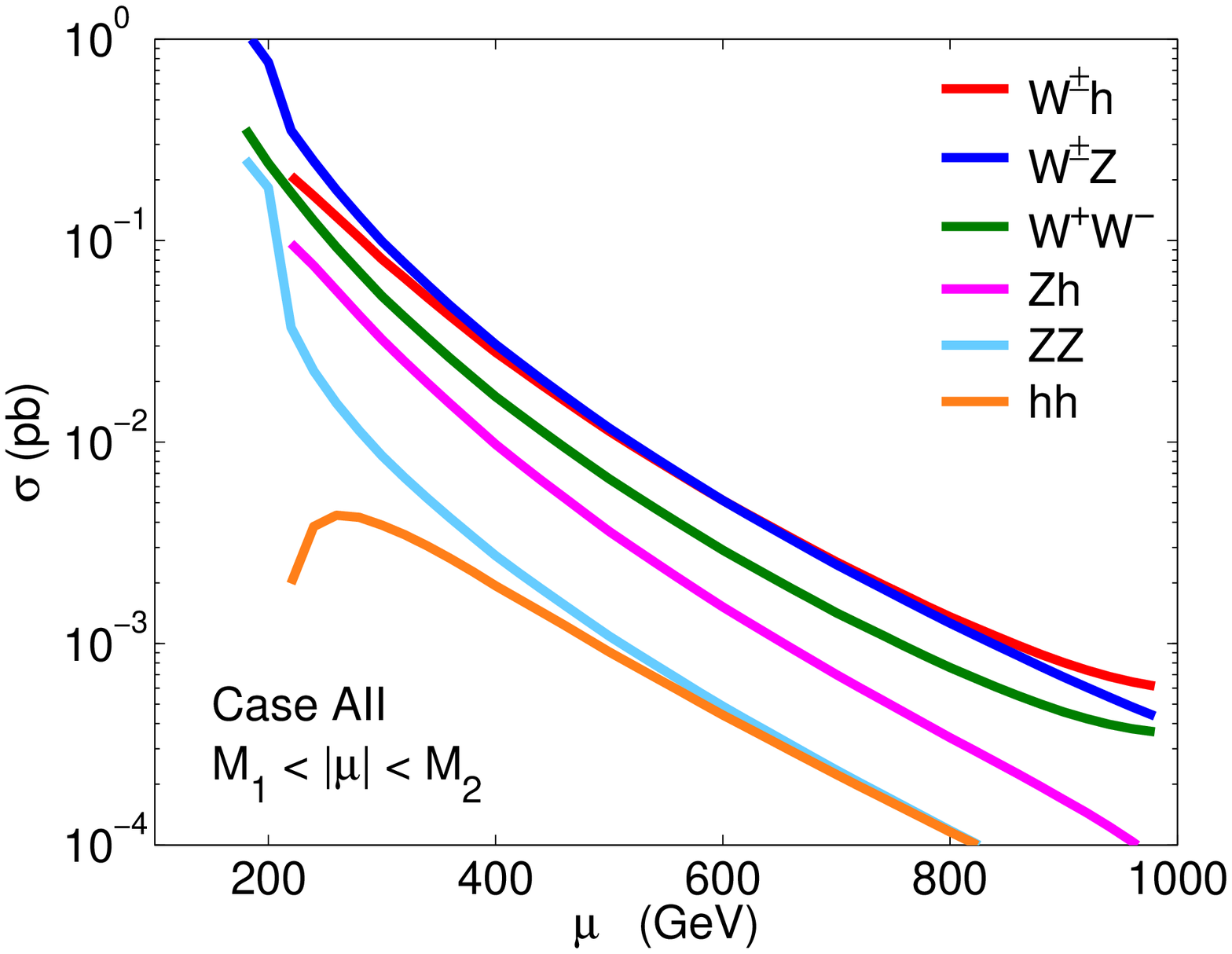}
\minigraph{8.1cm}{-0.2in}{(c)}{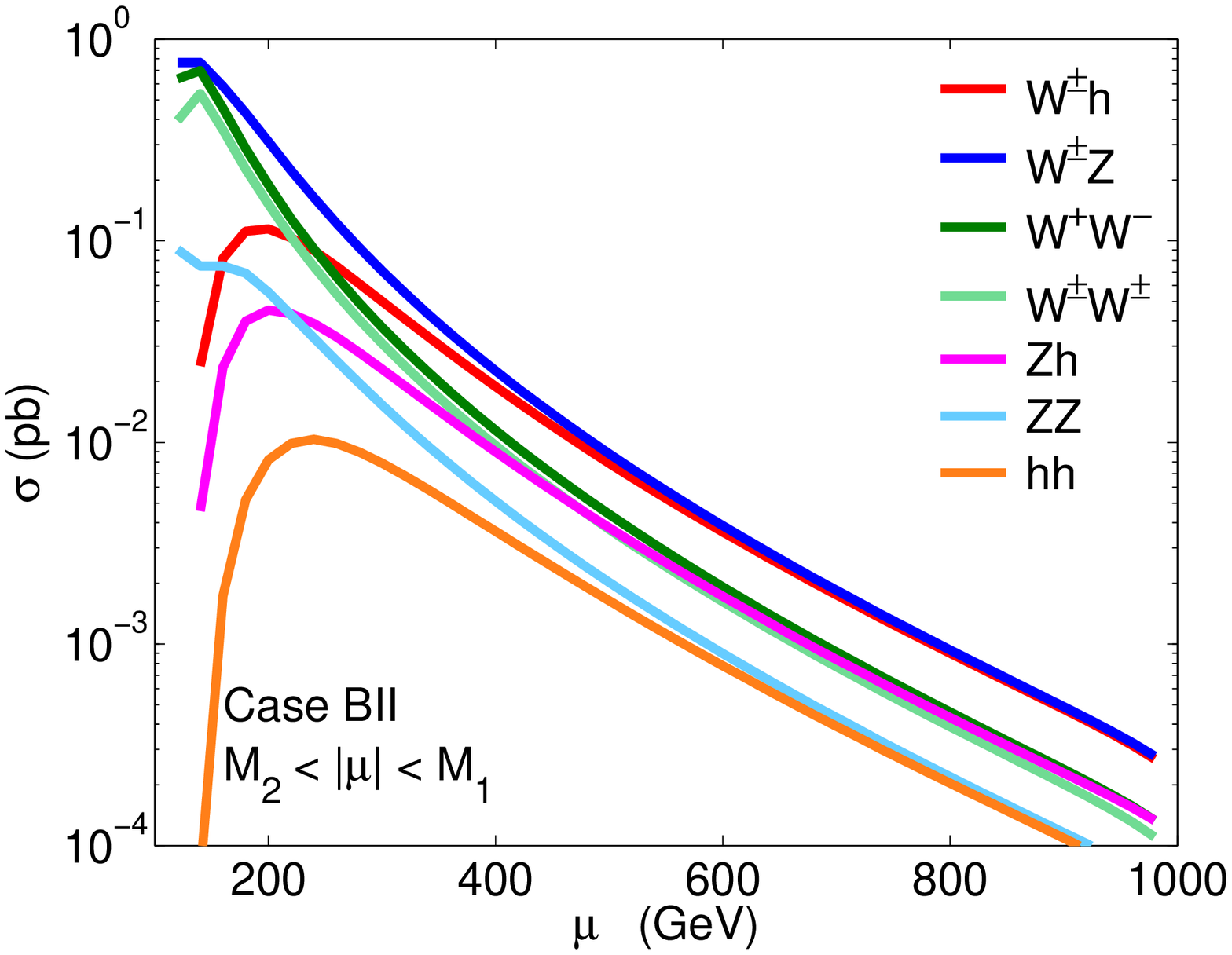}
\minigraph{8.1cm}{-0.2in}{(d)}{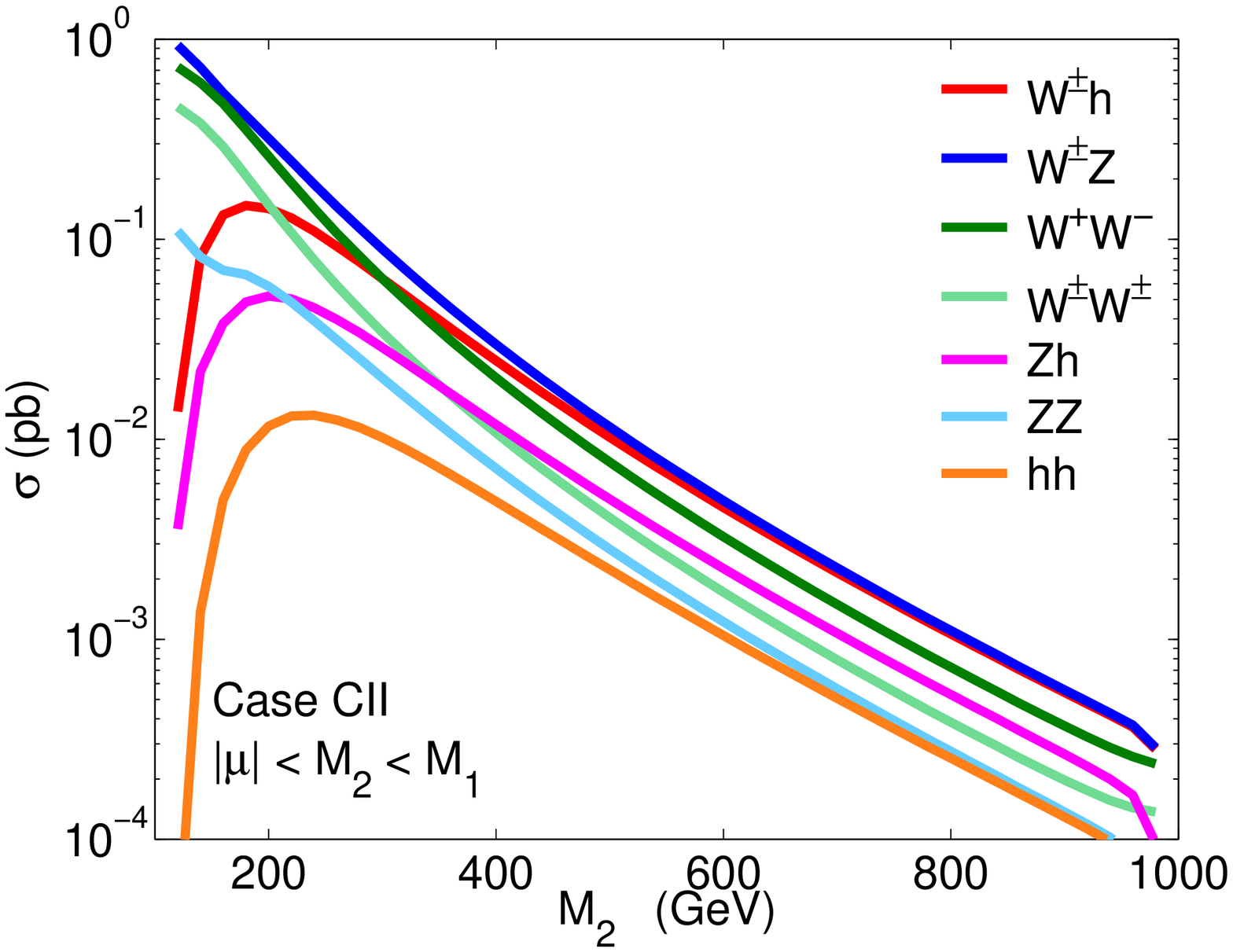}
\caption{  
Total cross sections for the chargino and neutralino pair production to specific final states at the 14 TeV LHC for the four cases relevant to direct searches with the NLPS production:
(a) Case AI: versus $M_2$ for $M_{1}=100$ and $\mu= 1$ TeV, 
(b) Case AII:  versus $\mu$ for $M_{1}=100$ and $M_{2}=1$ TeV,
(c) Case BII: versus $\mu$ for $M_{2}=100$ and $M_{1}=1$ TeV,
(d) Case CII: versus $M_{2}$ for $\mu=100$ and $M_{1}=1$ TeV.
}
\label{fig:CSBR} 
\end{figure}

To guide the searches at the LHC, we combine with the decay branching fractions of the corresponding NLSPs for each production mode, and show the total branching fraction into each particular final state 
\bea
XY=W^+W^-,\ W^\pm W^\pm,\ WZ,\ Wh,\ Zh,\ ZZ,\ {\rm and}\ hh,
\label{eq:XY}
\eea  
as in Table \ref{table:signal}. 
Note that all of the final states in addition include missing transverse energy introduced by the $\chi_1^0$ LSP, as well as soft jets and leptons that might appear from decays between nearly degenerate particles in LSP multiplet.  Since the same final states might come from different production processes, the total cross section of a particular final state is given by 
\beq
\sigma_{XY}^{\rm tot} = \sum_{i,j} \sigma(\chi_i\chi_j) \times Br(\chi_i \chi_j \rightarrow XY),
\label{eq:totalsignal}
\eeq
where the sum is over the dominant production modes listed in the table. 

Extending the above discussions, we present the total cross sections for the electroweakino pair production subsequently decaying to specific final states of the electroweak bosons $XY$ of Eq.~(\ref{eq:XY}) in Fig.~\ref{fig:CSBR}. Here we only show the four observationally relevant model cases to the LHC searches as laid out in Table \ref{table:signal}. Again, the leading signal rates can reach a few hundred of fb to a few tenth of fb with the mass parameters from 200 GeV to 1 TeV. 
It is important to note that one of the leading channels is $Wh$,  typically larger than the observationally clean channel $W^{+}W^{-}$ and comparable to (in Case AI, larger than) the conventionally considered leading channel $WZ$, except near the kinematical threshold at low $\mu$ or low $M_{2}$.
We thus emphasize that with unique decay $h \to b \bar b$ and reconstructable Higgs mass variable, this channel should serve as a ``standard candle'' for the signal of the electroweakino pair production, to be discussed in a later section. 
 
 %%%%%%%%%%%%%%%%%%%%%%%%%%%%%%%%%%%%
 
\section{Current bounds, the Higgs Boson Channel, and Future Perspectives}
 \label{sec:analyses}

\subsection{Bounds from LEP2 Experiments}

With the same mechanism as discussed in the last session, charginos $\chi_1^\pm$ could  be pair produced at LEP via $s$-channel exchange of $Z/\gamma^*$, as well as the $t$-channel exchange of $\tilde\nu_e$, with destructive interference. It decays to $f\bar{f}^\prime \chi_1^0$ via a real or virtual $W$ or a sfermion. 
 Results for the chargino mass lower bounds from standard searches at the LEP2 experiments are briefly summarized in Table \ref{tab:LEP}.

\begin{table}[t]
\begin{tabular}{|c|c|}
\hline
 Lower limit on the chargino mass &  Conditions \\ \hline
 $m_{\chi_1^\pm } > 103.5$ GeV & heavy $\tilde \nu$, large $m_{\chi_1^\pm}-m_{\chi_1^0}$ \cite{LEPchargino1} \\ \hline
  $m_{\chi_1^\pm } > 92.4$ GeV & ``deep Higgsino" region $|\mu| \ll M_{1,2}$ \cite{LEPchargino2} \\ \hline
    $m_{\chi_1^\pm } > 91.9$ GeV & degenerate gaugino region   \cite{LEPchargino2} \\ \hline
\end{tabular}
\caption{Chargino mass lower bounds  at $95\%$ C.L.  from the LEP2 experiments.  
}
\label{tab:LEP}
\end{table}

For lower sfermion mass, the bound is weaker due to the reduced pair production  cross section, as well as the reduction of selection efficiency due to the opening up of the two body decay. In particular, there is a so called ``corridor" region where $m_{\chi_1^\pm} - m_{\tilde{\nu}}$ is small and the lepton from $\chi_1^\pm 
\rightarrow \ell \tilde{\nu}$ is so soft that escape detection. Associated production of $\chi_1^0\chi_2^0$ can be adopted to improve the search in such case when the chargino search becomes ineffective.  Limits on chargino and neutralino masses for the light sfermion case, therefore,  depend on the sfermion spectrum.

As for the mass of  the lightest neutralino LSP, there is no general bound from LEP if the gaugino mass unification relation is relaxed. Production via $s$-channel  exchange of $Z/\gamma^*$ 
could be absent for a Bino-like neutralino, and $t$-channel production could be negligible for heavy selectrons.  Indirect mass limit on the neutralino LSP can be derived from chargino, slepton and Higgs boson searches, when guagino mass and sfermion mass unification relations are assumed.  A lower mass limit of 47 GeV can be obtained at large $\tan\beta$ \cite{LEPLSP}, while a tighter limit of 50 GeV can be derived in the mSUGRA scenario \cite{LEPLSPmSUGRA}.

%%%%%%%%%%%%%%%%%%%%%%%%%%

\begin{table}[t]
\begin{tabular}{|c|c|}
\hline
 Lower limit on the electroweakino mass &  Conditions \\ \hline
 $m_{\chi_1^\pm,\chi_2^0} > 350 - 740$ GeV &  $2\ell+\etmiss,\  3\ell+\etmiss, \ 4\ell+\etmiss$ \cite{ATLAS2013028, ATLAS2013035,ATLAS2013036,ATLAS2013049,CMS2013006} \\ 
 &  $m_{\tilde \ell} = (m_{\chi_1^\pm}+m_{\chi_1^0})/2,\ \ m_{\chi_1^\pm}-m_{\chi_1^0} >100$  GeV  \\ \hline
 
 $m_{\chi_1^\pm,\chi_2^0} > 300$ GeV &  $2\ell+{\rm jets} + \etmiss,\  3\ell+\etmiss$ \cite{ATLAS2013035, CMS2013006} \\ 
 &  $m_{\chi_1^0} =0,\ BR(\chi^\pm_{1} \rightarrow W^\pm \chi^0_1)=BR(\chi^0_2 \rightarrow Z \chi^0_1)=100\%$  \\ \hline

 $m_{\chi_1^\pm,\chi_2^0} > 204 -  287 $ GeV &  $\ell bb+\etmiss$,  $2\ell+{\rm jets} + \etmiss$, $\geq 3\ell+ \etmiss$\cite{ATLAS2013093, CMS2013017}  \\
  &  $m_{\chi_1^0} =0,\ BR(\chi^\pm_{1} \rightarrow W^\pm \chi^0_1)=BR(\chi^0_2 \rightarrow h \chi^0_1)=100\%$   \\ \hline
\end{tabular}
\caption{
Electroweakino mass lower bounds at $95\%$ C.L.~from the LHC experiments at 8 TeV with 21 fb$^{-1}$, with the assumption of $m_{\chi_1^\pm}\approx m_{\chi_2^0}$. }
 \label{tab:LHC}
\end{table}

\subsection{Current Bounds from the LHC Experiments}

The search for charginos and neutralinos are being actively pursued by the LHC experiments. 
The hadronic decay of $\chi_1^\pm$ and $\chi_2^0$  will give the fully hadronic mode under the usual assumption of Bino-like LSP and Wino-like NLSPs. The leptonic  decay of $\chi_{1}^{\pm}$ will lead to an isolated lepton, while the $\chi_{2}^{0}$ leptonic decay typically leads to the opposite-sign dilepton final state as well.
The $\chi^{+}_{1}\chi^{-}_{1}$  production gives opposite-sign dileptons  in their leptonic decay. 
The $\chi_1^\pm \chi_{2}^0$ production with decays via $W^{(*)}$ and $Z^{(*)}$ gives the clean signal of $3\ell+\etmiss$ (here and henceforth, $\ell=e,\mu$, and $\etmiss$ is the missing transverse energy), which has been the dominant search channel for neutralinos and charginos. 
 
The ATLAS and CMS collaborations recently performed searches for pair production of the electroweakinos through the conventional channels of multi-lepton plus $\etmiss$ \cite{ATLAS2013028, ATLAS2013035,ATLAS2013036,ATLAS2013049,CMS2013006}. 
The absence of signal put some bounds on the mass parameters under certain assumptions as collected 
in Table \ref{tab:LHC}.   
Note however that the decays included in their analyses via sleptons are only applicable for the slepton mass lighter than $\chi_2^0$, $\chi_1^\pm$. Limits from $W, Z$ channels assume a 100\% branching fraction to the gauge bosons, which is usually not realized in a realistic model.   Also shown in the last row are the latest results from the $Wh+\etmiss$ channel \cite{ATLAS2013093, CMS2013017}.

%%%%%%%%%%%%%%%%%%%%%%%%%%%%%%

\subsection{The Search for electroweakinos in the light of the Higgs Boson}

This section contains our key results. What we would like to emphasize here is the unique new signature due to $h\to b \bar b$. As discussed in the previous section, this channel is one of the leading channels. According to the production summary in Table \ref{table:signal}, there are significant fractions of the gaugino pair signal decaying to $Wh$ and $Zh$, leading to charged leptons plus $b \bar b$. Not only would this signal have the invariant mass peak $m_{bb}=m_{h}$ as a ``standard candle'' to discriminate against backgrounds, but also it reassures the clear non-SM origin of the Higgs boson from a SUSY parent. There is also the Higgs pair from the decay, but  this mode will be rather challenging due to the large background to the leading signal channel $b \bar b + b \bar b+ \etmiss$.

There exist some related studies on the electroweakino production with $\chi^0_{2,3} \rightarrow \chi^0_1 h$ \cite{Baer:2012ts}.
Our current work makes the most complete compilation for the channels in the MSSM and the comprehensive study for the Higgs boson in the decays, that is then combined with all the other channels to reach the final estimate for the LHC sensitivity. 

Monte Carlo simulations are used to estimate the SM backgrounds, as well as to calculate the efficiency for various electroweakino 
productions. In this study, events are generated using {\sc MADGRAPH} event generator \cite{madgraph} 
and  {\sc PYTHIA}~\cite{Sjostrand:2006za} for parton shower and hadronization. Next-to-leading-order (NLO) cross sections are used for background and signal normalization, calculated using {\sc MCFM}~\cite{Campbell:2010ff} and {\sc PROSPINO}~\cite{Beenakker:1996ed}, respectively. 
For both background and signal samples~\cite{Avetisyan:2013onh}, the events are processed through the
Snowmass detector~\cite{Anderson:2013kxz} using Delphes \cite{deFavereau:2013fsa} parameterized simulation and object reconstruction. Large statistics 
of background samples are generated using the Open Science Grid infrastructure~\cite{Avetisyan:2013dta}. Effects due to additional 
interactions (pile-ups) are studied and they are found to small for 300 fb$^{-1}$ luminosity scenario~\cite{Anderson:2013kxz}.
Jets are reconstructed using the anti-$k_T$ clustering algorithm~\cite{Cacciari:2008gp} with a distance parameter of 0.5, as implemented in the {\sc FASTJET} package~\cite{Cacciari:2011ma}. We have also assumed a systematic uncertainty of $20\%$ in this study.

%%%%%%%%%%%%%%%%

\begin{itemize}
\item{\bf $Wh$ channel: single lepton plus $h\to b \bar b$ analysis}
\end{itemize}

This study focuses on production modes such as $\chi^\pm_1 \chi^0_2$ and $\chi^\pm_1 \chi^0_3$ in the Bino-like LSP case, where $\chi^\pm_1 \rightarrow  \chi^0_1  W^{\pm}, \chi^0_{2,3} \rightarrow \chi^0_1 h$, with $h \rightarrow b \bar{b}$ in the 
final state, as listed in Table \ref{table:signal} and Fig.~\ref{fig:CSBR}. The $Wh$ mode may take place in all of the three cases of A, B, C as a leading production channel, although the LSPs may have rather  different properties.  Observationally, 
this is similar to the event topology of single lepton channel: $\ell^{\pm} + {\rm jets} +\etmiss $, in which there is a resonant 
production of $h\to b \bar b$. We consider the following event selection for this study:  
    \begin{enumerate}
    \item Exactly one lepton with $p_T^\ell > 25$ GeV, $|\eta^{\ell}| < 2.5$ and veto any isolated track with $p_T > 10$ GeV 
within the tracker acceptance of $|\eta| < 2.5$ as well as hadronic $\tau$'s with $p_T > 20$ GeV and $|\eta| < 2.5$.
    \item Exactly two $b$-tag jets with $p_T^{b_1,b_2} > 50, 30$ GeV, $|\eta^b| < 2.5$ and are expected to be in one hemisphere of the transverse plane.
    \item Invariant mass of the $b$-jets must be within 100 GeV$<m_{bb}< 150$ GeV.
    \item Transverse mass ($M_{T}^{\etmiss,h}$) between  $\etmiss$  and the Higgs    $>$ 200 GeV and $\etmiss > 100$ GeV.  
    \item Difference in azimuthal angle $\Delta\phi^{\etmiss,h} > 2.4$ between  $\etmiss$ and the Higgs boson.
    \end{enumerate}
Several signal regions are defined using combination of variables, including $\etmiss$, $p_T$-axis of hemisphere containing the lepton, 
$m_{eff}$ as the scalar sum of $p_T^\ell$, $p_T^b$ and $\etmiss$, and $M_{T2}^{b\ell}$ variable.
%~\cite{Bai:2012gs}. 
We use the best signal significance from all of the signal regions to determine the sensitivity. The dominant SM backgrounds for this signal come  from $t\bar{t}$, single tops, $Wb\bar{b}$ and dibosons productions.

\begin{figure}
\minigraph{8.1cm}{-0.2in}{(a)}{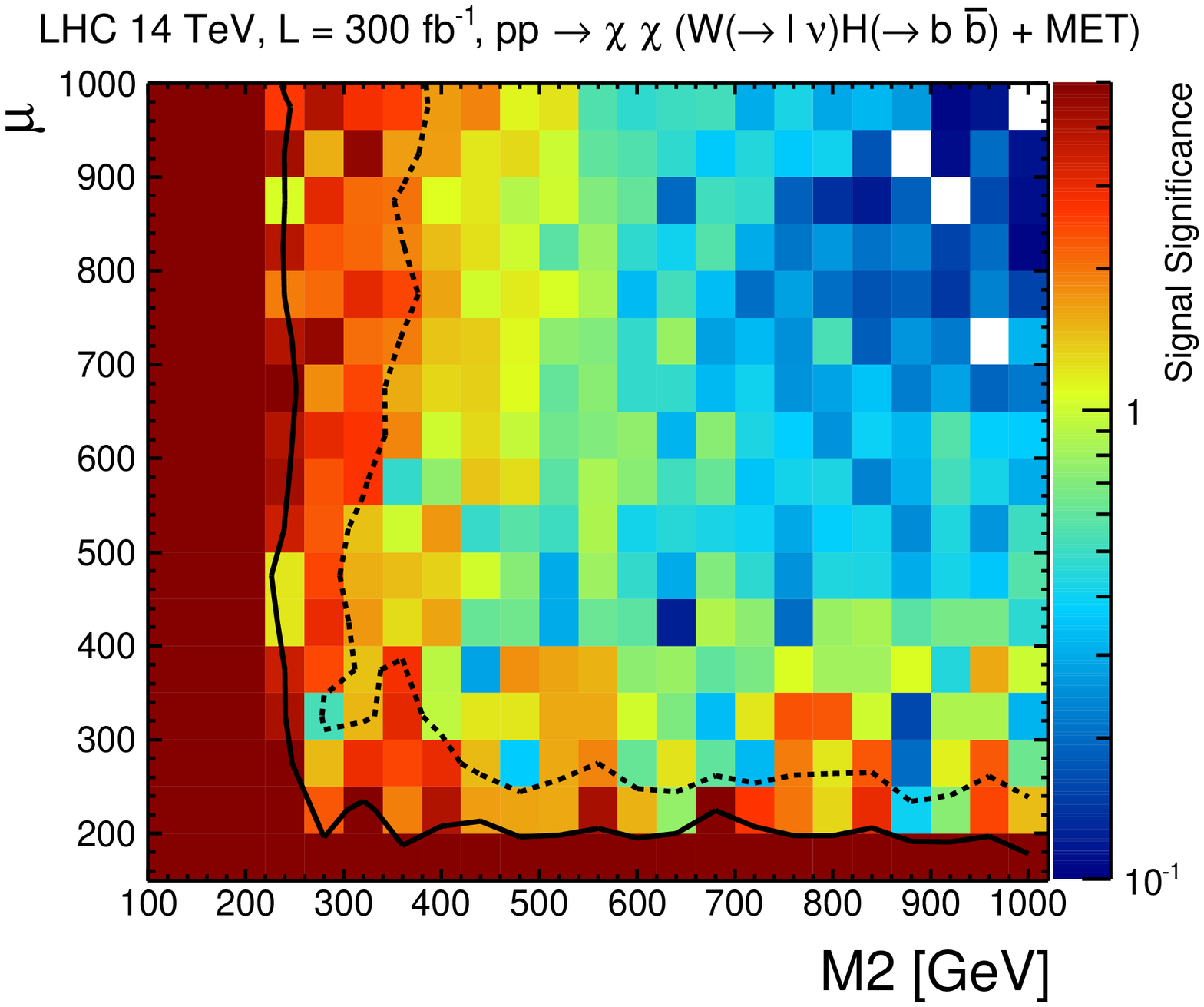}
\minigraph{8.1cm}{-0.2in}{(b)}{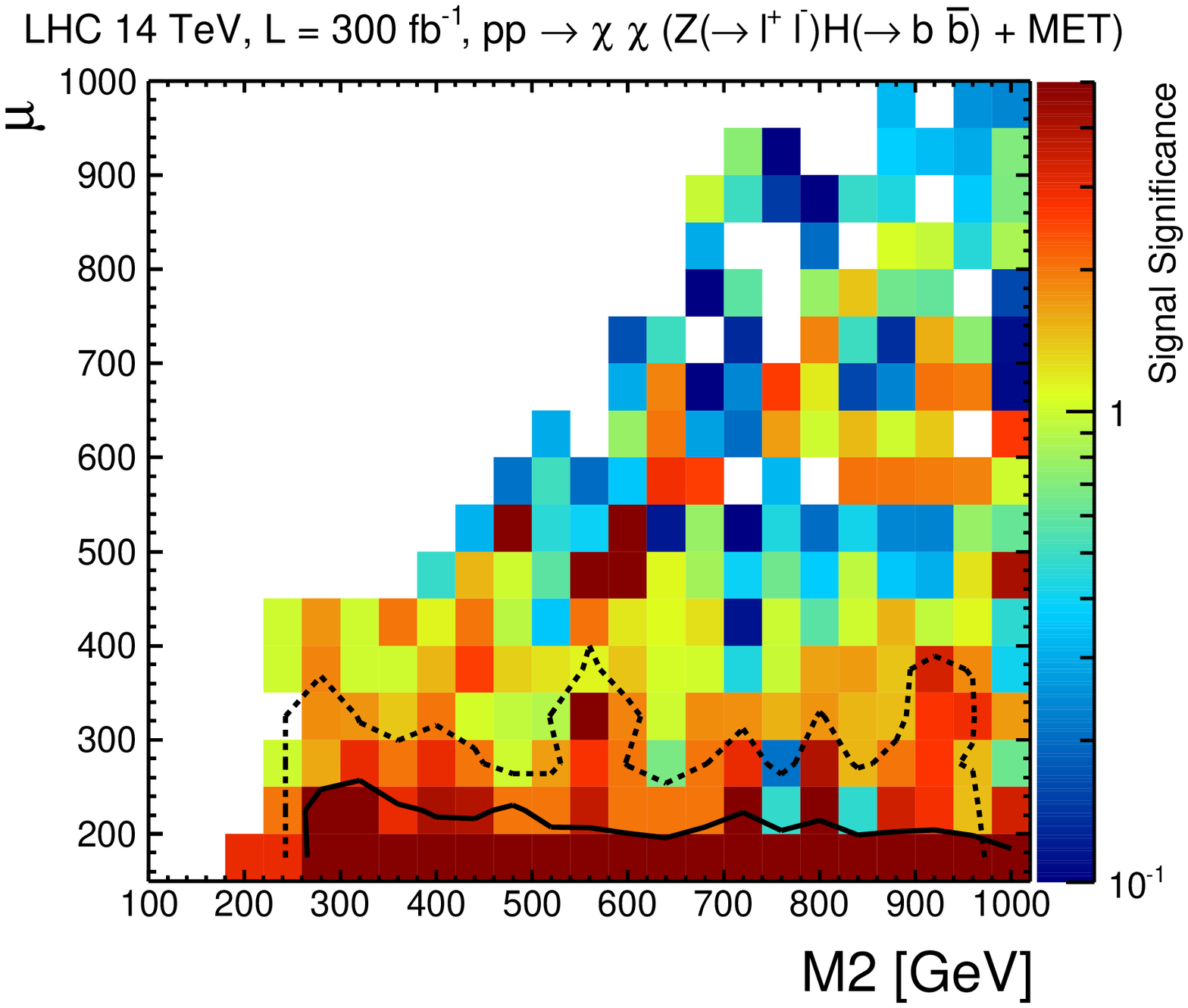}
\caption{
Sensitivity reach at the 14 TeV LHC with 300 fb$^{-1}$  for Case A Bino-like LSP in $\mu-M_{2}$ plane,  (a) for $Wh$ ($\ell bb+\etmiss$) and (b) for $Zh$ ($\ell\ell bb+\etmiss$) channels.   The statistical significance is labelled by the color code on the right-hand side. The solid and dashed curves indicate the $5 \sigma$ discovery and  95\% C.L. exclusion   reach. The other MSSM parameters are set to be $M_1=0$ GeV, $\tan\beta=10$ and $\mu>0$.   
}
\label{fig:LHC14bb}
\end{figure}

The sensitivity reach for $Wh \to \ell bb+\etmiss$ is shown in Fig.~\ref{fig:LHC14bb}(a) for Case A Bino-like LSP at the 14 TeV LHC with 300 ${\rm fb^{-1}}$. We take $M_{1}=0,\ \mu>0,\ \tan\beta = 10$, but with arbitrary mixing in $\mu-M_{2}$ plane. We see that the 95\% C.L.~($5\sigma$) reach for $M_2$ is about 400 GeV (250 GeV). 
The asymptotic reach in $\mu$ is slightly less comparing to that of $M_2$, giving about 250 GeV (200 GeV) for $95\%$ C.L.~($5\sigma$). This is due to that $\chi_{2,3}^0$ decays to $\chi_1^0 h$ only half of the time in Case AII, while $\chi_2^0$ dominantly decays via $h$-channel in Case AI.  

%%%%%%%%%%%%%%%%
 
\begin{itemize}
\item{ \bf $Zh$ channel: di-lepton plus $h\to b\bar b$ analysis}
\end{itemize}

    This study focuses on production modes such as $\chi_2^0\chi_3^0$ in the Bino-like LSP, Higgsino-like NLSPs case, where $\chi_{2,3}^0 \rightarrow \chi_1^0h, \chi_1^0 Z$ as listed in Table \ref{table:signal} and Fig.~\ref{fig:CSBR}.   The $Zh$ mode may also take place  in Cases BII and CII.  This channel is similar to the event topology of   opposite sign di-lepton channel:   $\ell^{+} \ell^{-} + {\rm jets} + \etmiss$, again with the di-jet as $h \to b \bar b$. We consider the following event selection for this study:  
    \begin{enumerate}
    \item Exactly two opposite sign same flavor leptons (OSSF) with $p_T^{\ell_1, \ell_2} > 50, 20$ GeV, $|\eta^{\ell}| < 2.5$ and veto any isolated track 
      with $p_T > 10$ GeV within the tracker acceptance of $|\eta| < 2.5$  as well as hadronic $\tau$'s with $p_T > 20$ GeV and $|\eta| < 2.5$.
    \item Exactly two $b$-tag jets with $p_T^{b_1, b_2} > 50, 30$ GeV, $|\eta^b| < 2.5$ and are expected to be in one hemisphere of the transverse plane.
    \item Invariant mass of the $b$-jets must be within 100 GeV$<m_{bb}< 150$ GeV.
    \item Invariant mass of OSSF dileptons be within 76 GeV$<m_{\ell^+ \ell^-}< 106$ GeV.
    \item $\etmiss > 50$ GeV.  
    \item Difference in azimuthal angle $\Delta\phi^{\etmiss,h} > 1.0$ between  $\etmiss$ and the Higgs boson.
     \end{enumerate}
    Several signal regions are defined using combination of variables, including $\etmiss$, leading lepton $p_T^\ell$, 
$p_T$-axis of hemisphere containing di-lepton, $m_{eff}$, $M_{T}^{\etmiss, h}$, and $M_{T2}^{Zh}$. 
The dominant SM backgrounds for this signal are from $t\bar{t}$, single top associated with a boson, $Zb\bar{b}$ and dibosons.

%%%%

The $Zh \to \ell\ell bb+\etmiss$ channel has less SM background than the $Wh$ mode, and is promising in the region of  $|\mu|<M_2$. The sensitivity reach is shown in Fig.~\ref{fig:LHC14bb}(b). The 95\% C.L.~($5\sigma$) reach is about 
$\mu \sim 300$ GeV (200 GeV). In Fig.~\ref{fig:LHC14bb}, the white spots indicate the region where the sensitivity is 
weaker than approximately 0.1 as we plotted. Note that no sensitivity in $Zh$ channel is obtained for $M_2<\mu$ (Case AI) since such final states do not appear, as shown in Table \ref{table:signal}.  
We combine the Higgs boson channels $Wh$ and $Zh$ together and present the sensitivity reach in Fig.~\ref{fig:LHC14_all}(a). The summary results for their mass reach are shown in the first column in Table \ref{tab:LHC14}.  

%%%%%%%%%%

\begin{figure}
\minigraph{8.1cm}{-0.2in}{(a)}{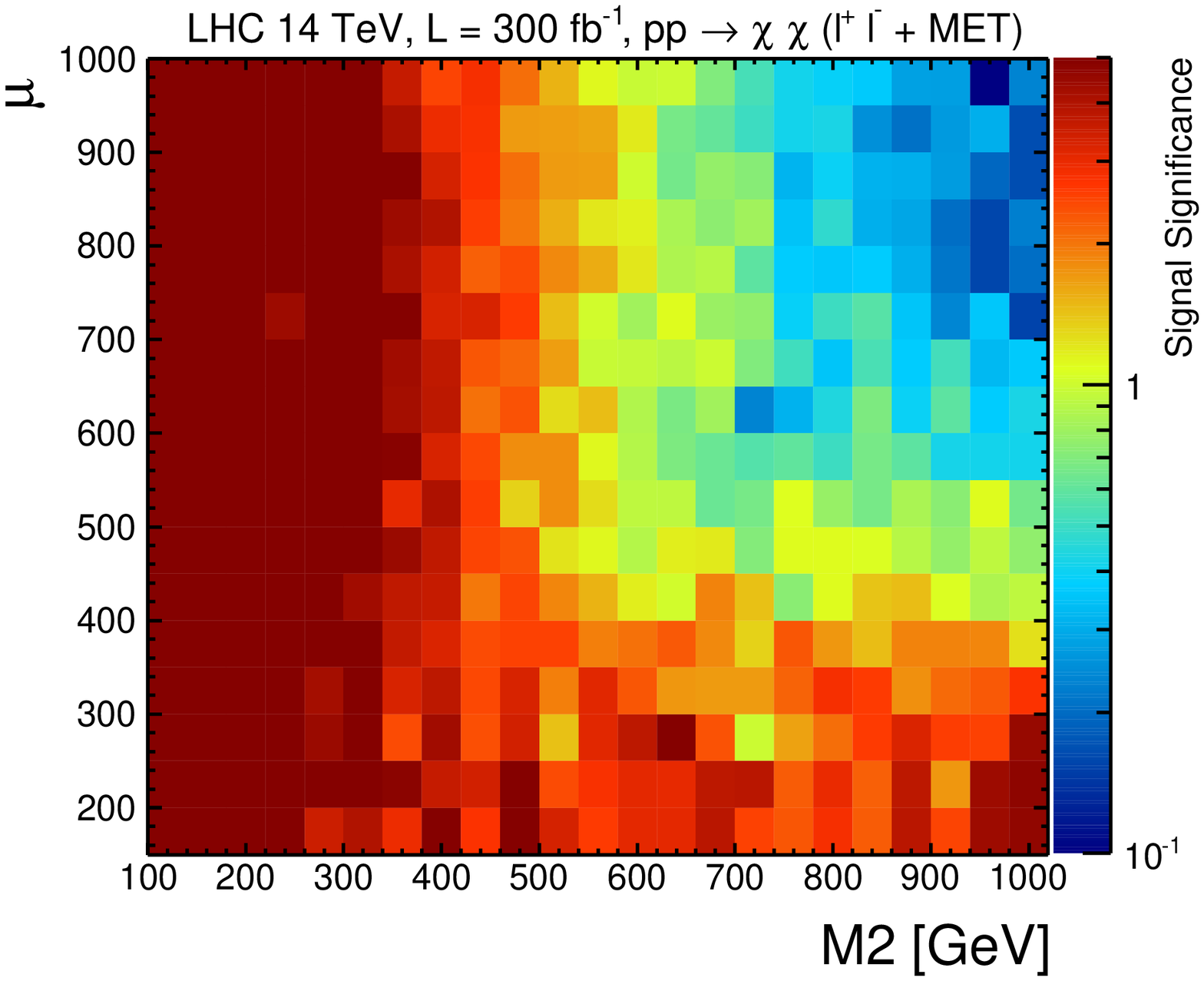}
\minigraph{8.1cm}{-0.2in}{(b)}{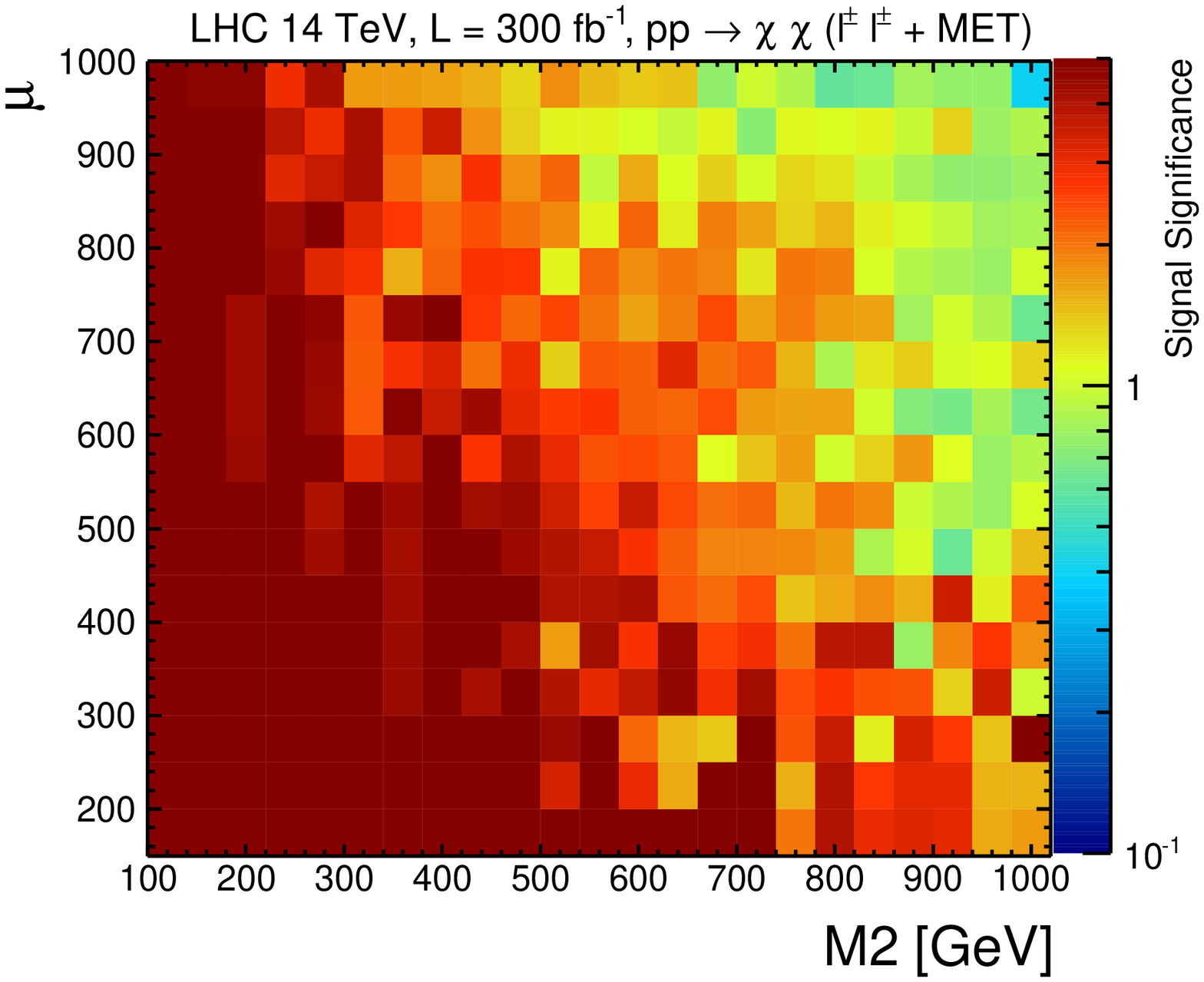}
\minigraph{8.1cm}{-0.2in}{(c)}{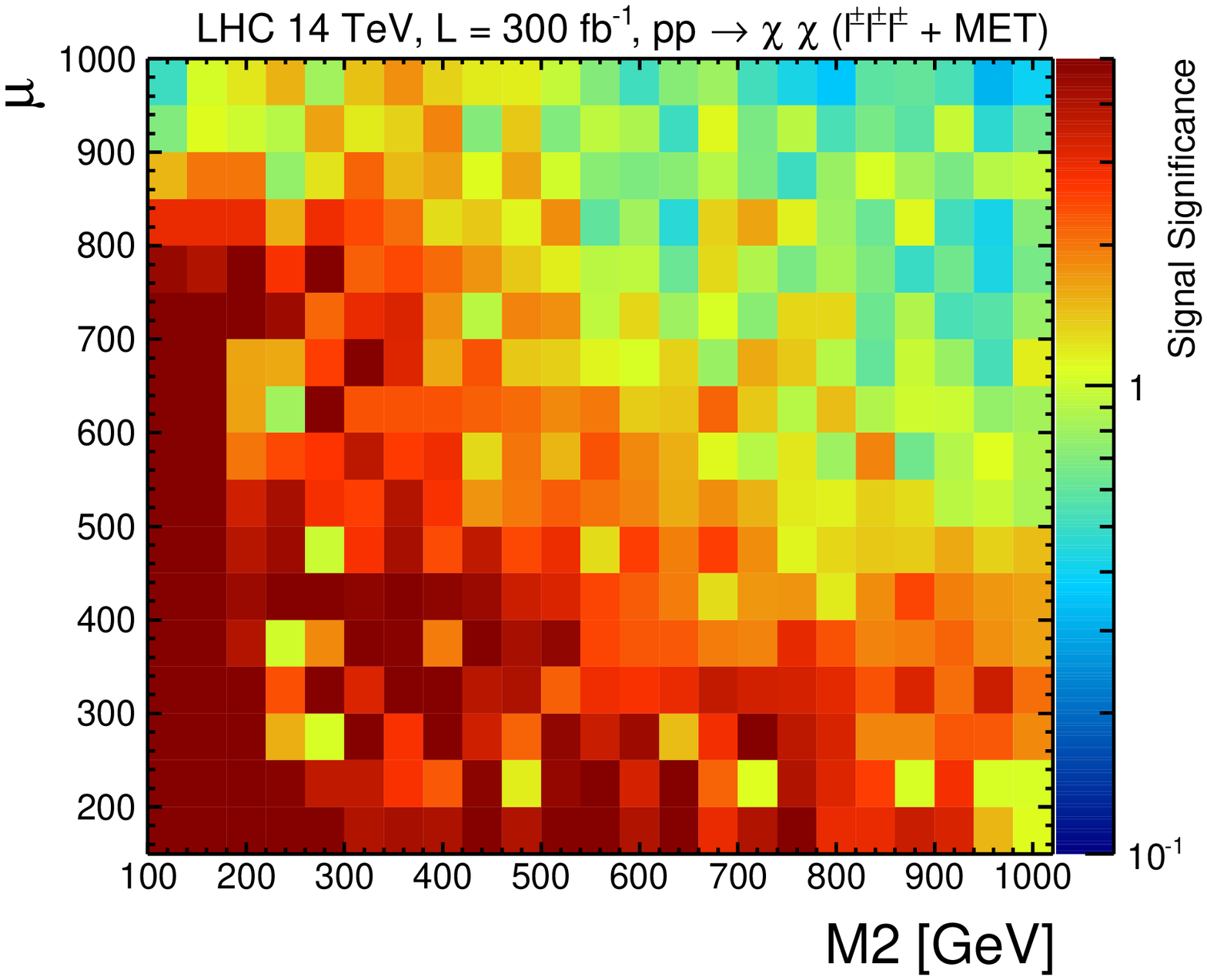}
\minigraph{8.1cm}{-0.2in}{(d)}{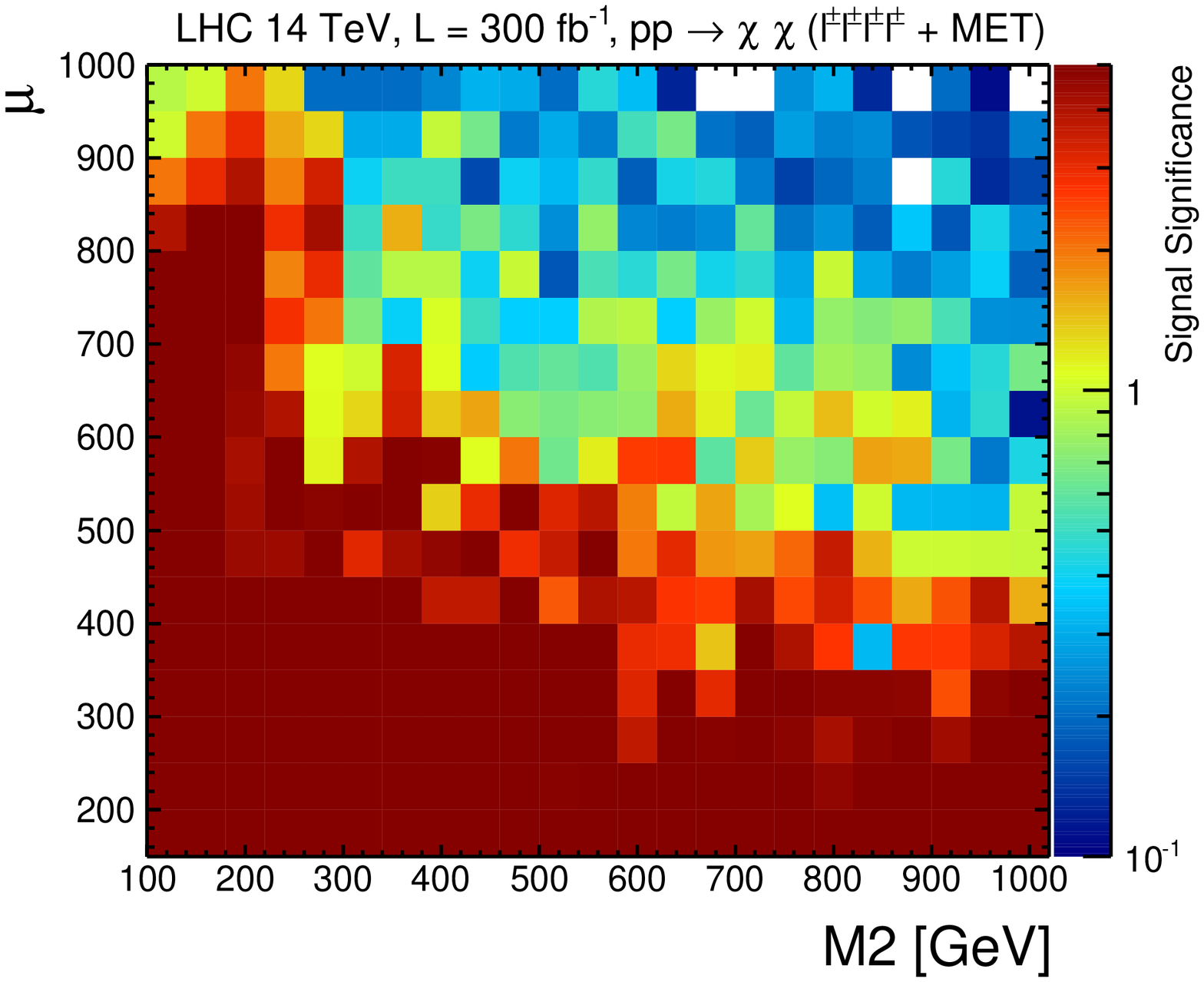}
 \caption{
Sensitivity reach at the 14 TeV LHC with 300 fb$^{-1}$  for Case A Bino-like LSP in $\mu-M_{2}$ plane,  for 
(a) OSWW, (b) SSWW, (c) 3L and (d) 4L channels.   The statistical significance is labelled by the color code on the right-hand side. 
%The solid and dashed curves indicate the  $5 \sigma$  discovery and 95\% C.L. exclusion   reach. 
The other MSSM parameters are set to be $M_1=0$ GeV, $\tan\beta=10$ and $\mu>0$.  
}
 \label{fig:LHC14_rest}
\end{figure}

%%%%%%%%%%%%%%%%%%%%%%%%%

\subsection{Combined Results for All Channels}

For completeness, we combine the Higgs channels studied above with the other conventional electroweakino search channels, in which we have also included the contributions from $h\to WW^{*},\ ZZ^{*}$ in the due course.  It would be informative to first compare the signal significance involving $h\to b \bar b$ with the other  channels. We show this in Fig.~\ref{fig:LHC14_rest} again for Case A Bino-like LSP with $M_{1}=0,\ \mu>0,\ \tan\beta = 10$, but with arbitrary mixing in $\mu-M_{2}$ plane. 
\begin{itemize}
\item{OSWW}: $\ell^+\ell^- + \etmiss$ with  jet veto and $Z$ veto.  While signal dominantly comes from $W^+W^-$ final states,  $Wh(\rightarrow WW^*)$ with one missing lepton could also contribute as well. We use the same event selection as in CMS $h \rightarrow W^+ W^-$ study \cite{CMSWW}
with jet veto. Signal regions are defined using $\etmiss$, $m_{eff}$, $M_{T2}^{W}$ and $M_{T}^{\etmiss, \ell}$.
\item{SSWW}: $\ell^\pm\ell^\pm + {\rm jets} + \etmiss$ with signal dominantly from $W^\pm W^\pm$ final states, or $Wh\rightarrow WWW$ with two $W$ decay leptonically and one $W$ decay hadronically.  We select same sign di-leptons with veto on b-tagged jets as well as any additional lepton.
Signal regions are based on $\etmiss$, $p_T^{\ell_1}$ , $m_{eff}$ and $M_{T2}^{W}$.
\item{3L}: $\ell\ell\ell+{\rm jets}+ \etmiss$, with signals dominantly  from $WZ$ final states, or $Wh, Zh$ with $h\rightarrow WW^*, ZZ^*$.  
We select tri-leptons with $p_T^{\ell_1, \ell_2, \ell_3} > 20, 20, 7(5)$ GeV using electrons(muons) with veto on  b-tagged jets. Signal regions are based on 
$\etmiss$, $m_{eff}$, $M_{T2}^{W}$ and on-shell $Z$ in case of opposite sign same flavor leptons with invariant mass within 
$60 < m_{\ell^+ \ell^-} < 120$ GeV. If on-shell $Z$ boson is found, asymmetric $M_{T2}$ is computed using $Z$, $\etmiss$ and the 3rd lepton.
 \item{4L}: $\ell\ell\ell\ell +{\rm jets} + \etmiss$, with signals dominantly   from $ZZ$ final states, or $Wh, Zh$  with $h\rightarrow WW^*, ZZ^*$. 
\end{itemize}
  
As expected, we see that the OSWW mode is more sensitive to Case AI with $M_{2}< \mu$ reaching $M_2\sim 500$ GeV (400 GeV) for $95\%$ C.L.~(5$\sigma$) for any value of $\mu$.
Similar feature appears for SSWW channel sensitive to the small $M_2$ region, with the dominant contributing channel from $Wh$ with $h\rightarrow WW, ZZ$ and $\tau\tau$. The more interesting probe from this channel occurs when $M_{2}\approx \mu$ where a 5$\sigma$ sensitivity for a 500 GeV mass scale can be achieved.    
The 3L and 4L modes, on the other hand, are more sensitive to Case AII with $\mu < M_{2}$.
The 3L mode can reach $\mu\sim 350$ GeV at $95\%$ C.L.~for asymptotic value of $M_{2}$.\footnote{We note that our results for the 3L mode in the Case AI are less sensitive comparing to the ATLAS and CMS studies \cite{ATLAS_CMS_3L_future}, in which a 5$\sigma$ sensitivity was expected for 500 $-$ 600 GeV for 300 ${\rm fb}^{-1}$ luminosity. This is due to the fact that their results were obtained under the assumption of 100\% branching fraction for the $WZ+\etmiss$ final state, while in the realistic case with $\mu>0$, such branching fraction is only about 20\% or less. }
The 4L channel has the lowest SM backgrounds, and a 5$\sigma$ reach in the $\mu$ parameter can be obtained  around 350 GeV.    

%    
% The 95\% C.L. exclusion limit for  the NLSP mass reach can be about $M_{2}\sim xxx$ GeV,  $\mu \sim xxx$ GeV 
%{\bf more...} 
  
\begin{figure}
 \minigraph{8.1cm}{-0.2in}{(a)}{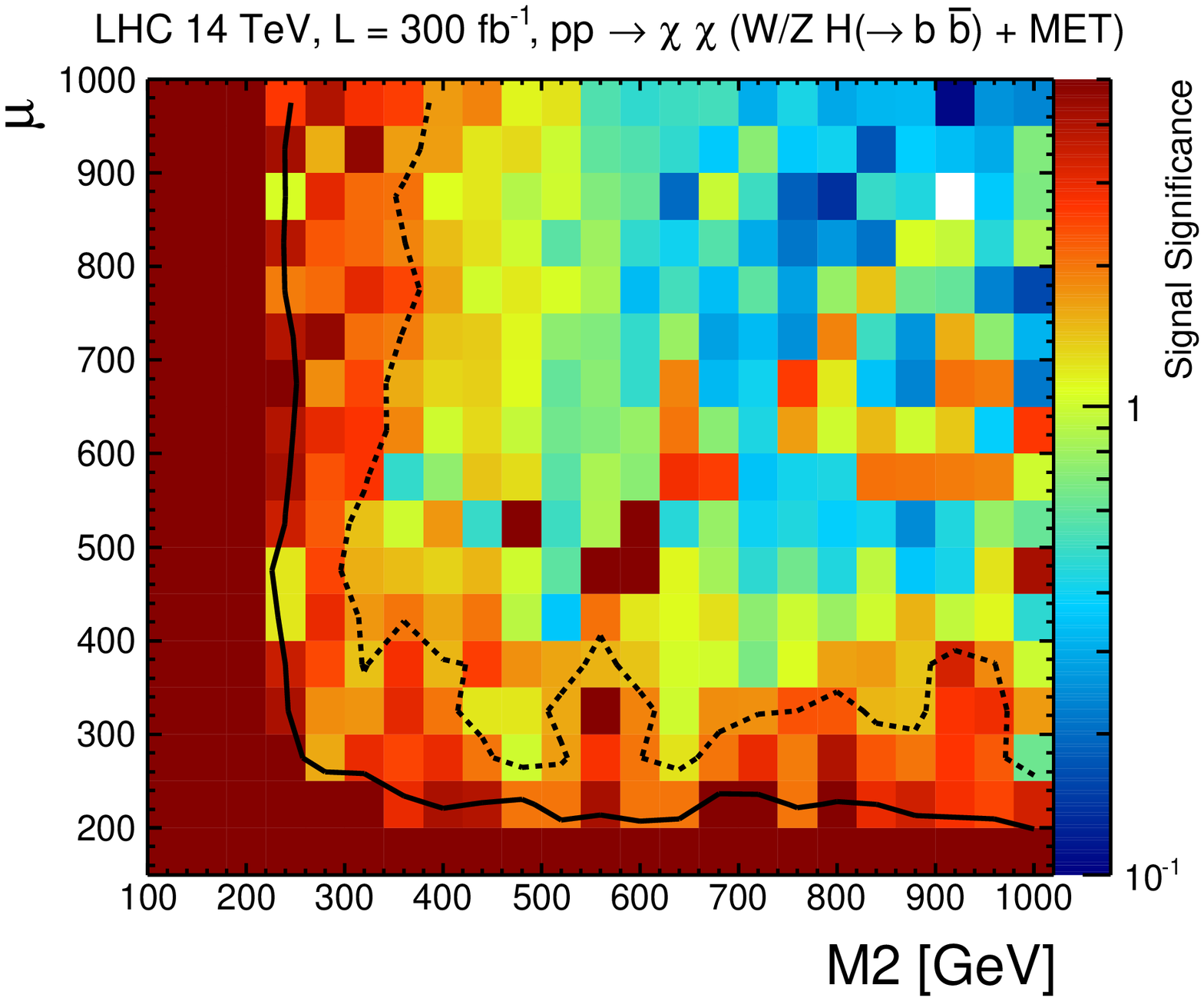}
  \minigraph{8.1cm}{-0.2in}{(b)}{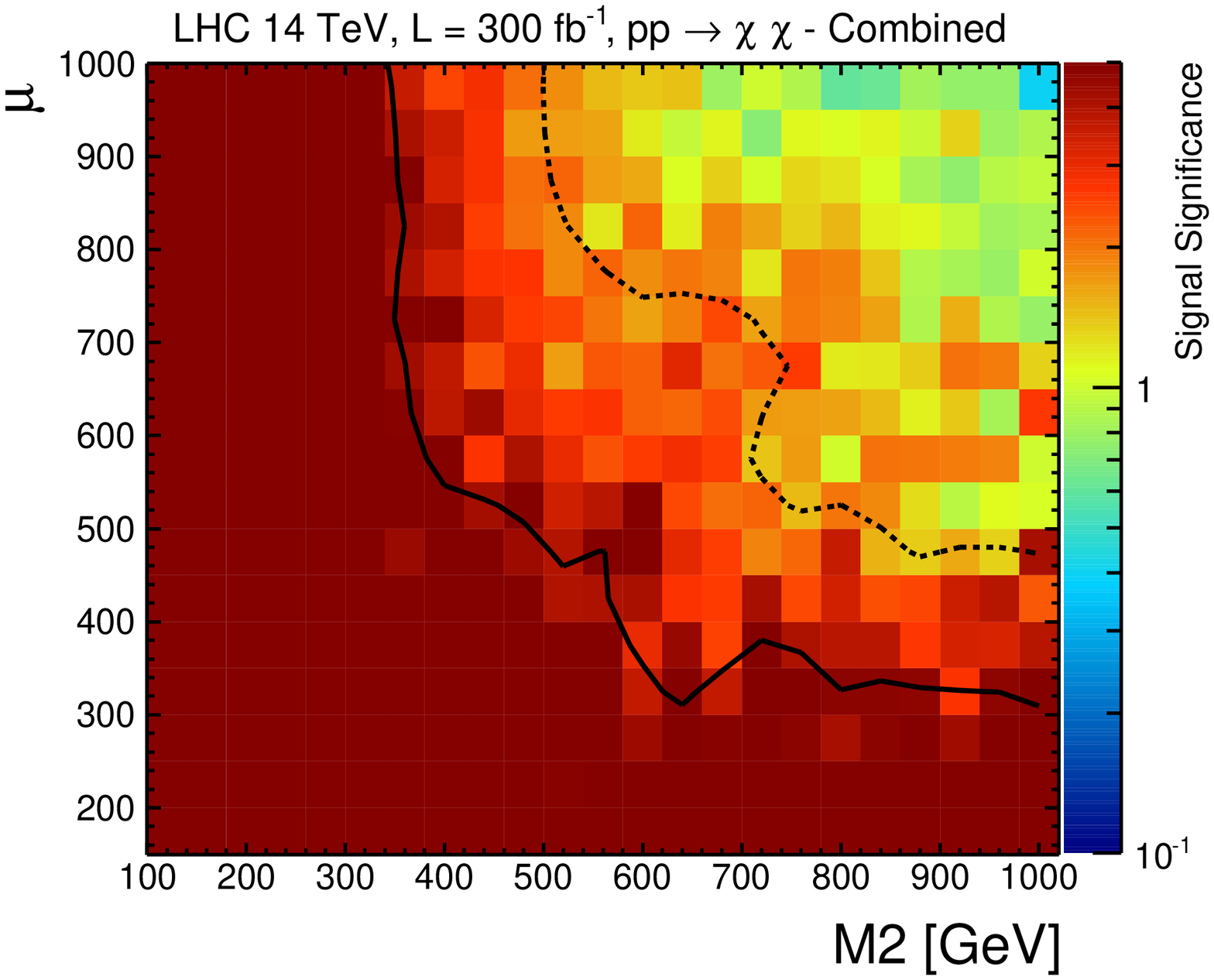}
  \caption{
Combined sensitivity reach at the 14 TeV LHC with 300 fb$^{-1}$ for Case A Bino-like LSP in $\mu-M_{2}$ plane, (a) for the Higgs final states $Wh+Zh$ with $h\to b\bar b$, and (b) for all final states. 
The statistical significance  is labelled by the color code on the right-hand side. 
The solid and dashed curves indicate the  $5 \sigma$ discovery  and 95\% C.L. exclusion reach. 
The other MSSM parameters are set to be $M_1=0$ GeV, $\tan\beta=10$ and $\mu>0$.  
} 
\label{fig:LHC14_all}
\end{figure}

Based on those detailed analyses above, we show the combined sensitivity reach in Fig.~\ref{fig:LHC14_all}(b) in the $\mu-M_2$ plane using all the six channels (two from $Wh/Zh,\ h\to b \bar b$ and four from the conventional multi-lepton searches), again for Case A Bino-like LSP at the 14 TeV LHC  with 300 fb$^{-1}$ integrated luminosity.  The reach for 95\% C.L. exclusion and 5$\sigma$ discovery based on Fig.~\ref{fig:LHC14_all} are summarized in Table \ref{tab:LHC14}. The robust search results from $Wh,\ Zh$ with $h \to b \bar b$ are separately listed in the first column. The final results for the combined channels are summarized in the second column. 
%%%%%%%%%%%%%%%%%%%%%%%%%%%%%

\begin{table}[t]
\begin{tabular}{|c|c|c|}
\hline
  Mass parameters &  $95\%$ C.L.  $(5\sigma)$ reach & $95\%$ C.L. $(5\sigma)$ reach  
  \\  
 &  $2b$-tag from $h\to b \bar b$  &  combined 
 \\ \hline 
Case AI:\ $\mu \gg M_{2} \sim m_{\chi_1^\pm,\chi_2^0}$  &  380 GeV (250 GeV) & 500 GeV (350 GeV)
\\  \hline 
Case AII:\ $M_{2}\gg \mu \sim m_{\chi_1^\pm,\chi_{2,3}^0}$  & 350 GeV (220 GeV)  & 480 GeV (320 GeV)
\\  \hline 
Case A:\  $M_{2} \approx \mu \sim m_{\chi_1^\pm,\chi_{2,3}^0}$  & 400 GeV (270 GeV) &  700 GeV (500 GeV)
\\ \hline  \hline 
  \end{tabular}
\caption{
NLSP electroweakino mass lower bounds at $95\%$ C.L. $(5\sigma)$ from the LHC experiments at 14 TeV and 300 fb$^{-1}$. The results of sensitivity in the first column are from the Higgs final states $Wh+Zh$ with $h\to b\bar b$ as in Fig.~\ref{fig:LHC14_all}(a), and those in the second column are from all the six channel combination as in Fig.~\ref{fig:LHC14_all}(b). Case A with a light Bino-like LSP is assumed. 
}
\label{tab:LHC14}
\end{table}

%%%%%%%%%%%%%%%%%%%%%%%%%%%%%%%%%%%%%%%

\section{Electroweakinos at the ILC}
\label{sec:ilc}

\begin{figure}
\minigraph{7.8cm}{-0.2in}{(a)}{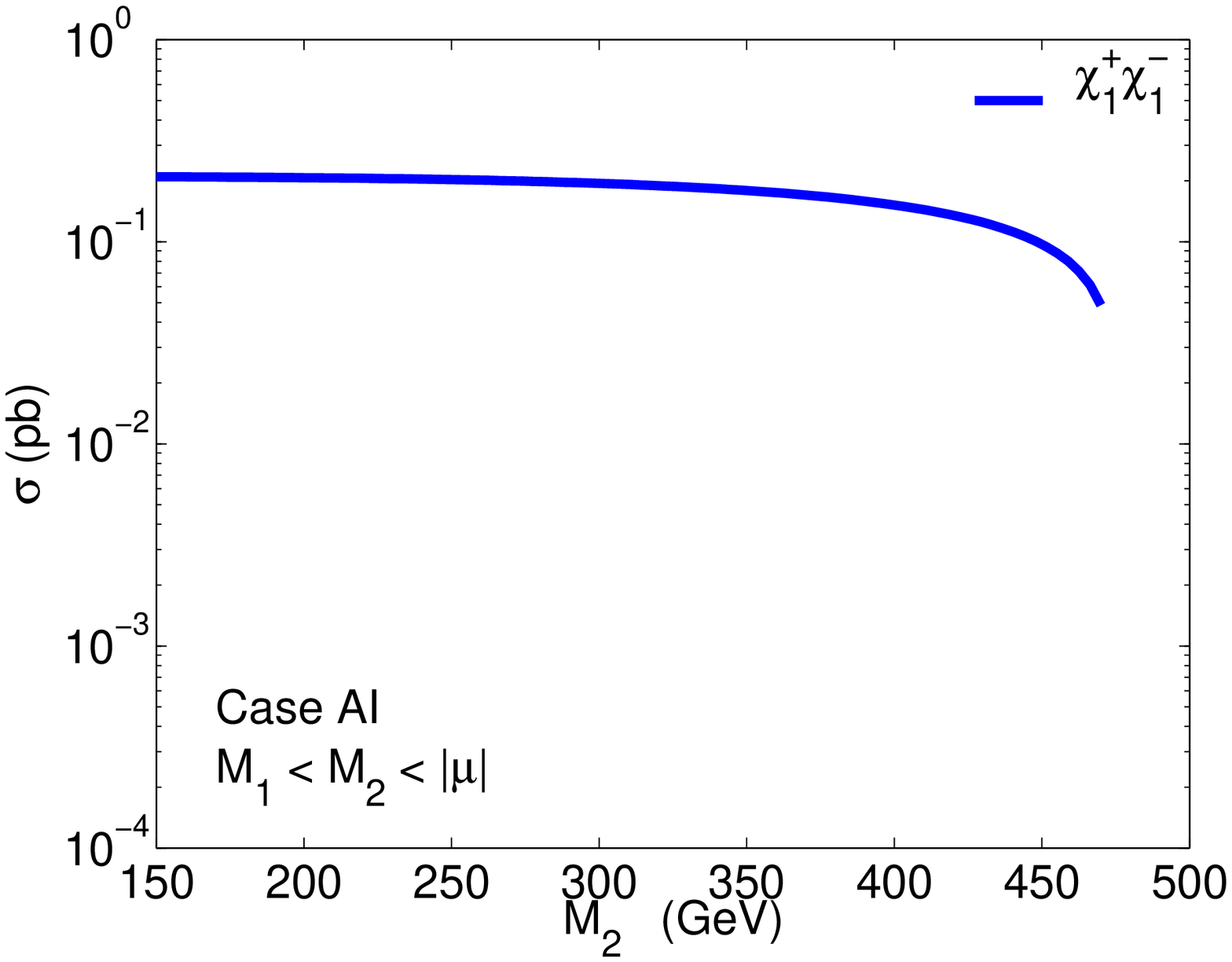}
\minigraph{7.8cm}{-0.2in}{(b)}{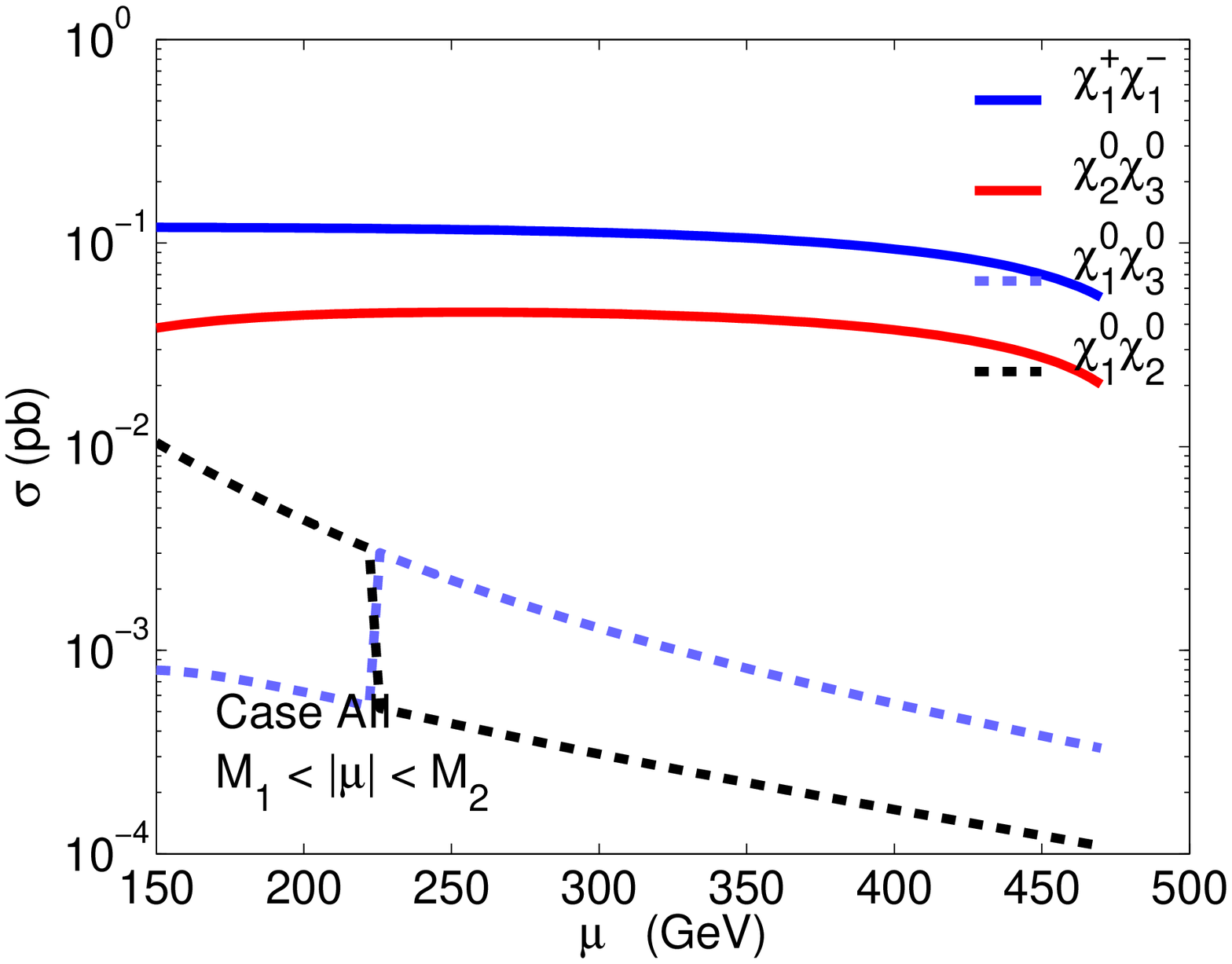}
\minigraph{7.8cm}{-0.2in}{(c)}{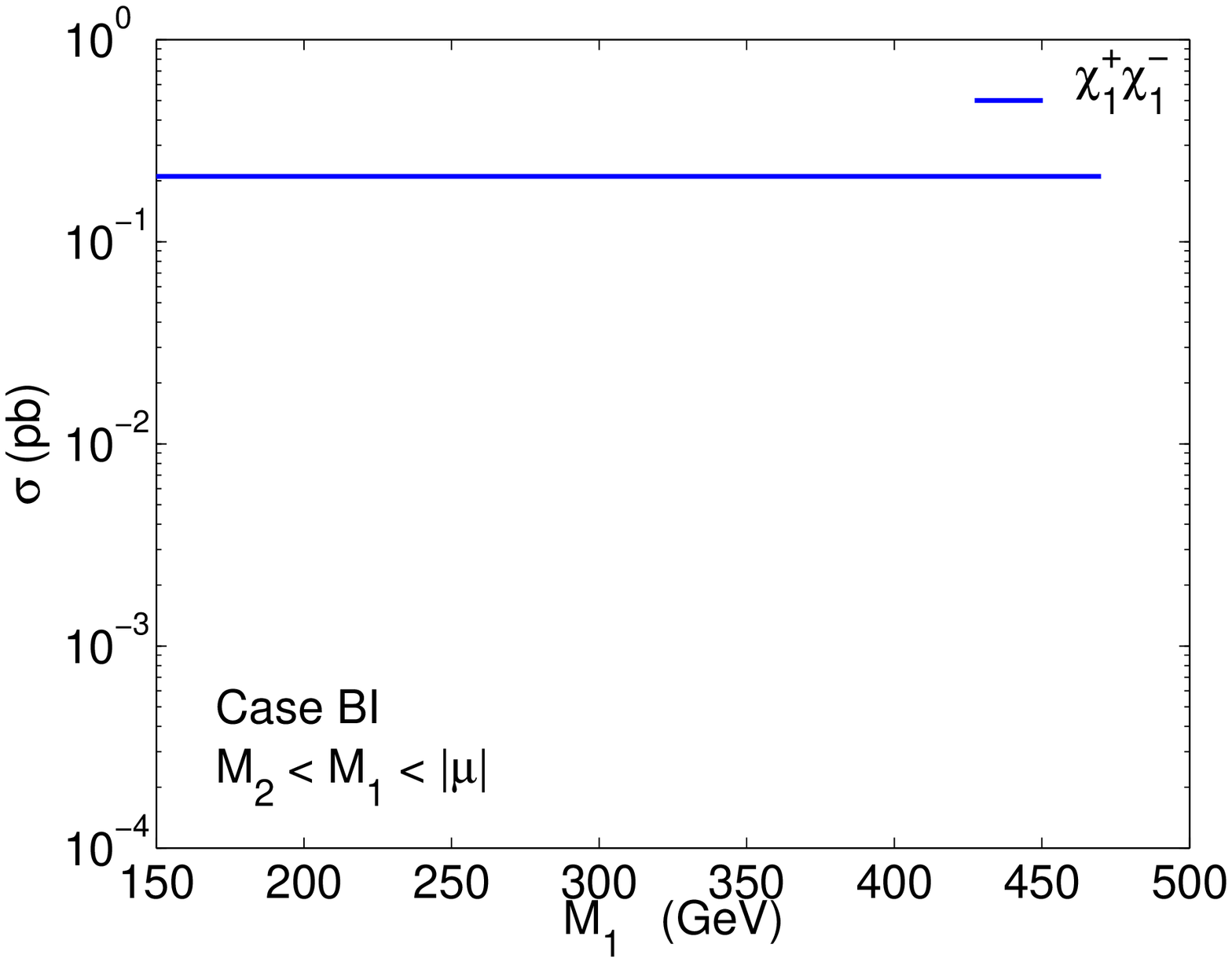}
\minigraph{7.8cm}{-0.2in}{(d)}{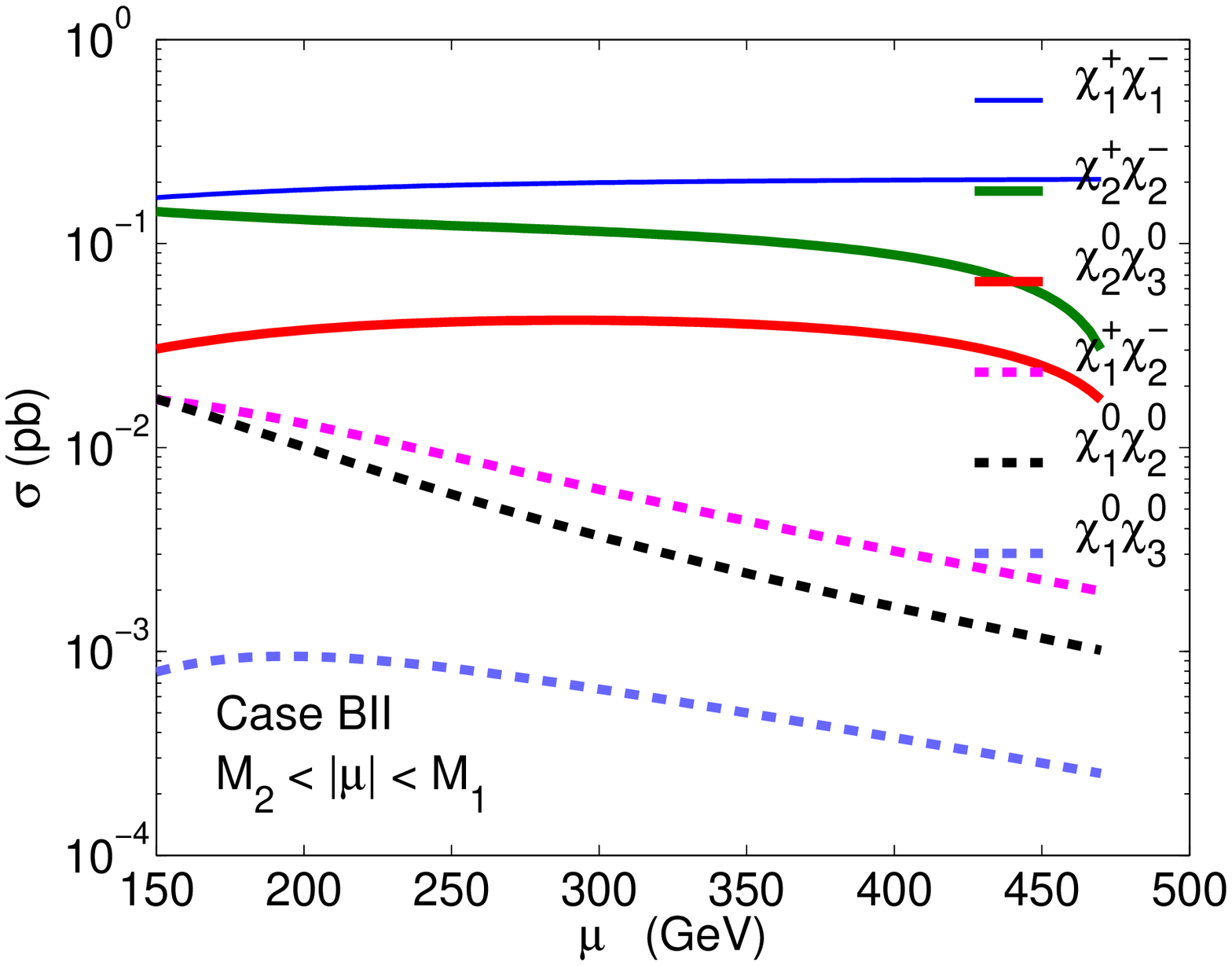}
\minigraph{7.8cm}{-0.2in}{(e)}{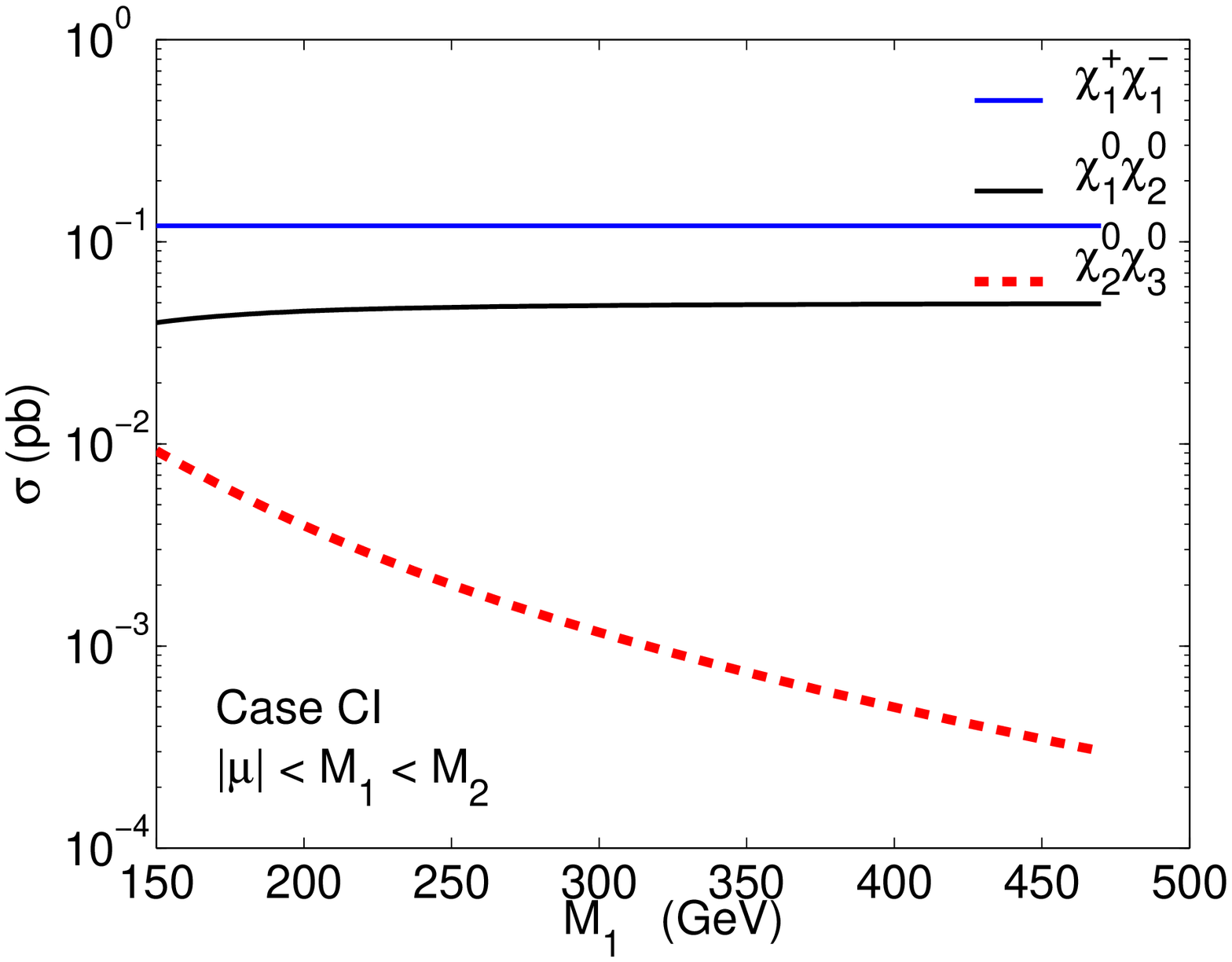}
\minigraph{7.8cm}{-0.2in}{(f)}{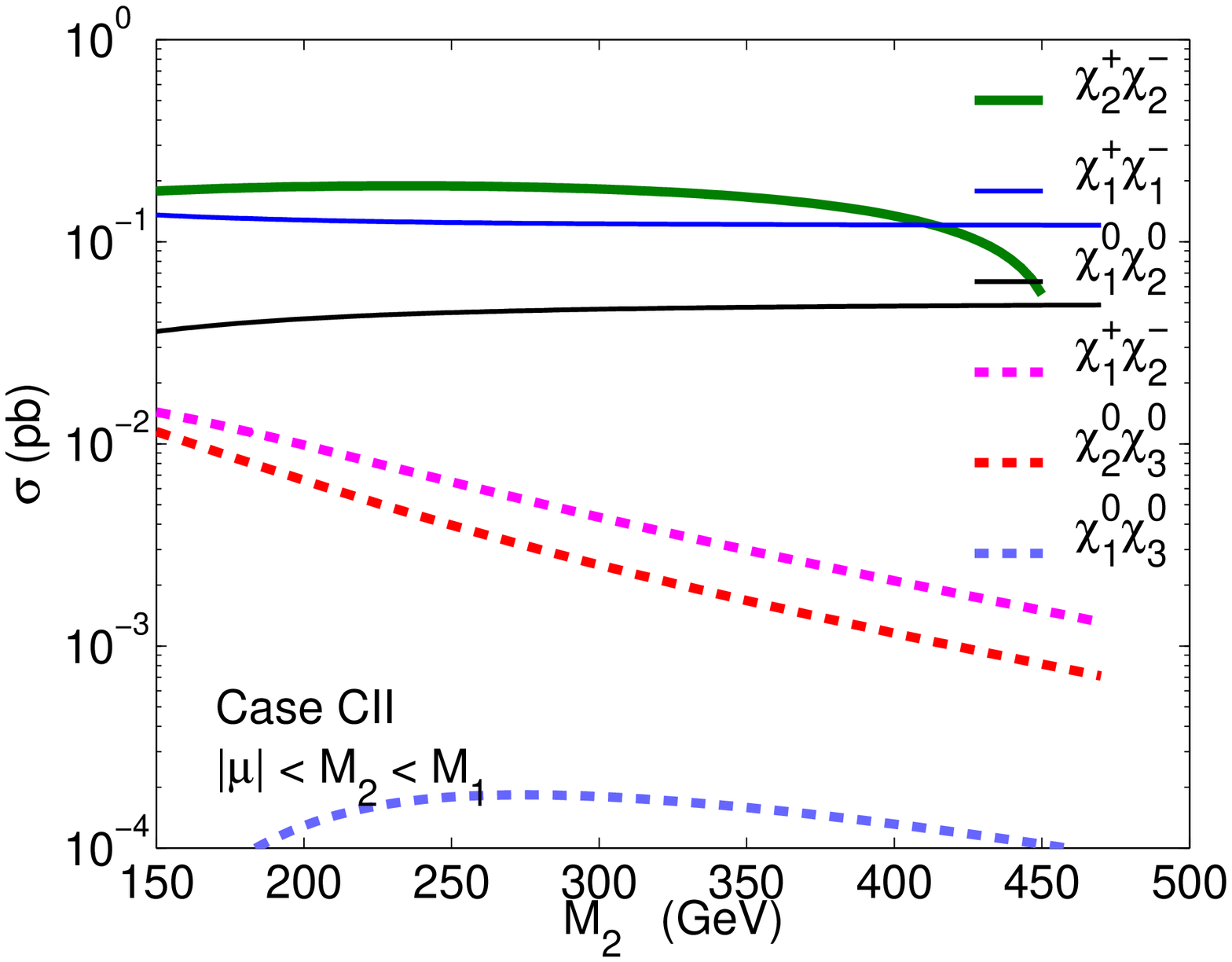}
\caption{
Total cross sections for the chargino and neutralino pair production at the ILC for $\sqrt s = 1$ TeV for all the six cases. 
%(a) Case AI: versus $M_2$ for $M_{1}=100$ and $\mu= 1$ TeV, (b) Case AII: versus $\mu$ for $M_{1}=100$ and $M_{2}=1$ TeV, (c) Case BI: versus $M_1$ for $M_{2}=100$ and $\mu= 1$ TeV, (d) Case BII: versus $\mu$ for $M_{2}=100$ and $M_{1}=1$ TeV, (e) Case CI: versus $M_1$ for $\mu =100$ and $M_{2}= 1$ TeV, (f) Case CII: versus $M_{2}$ for $\mu=100$ and $M_{1}=1$ TeV. 
The parameter choices are the same as in Fig.~\ref{fig:m2mu}.
}
\label{fig:ILC} 
\end{figure}

%%%%%%%%%%%%

Due to the rather small electroweak production cross sections and large SM backgrounds at the LHC, the discovery of the electroweakinos via direct production would be very challenging as discussed in the previous section. Exploiting the additional feature of the Higgs in the final state, the signal observability and identification can be improved. Even if the signal is observed, the determination of the gaugino properties would be very difficult. This is where an ILC would show the major advantage. Similar to the mechanism in Fig.~\ref{fig:Feyn_diam}, the electroweakinos can be produced via the $s$-channel $\gamma/Z$ exchange as in shown in Fig.~\ref{fig:Feyn_diam}(b) and (c). 

The total cross section for the electroweakino pair production at a 1 TeV ILC is shown in Fig.~\ref{fig:ILC} versus the appropriate mass, with (a) and (b) the Bino-like LSP, (c) and (d) the Wino-like LSPs, and (e) and (f) the Higgsino-like LSPs. 
The typical cross sections are quite sizable and are of the order of 100 fb. Once crossing the kinematical threshold, the fermionic pair production reaches the maximum rather soon, while the cross section falls off above the threshold like $1/s$. This scaling law also leads to an estimate at different energies. With the designed annual luminosity of the order 100 fb$^{-1}$, there are plenty of signal events produced, without the major background problem.   Even the sub-leading channels of the NLSPs produced in association with the LSP could be observed.

\begin{figure}
\minigraph{8.1cm}{-0.2in}{(a)}{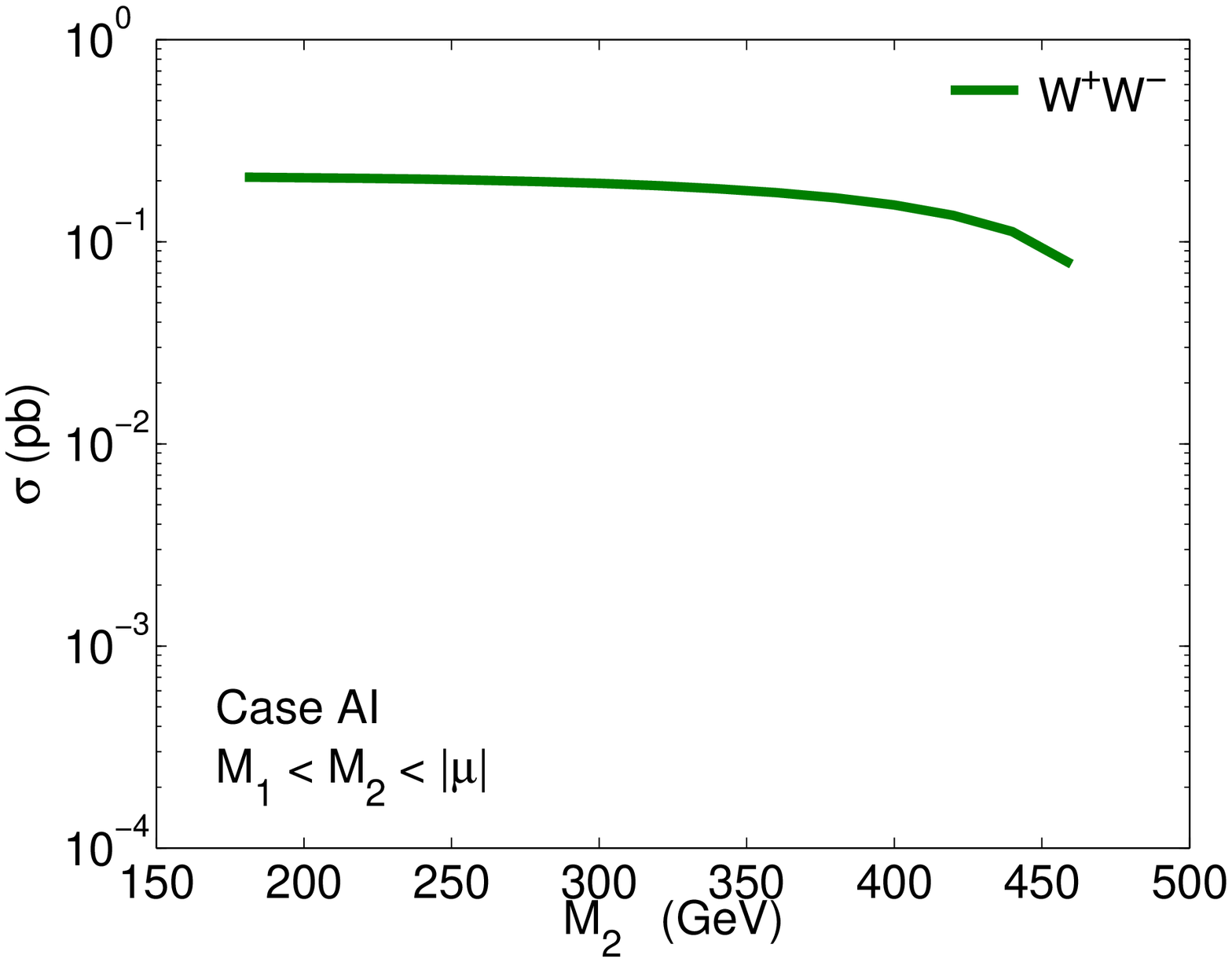}
\minigraph{8.1cm}{-0.2in}{(b)}{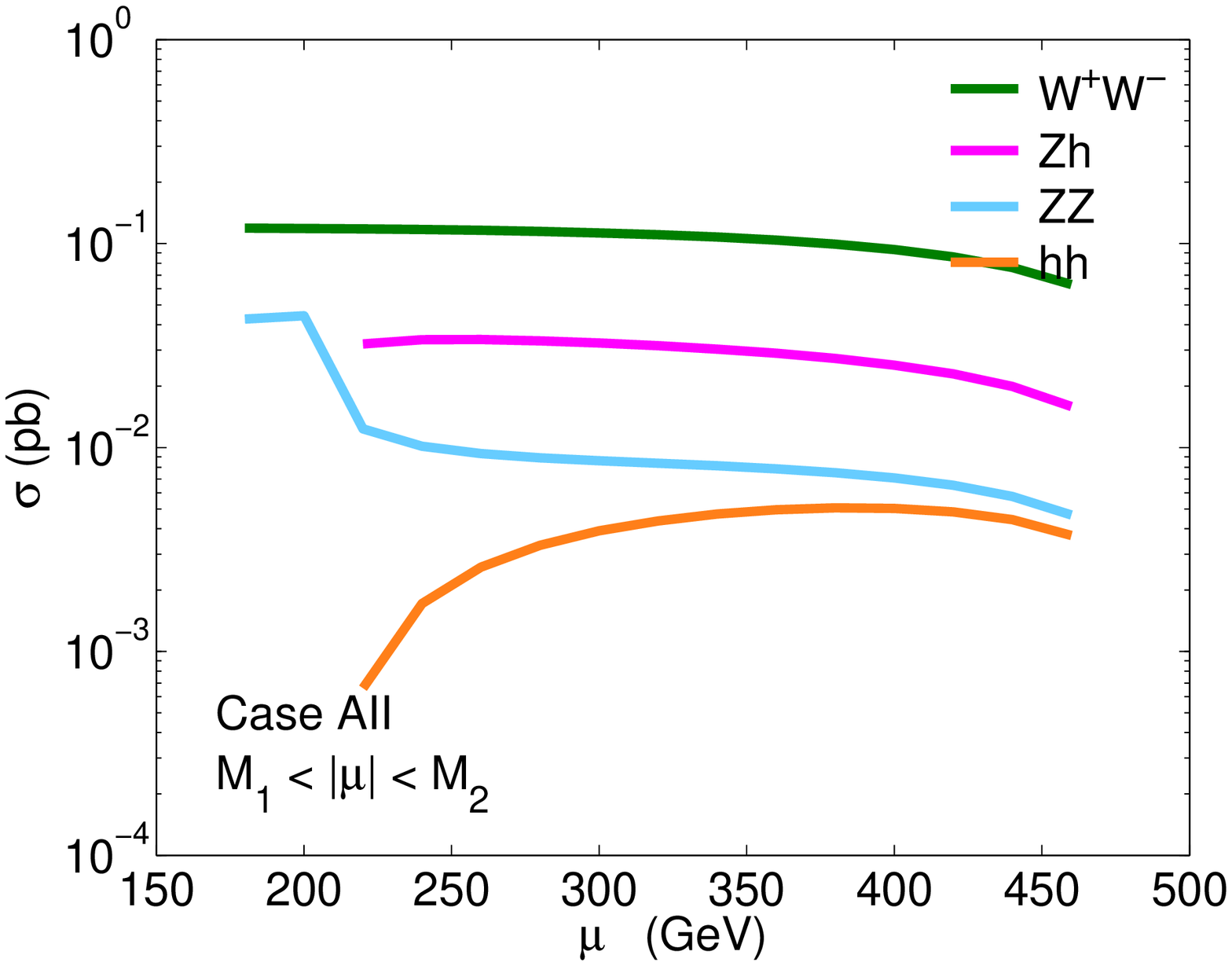}
\minigraph{8.1cm}{-0.2in}{(c)}{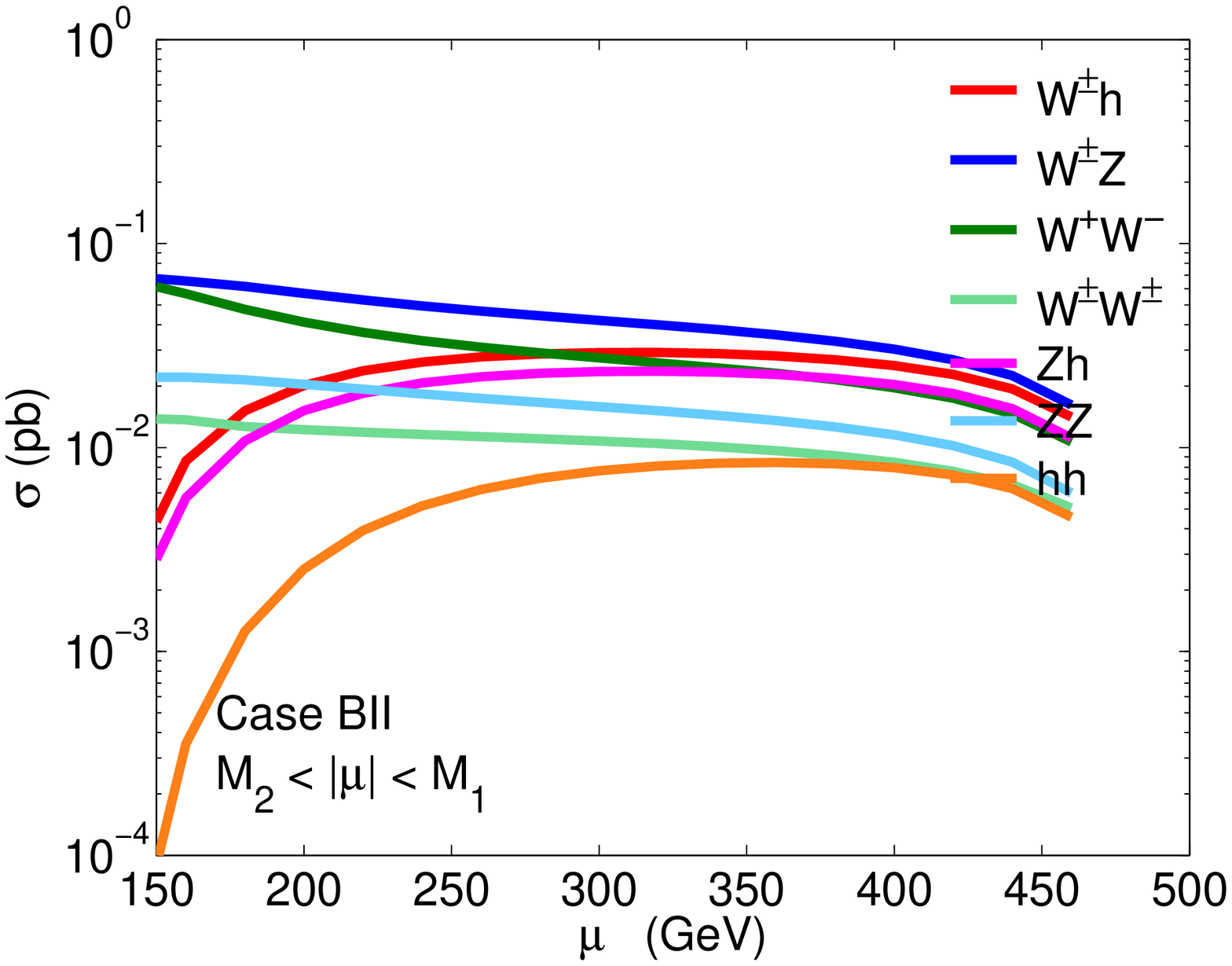}
\minigraph{8.1cm}{-0.2in}{(d)}{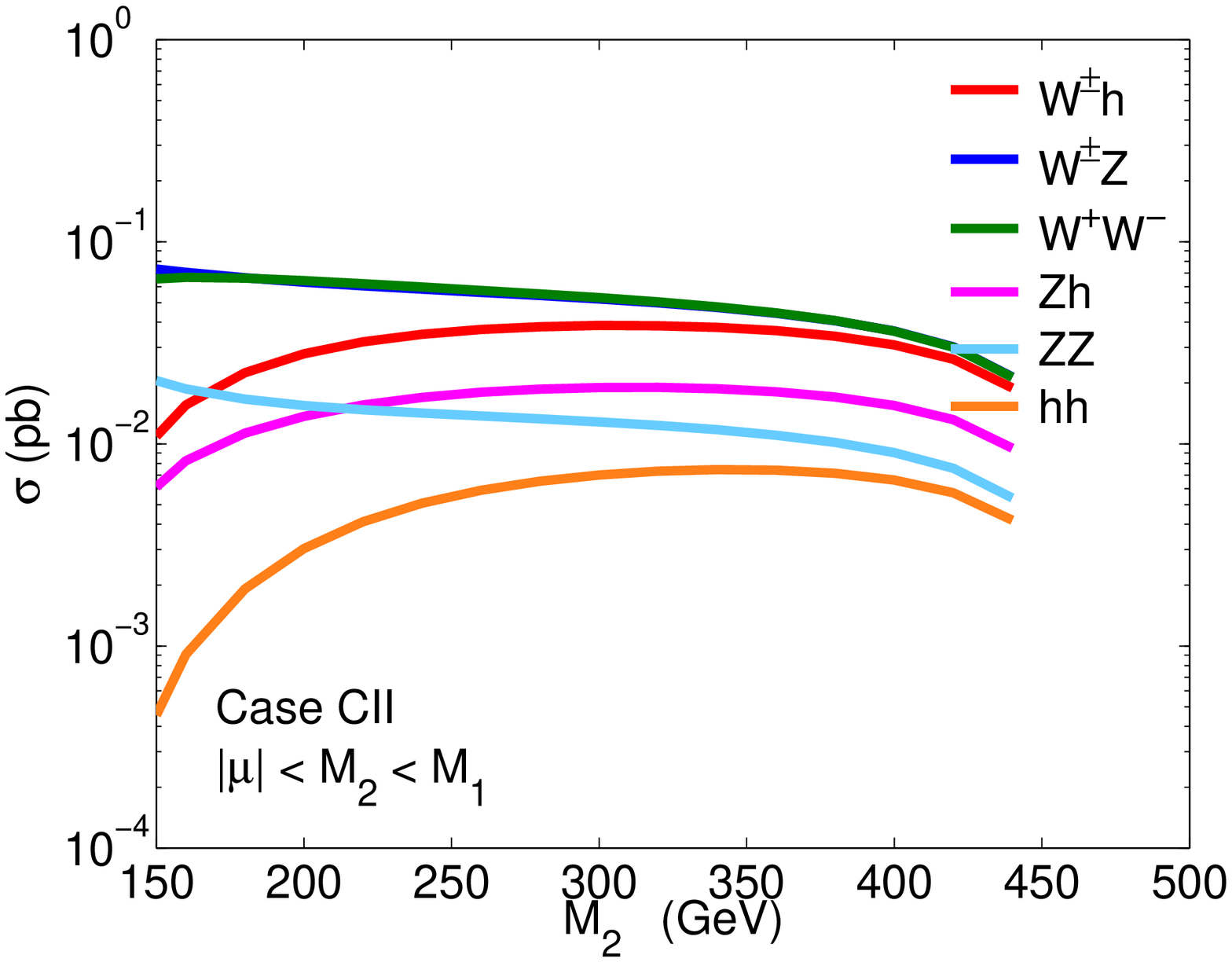}
 \caption{
 Total cross section for the chargino and neutralino pair production at the ILC for $\sqrt s = 1$ TeV  to specific final states for the four cases for the NLPS production.
 %,  (a) Case AI: versus $M_2$ for $M_{1}=100$ and $\mu= 1$ TeV, (b)  Case AII: versus $\mu$ for $M_{1}=100$ and $M_{2}=1$ TeV,
%(c) Case BII:  versus $\mu$ for $M_{2}=100$ and $M_{1}=1$ TeV,
%(d) Case CII:  versus $M_{2}$ for $\mu=100$ and $M_{1}=1$ TeV.
The parameter choices are the same as in Fig.~\ref{fig:CSBR}. 
}
\label{fig:ILC2} 
\end{figure}

Extending the above discussions, we present the total cross section for the electroweakino pair production subsequently decaying to specific final states of the electroweak bosons $XY$ of Eq.~(\ref{eq:XY}) in Fig.~\ref{fig:ILC2}. 
Once again, we note that besides observationally clean channels $W^{+}W^{-}$, $W^\pm W^\pm$,  and $WZ$, $Wh$ and $Zh$ channels contribute significantly as well. 
%Given the clean experimental environment at the ILC, 
Even the sub-dominant $hh$ mode could be identifiable.

Although not shown, one would expect that the ILC will be able to uncover the challenging decay modes with rather soft (10 GeV or less) leptons and jets in the final state, such as in the difficult cases of BI and CI with compressed mass spectrum of Wino- or Higgsino-like LSPs, because of the clean experimental environment for event reconstruction. The situation with very soft final states may be further improved by making use of the hard photon radiation (ISR) plus large missing energy, to identify the SUSY signal \cite{Tanabeetal}.
 For the same reason, the large rate signal, such as 4-jets$+\etmiss$ events could be fully utilized. The effective kinematical reconstruction and unambiguous final state identification will help to determine the properties of the electroweakinos \cite{Feng:1995zd}, and the missing LSP (dark matter) mass \cite{Baltz:2006fm}.

%%%%%%%%%%%%%%%%%%%%%%%%%%%%%%%%%
 
\section{Summary and Outlook}
\label{sec:conclusion}

Given the current null results on the SUSY searches at the LHC, namely, the non-observation of
gluinos and squarks under naive assumptions below 1 TeV, it is strongly motivated to consider the situation in which the only accessible SUSY states are the electroweakinos.
 
Within the constraints from collider searches,  we explored the gaugino and Higgsino mass parameter space and categorized the general EW SUSY parameter relations into three scenarios with six distinctive cases, as presented in Sec.~\ref{sec:cases} and for the mass relations in Fig.~\ref{fig:masses}. The four cases in B and C would naturally result in a compressed spectrum of nearly degenerate LSPs. We outline the decay patterns for the NLSPs as depicted in Fig.~\ref{fig:CaseABC_decay}, and discussed in great detail their decay branching fractions, as shown in Figs.~\ref{fig:decays_winoNLSP}$-$\ref{fig:decays_M2NLSP_muLSP}. 
In particular, we provide some insightful understanding about the decay modes in connection with the Goldstone-boson Equivalence Theorem, shown in the Appendix.

We presented the pair production cross sections for the electroweakinos via the 
DY processes (Fig.~\ref{fig:Feyn_diam}) at NLO in QCD for the 14 TeV LHC in Fig.~\ref{fig:m2mu}. The production rate can typically be of a few hundred of fb at the 200 GeV mass scale, but drop to about a few tenth of fb at a higher mass scale of $500-1000$ GeV. 
Unfortunately, the LSP multiplet production, such as in Cases BI and CI, will be difficult to observe at hadron colliders because of the mass degeneracy and the soft decay products \cite{WinoLSP_collider,HiggsinoLSP_collider}. We will thus leave them for the future exploration.
We reiterate that the electroweakino phenomenology and their searches at the LHC are largely dictated by the NLSP production and decays. 
Incorporating the dominant decays to the observable final states of a pair of gauge bosons and Higgs boson as listed in Eq.~(\ref{eq:XY}), we summarized the leading channels and their branching fractions in Table \ref{table:signal}, and showed the corresponding production cross sections in Fig.~\ref{fig:CSBR}. 
Again, the leading signal rates can reach a few hundred of fb to a few tenth of fb with the mass parameters from 200 GeV to 1 TeV. 

Of particular interest is the SM-like Higgs boson in the final state, that turned out to be one of the leading channels. We thus emphasize that with unique decay $h \to b \bar b$ and reconstructable Higgs mass variable, this channel may serve as a ``standard candle'' for the signal of the electroweakino pair production since it is clearly of a non-SM origin. The decays to gauge bosons $h\rightarrow WW^{*}, ZZ^{*}$ can also help to enhance the signal rate for the conventional SSWW and 4L search channels, although the identification to the Higgs contribution is less obvious. 
 
The current experimental bounds on the masses of the electroweakinos from the direct searches at the LEP2 (Table \ref{tab:LEP}) and the LHC (Table \ref{tab:LHC}) are summarized in Sec.~\ref{sec:analyses}. 
Extending the existing work, we explored the potential observability for future LHC run at 14 TeV with an integrated luminosity of 300 fb$^{-1}$. We first showed in Fig.~\ref{fig:LHC14bb} the sensitivities for the robust Higgs channels $Wh$ and $Zh$ with the identifiable $h\to b \bar b$ decay. The combined results for the Higgs channels were shown
in Fig.~\ref{fig:LHC14_all}(a).  For completeness, we also presented our studies in Fig.~\ref{fig:LHC14_rest}
for the four conventional channels, and combined all the results in Fig.~\ref{fig:LHC14_all}(b). 
We conclude that, for the case of a light Bino-like LSP, with the Higgs channels, we may reach the electroweakino mass scale about $M_{2},\ \mu \sim 220-270$ GeV at a $5\sigma$ sensitivity, and about $350-400$ GeV for $95\%$ C.L.~exclusion.
Combining with all the other channels, we may expect to extend the reach to the mass scale about $M_{2},\ \mu \sim 320-500$ GeV at a $5\sigma$ sensitivity, and about $480-700$ GeV for $95\%$ C.L.~exclusion.
The summary table for the achievable mass values were given in Table \ref{tab:LHC14}.
Although we only carried out the detailed analyses for Case A with a light Bino-like LSP as above, we expect that our results are equally applicable to Case BII (with light Wino-like LSPs) and Case CII (with light Higgsino-like LSPs), where the NLSPs have similar production rates and decay patterns to Case A as demonstrated in Fig.~\ref{fig:CSBR} and in Table \ref{table:signal}, with the only trade off between $W^{+}W^{-}$ and  $W^{\pm}W^{\pm}$ channels.

Due to the rather low production cross sections and large SM backgrounds, it would be nevertheless challenging for the SUSY searches at the LHC for the electroweakinos to extend the mass reach beyond what obtained above. It would be particularly difficult if the LSPs are nearly mass-degenerate while the NLSP pair production is  suppressed, like in Cases BI and CI. 
This motivates the complementary experiments at future lepton colliders with low backgrounds and easy signal reconstruction, in which the electroweakinos can be readily discovered as long as the kinematical threshold is crossed, as illustrated in Figs.~\ref{fig:ILC} and \ref{fig:ILC2}.
The electroweakino pair signals would be easily identified and their properties thoroughly studied.  

Our conclusions should still be viewed on the conservative side. First, we have not taken into account the possible contributions from the other electroweak states such as the sleptons and the heavier Higgs bosons. 
Second, there may exist other additional search channels such as the vector boson fusion (VBF) for the electroweakino production, in which its characteristic kinematics and the $t$-channel production mechanism may provide additional handles to complement the leading searches considered here.

Looking forward, the high luminosity LHC with 3000 fb$^{-1}$ would be expected to extend the 5$\sigma$ electroweakino reach to a mass generically of 800 GeV assuming a $100\%$ branching fraction to the gauge bosons \cite{ATLAS_CMS_3L_future}. 
It would be a pressing issue to address to what extent one would be able to uncover the observationally difficult scenarios like Cases BI and CI, where the lower lying electroweakinos are in a compressed LSP spectrum and the NLSPs may not be copiously produced. 
Furthermore, if a multiple TeV lepton collider is ever available \cite{Lebrun:2012hj,Alexahin:2013ojp}, it would readily cover to a mass scale about a half of the center of mass energy. 
 
\acknowledgments
We thank Xerxes Tata for a careful reading of the manuscript, and Howard Baer and Kuver Sinha for discussions. 
We also thank the Aspen Center for Physics for hospitality when this work was initiated. ACP is supported by NSF under grant  1066293.
The work of T.H.~was supported in part by the U.S.~Department of Energy  under grant DE-FG02-95ER40896 and in part by the PITT PACC, the work of S.P.~was supported in part by the Department of Energy grant DE-FG02-90ER40546 and the FNAL LPC Fellowship, and that of S.S.~was supported by the  Department of Energy  under Grant~DE-FG02-04ER-41298. 

 %%%%%%%%%%%%%%%%%%%%%%%%%%%%%%%%%%%%

\appendix

\section{NLSP Decays and the Goldstone-boson Equivalence Theorem}

When the NLSP mass is large in comparison to $m_Z$, the Goldstone-boson Equivalence theorem \cite{Lee:1977eg} becomes an adequate tool to understand the nature of the NLSP decays. We present some approximate formulae and provide some discussions.  A collection of the partial decay widths of neutralinos and charginos can be found in the earlier works \cite{Gunion:1987kg}.
%\cite{Gunion:1987yh, Djouadi:2001fa}.

\begin{itemize}
{\bf \item{Scenario A:}  $M_{1} < M_{2},\ |\mu|$ }
\end{itemize}

The relative size of   ${\rm Br}(\chi_2^0 \rightarrow \chi_1^0 h)$ and ${\rm Br}(\chi_2^0 \rightarrow \chi_1^0 Z)$ can be understood with the help of the Goldstone-boson Equivalence Theorem. 
%Goldstone equivalent principle.   
In Case AI with $M_2 - M_1 \gg m_{Z}$,
the decay of $\chi_2^0 \rightarrow \chi_1^0 Z$ is dominantly to the longitudinal polarization of the $Z$ boson, which is related to the Goldstone modes of $H_u^0$ and $H_d^0$. 
For $M_2 - M_1 \gg m_Z$ and $|\mu \pm M_{1,2}| \gg m_Z$,   the partial decay widths of $\chi_2^0 \rightarrow \chi_1^0 h$ and $\chi_2^0 \rightarrow \chi_1^0 Z$ are given approximately by the following formulae:   
\bea
&&{\Gamma(\chi_2^0 \rightarrow \chi_1^0 h)} \approx C_{AI} \frac{1}{8 \pi} \frac{p_h}{M_2^2} { \left( 2 s_{2 \beta} + \frac{M_2}{\mu} \right)^2}
\left[ {(M_2 + M_1)^2 - m_h^2}\right],  
\label{eq:AI_chi20h}\\
&&{\Gamma(\chi_2^0 \rightarrow \chi_1^0 Z)} \approx C_{AI} \frac{1}{8 \pi} \frac{p_Z}{M_2^2}  {\left(c_{2 \beta}\frac{M_2}{\mu}\right)^2}\left[{(M_2 - M_1)^2 - m_Z^2}\right], 
\label{eq:AI_chi20Z}
\eea
where  $C_{AI}= \frac{e^2}{4} ( \frac{m_Z}{\mu} )^2$ and $p_{h}$ ($p_{Z}$) is the momentum for $h$ ($Z$) in the rest frame of $\chi_2^0$.
 For large $\tan\beta\gg 4 \mu/M_2$ such that $2 s_{2 \beta} \ll M_2/\mu$,  
the second term in the parenthesis of Eq.~(\ref{eq:AI_chi20h}) dominates for decay of $\chi_1^0h$ channel.   Relative size of the $h$ and $Z$ decay channel is almost independent of $\tan\beta$, determined completely by the ratio
$\left[ {(M_2 + M_1)^2 - m_h^2}\right]/\left[ {(M_2 - M_1)^2 - m_Z^2}\right]$.  For relatively small $1 \lesssim \tan\beta  \ll  4 \mu/M_2$, the first term in the parenthesis dominants.  The additional suppression of $(M_2/\mu)^2$ in $Z$-channel decrease the size of $\chi_2^0 \rightarrow \chi_1^0 Z$ channel.   Note however, for the case of negative $\mu$,  two term in the parenthesis of Eq.~(\ref{eq:AI_chi20h}) could cancel each other, leading to the suppression of the branching fraction for the  $\chi_1^0 h$ channel.

In Case AII with Higgsino NLSPs, the decay of $\chi_2^0 \rightarrow \chi_1^0 h$ occurs at the leading order via unsuppressed $\tilde{H}_{u,d}^0 - \tilde{B}^0 -H_{u,d}^0$ coupling.
For $|\mu| - M_1 \gg m_{Z}$, 
 $\chi_2^0 \rightarrow \chi_1^0 Z$ again is dominated by the longitudinal mode of the $Z$ boson. Under the limit of $|\mu \pm M_1| \gg m_Z$, the Goldstone-boson Equivalence theorem relates the partial decay widths of $\chi_2^0 \approx \frac{1}{\sqrt{2}} ({\tilde{H}}_d^0-{\tilde{H}}_u^0)$  as
 \bea
&&{\Gamma(\chi_2^0 \rightarrow \chi_1^0 h)} \approx    C_{AII}  \frac{1}{8 \pi} \frac{p_h}{\mu^2} ({s_\beta + c_\beta})^2
\left[ {(\mu + M_1)^2 - m_h^2}\right],
\label{eq:AII_chi20h}\\
&&{\Gamma(\chi_2^0 \rightarrow \chi_1^0 Z)} \approx    C_{AII}  \frac{1}{8 \pi} \frac{p_Z}{\mu^2}   ({s_\beta - c_\beta} )^2 \left[{(\mu - M_1)^2 - m_Z^2}\right], 
\label{eq:AII_chi20Z}
\eea
where  $C_{AII}= \frac{e^2}{8 c_W^2} $.  
%The relative sign difference in $s_\beta \pm c_\beta$ in $h$ and $Z$ channel can be traced back to the composition of $h$ and $G^0$ (the Goldstone mode being absorbed by $Z$) in terms of the real and imaginary part of $H_u^0$ and $H_d^0$:
%\bea
%h = -\sqrt{2} &&(s_\beta \ {\rm Re} (H_u^0) + c_\beta\  {\rm Re}(H_d^0) ), \\
%G^0= \sqrt{2} &&(s_\beta\  {\rm Im}(H_u^0) - c_\beta\  {\rm Im}(H_d^0) ).
%\eea
For $\tan\beta>1$ and positive $\mu$, $M_1$,  the $\chi_1^0h$ channel is enhanced relatively to the $Z$ channel by both the $ (s_\beta + c_\beta)^2/(s_\beta - c_\beta)^2$ factor, as well as the mass terms inside the square bracket.

The third neutralino $\chi_3^0 \approx \frac{1}{\sqrt{2}} ({\tilde{H}}_d^0+{\tilde{H}}_u^0)$ exhibits a similar decay pattern, with the role of $h$ and $Z$ switched: 
\bea
&&{\Gamma(\chi_3^0 \rightarrow \chi_1^0 h)} \approx C_{AII}  \frac{1}{8 \pi} \frac{p_h}{\mu^2}  ({s_\beta - c_\beta})^2
\left[ {(\mu - M_1)^2 - m_h^2}\right],
\label{eq:AII_chi30h}\\
&&{\Gamma(\chi_3^0 \rightarrow \chi_1^0 Z)} \approx C_{AII} \frac{1}{8 \pi} \frac{p_Z}{\mu^2}  ({s_\beta + c_\beta} )^2 \left[{(\mu + M_1)^2 - m_Z^2}\right].
\label{eq:AII_chi30Z}
\eea
The exchange of $s_\beta \pm c_\beta \leftrightarrow s_\beta \mp c_\beta$ in $\chi_{2,3}^0$ decay   is due to the  composition of $\chi_{2,3}^0$ as $\frac{1}{\sqrt{2}} ({\tilde{H}}_d^0 \mp {\tilde{H}}_u^0)$.
The exchange of $\mu \pm M_1 \leftrightarrow \mu \mp M_1$ can be traced back to the mass eigenvalues of $\chi_{2,3}^0$ being $\pm \mu$, respectively.    In the limit of large $\tan\beta$ and $|\mu \pm M_1| \gg m_Z$ such that all final states particles are effectively massless comparing to the parent particle,  ${\rm Br}(\chi_{2,3}^0 \rightarrow \chi_1^0 h) \approx {\rm Br}(\chi_{2,3}^0 \rightarrow \chi_1^0 Z) \approx $ 50\%.  While for $\tan\beta \rightarrow 1$, one of the $h$ or $Z$ channel is highly suppressed while the other channel is greatly enhanced.

%%%%%%%%%%%

\begin{itemize}
\item{\bf Scenario B:} $M_{2} < M_{1},\ |\mu|$
\end{itemize}

Under the limit of $M_1 - M_2 \gg m_Z$, $|\mu \pm M_{1,2}| \gg m_Z$, the partial decay widths to various final states in Case BI follow the  simplified formulae: 
  \bea
\Gamma(\chi_{2}^0 \rightarrow \chi_1^0 h)  &\approx& C_{BI}   \frac{1}{8 \pi} \frac{p_h}{M_1^2}  { \left( 2 s_{2 \beta} + \frac{M_1}{\mu} \right)^2}   \left[(M_1 + M_2)^2 - m_h^2\right],  
\label{eq:BI_chi20h}
\\
\Gamma(\chi_{2}^0 \rightarrow \chi_1^0 Z)&\approx& C_{BI}   \frac{1}{8 \pi} \frac{p_Z}{M_1^2}  {\left(c_{2 \beta}\frac{M_1}{\mu}\right)^2}   \left[(M_1 - M_2)^2 - m_Z^2\right],  
\label{eq:BI_chi20Z}
\\
\Gamma(\chi_{2}^0 \rightarrow \chi_1^+ W^-)&=&\Gamma(\chi_{2}^0 \rightarrow \chi_1^- W^+)  \\ \nonumber 
&\approx& C_{BI}   \frac{1}{8 \pi} \frac{p_W}{M_1^2}  (c_{\beta}^4 + s_\beta^4)  {\left(\frac{M_1}{\mu}\right)^2}    2 \left[M_1^2 + M_2^2 - m_W^2\right], 
\label{eq:BI_chi20W}
\eea
where    $C_{BI}= \frac{e^2}{4} ( \frac{m_Z}{\mu} )^2$.
In the limit of large $\tan\beta$, the approximate relation holds: 
\beq
\Gamma_{\chi_1^+W^-}=\Gamma_{\chi_1^-W^+}\approx \Gamma_{\chi_1^0Z} +  \Gamma_{\chi_1^0h}.
\eeq
%Flipping the sign of $\mu$ switches the relative decay branching fractions of the $Z$ and $h$ channel for small $\tan\beta$, similar to Case AI.

%%%%%%%%%%%%%%%%

In Case BII under the limit of   $|\mu \pm M_2| \gg m_Z$, 
the partial decay widths of $\chi_2^\pm$ to  various final states  follow the  simplified formulae: 
 \bea
&&\Gamma(\chi_{2}^\pm \rightarrow \chi_1^\pm h)\approx  C_{BII}  \frac{1}{8 \pi} \frac{p_h}{\mu^2}  2 \left[ \mu^2 + M_2^2 - m_h^2\right],  
\label{eq:BII_chi2pmh}
\\
&&\Gamma(\chi_{2}^\pm \rightarrow \chi_1^\pm Z)\approx C_{BII} \frac{1}{8 \pi} \frac{p_Z}{\mu^2}  2 \left[ \mu^2 + M_2^2 - m_Z^2\right],  
\label{eq:BII_chi2pmZ}
\\
&&\Gamma(\chi_{2}^\pm \rightarrow \chi_1^0 W^\pm) \approx C_{BII}  \frac{1}{8 \pi} \frac{p_W}{\mu^2}  2 \left[ \mu^2 + M_2^2 - m_W^2\right], 
\label{eq:BII_chi2pmW}
\eea
where $C_{BII}=\frac{e^2}{8 s_W^2}$.    In the   limit of large Higgsino mass, 
${\rm Br}(\chi_{2}^\pm \rightarrow \chi_1^\pm h) \approx {\rm Br}(\chi_{2}^\pm \rightarrow \chi_1^\pm Z) \approx {\rm Br}(\chi_{2}^\pm \rightarrow \chi_1^0 W^\pm) \approx$ 33 \%.

The partial decay widths of $\chi_{2,3}^0\approx \frac{1}{\sqrt{2}} ({\tilde{H}}_d^0 \pm {\tilde{H}}_u^0)$ to various final states follow the  simplified formulae: 
\bea
&&\Gamma(\chi_{2,3}^0 \rightarrow \chi_1^0 h)\approx C_{BII}  \frac{1}{8 \pi} \frac{p_h}{\mu^2} (s_\beta \mp c_\beta)^2\left[(\mu \mp M_2)^2 - m_h^2\right],  
\label{eq:BII_chi230h}
\\
&&\Gamma(\chi_{2,3}^0 \rightarrow \chi_1^0 Z)\approx C_{BII}  \frac{1}{8 \pi} \frac{p_Z}{\mu^2}(s_\beta \pm c_\beta)^2\left[(\mu \pm M_2)^2 - m_Z^2\right],  
\label{eq:BII_chi230Z}
\\
&&\Gamma(\chi_{2,3}^0 \rightarrow \chi_1^+ W^-)=\Gamma(\chi_{2,3}^0 \rightarrow \chi_1^- W^+)\approx C_{BII}  \frac{1}{8 \pi} \frac{p_W}{\mu^2} 2 \left[\mu^2 + M_2^2 - m_W^2\right].
\label{eq:BII_chi230W}
\eea
In the limit of large $\tan\beta$ and very heavy Higgsino mass, ${\rm Br}(\chi_{2,3}^0 \rightarrow \chi_1^0 h) \approx {\rm Br}(\chi_{2,3}^0 \rightarrow \chi_1^0 Z) \approx \frac{1}{4} {\rm Br}(\chi_{2,3}^0 \rightarrow \chi_1^\pm W^\mp) \approx$ 16.7 \%.

%%%%%%%%%%%%%%%%

\begin{itemize}
\item{\bf Scenario C:} $|\mu| < M_{1},\ M_{2}$
\end{itemize}

Under the limit of $|M_1 \pm \mu|\gg m_Z$ for Case CI, the partial decay widths to various final states follow the simplified formulae for $\chi_{1,2}^0 \approx \frac{1}{\sqrt{2}} ({\tilde{H}}_d^0 \mp{\tilde{H}}_u^0)$:
\bea
&&\Gamma(\chi_{3}^0 \rightarrow \chi_{1,2}^0 h)\approx C_{CI}  \frac{1}{8 \pi} \frac{p_h}{M_1^2} (s_\beta \pm c_\beta)^2\left[(M_1 \pm \mu)^2 - m_h^2\right],  
\label{eq:CaseCI_chi30h}
\\
&&\Gamma(\chi_{3}^0 \rightarrow \chi_{1,2}^0 Z) \approx C_{CI}  \frac{1}{8 \pi} \frac{p_Z}{M_1^2}(s_\beta \mp c_\beta)^2\left[(M_1 \mp \mu)^2 - m_Z^2\right],  
\label{eq:CaseCI_chi30Z}
\\
&&\Gamma(\chi_{3}^0 \rightarrow \chi_1^+ W^-)=\Gamma(\chi_{3}^0 \rightarrow \chi_1^- W^+) \approx C_{CI}  \frac{1}{8 \pi} \frac{p_W}{M_1^2} 2 \left[M_1^2 + \mu^2 - m_W^2\right], 
\label{eq:CaseCI_chi30W}
\eea
where $C_{CI}= \frac{e^2}{8 c_W^2}$. 
The following relation between the partial decay width (and decay branching fractions as well)  holds for $\chi_3^0$:
\beq
\Gamma_{\chi_1^+W^-}=\Gamma_{\chi_1^-W^+}\approx\Gamma_{\chi_1^0Z}+\Gamma_{\chi_1^0h}
\approx \Gamma_{\chi_2^0Z}+\Gamma_{\chi_2^0h} \approx \Gamma_{\chi_1^0h}+\Gamma_{\chi_2^0h} \approx \Gamma_{\chi_1^0Z}+\Gamma_{\chi_2^0Z}. 
\eeq 
For large Bino mass $M_1$, the branching fractions approach the asymptotic value  
 ${\rm Br}(\chi_{3}^0 \rightarrow \chi_{1}^0 h) +{\rm Br}(\chi_{3}^0 \rightarrow\chi_{2}^0 h ) \approx {\rm Br}(\chi_{3}^0 \rightarrow \chi_{1}^0 Z)+{\rm Br}(\chi_{3}^0 \rightarrow\chi_{2}^0 Z) \approx \frac{1}{2} {\rm Br}(\chi_{3}^0 \rightarrow \chi_1^\pm W^\mp) \approx$ 25 \%.

The approximate expression for $\chi_2^\pm$ decay in Case CII under the limit of $|M_2 \pm \mu| \gg m_Z $ is:
\bea
&&\Gamma(\chi_{2}^\pm \rightarrow \chi_1^\pm h)\approx  C_{CII}  \frac{1}{8 \pi} \frac{p_h}{\mu^2}  2 \left[  M_2^2 + \mu^2 - m_h^2\right],  
\label{eq:CII_chi2pmh}
\\
&&\Gamma(\chi_{2}^\pm \rightarrow \chi_1^\pm Z)\approx C_{CII} \frac{1}{8 \pi} \frac{p_Z}{\mu^2}  2 \left[ M_2^2 + \mu^2  - m_Z^2\right],  
\label{eq:CII_chi2pmZ}
\\
&&\Gamma(\chi_{2}^\pm \rightarrow \chi_1^0 W^\pm)=\Gamma(\chi_{2}^\pm \rightarrow \chi_2^0 W^\pm) \approx C_{CII}  \frac{1}{8 \pi} \frac{p_W}{\mu^2}  2 \left[ M_2^2 + \mu^2  - m_W^2\right],  
\label{eq:CII_chi2pmW}
  \eea
where   $C_{CII}= \frac{e^2}{8 s_W^2}$ .
For large Wino mass, the branching fractions approach the asymptotic value  
 ${\rm Br}(\chi_{2}^\pm \rightarrow \chi_1^\pm h) \approx {\rm Br}(\chi_{2}^\pm \rightarrow \chi_1^\pm Z) \approx  \frac{1}{2} ({\rm Br}(\chi_{2}^\pm \rightarrow \chi_{1}^0 W) +{\rm Br}(\chi_{2}^\pm \rightarrow \chi_{2}^0 W))    \approx$ 25 \%.

The expression for the $\chi_3^0$ decay in the Case CII is very similar to that in Case CI, with $C_{CII}= \frac{e^2}{8 s_W^2}$   and the replacement of $M_1 \leftrightarrow M_2$.

%%%%%%%%%%%%

\end{document}